\documentclass[aps,showpacs,twocolumn,dvips]{revtex4}
\usepackage{feynmp}
\usepackage{amssymb}
\usepackage{epsfig}
\usepackage{graphicx}
\usepackage{subfigure}
\usepackage{mathrsfs}
\usepackage{hyperref}
\usepackage{cleveref}
\usepackage{cases}
\usepackage{bbm}
\usepackage{xcolor}
\usepackage{tabularx}
\usepackage{array}
\usepackage{multirow}
\usepackage{CJK}
\usepackage{lmodern}
\usepackage{textcomp}
\usepackage{makecell}
\usepackage{threeparttable}

\begin{document}



\title{Fermion-fermion interaction driven instability and criticality of quadratic
band crossing systems with the breaking of time-reversal symmetry}

\date{\today}

\author{Ya-Hui Zhai}
\affiliation{Department of Physics, Tianjin University, Tianjin 300072, People's Republic of China}

\author{Jing Wang}
\altaffiliation{Corresponding author: jing$\textunderscore$wang@tju.edu.cn}
\affiliation{Department of Physics, Tianjin University, Tianjin 300072, People's Republic of China}

\begin{abstract}
We carefully study how the fermion-fermion interactions
affect the low-energy states of a two-dimensional spin-$1/2$ fermionic system on the
kagom\'{e} lattice with a quadratic band crossing point. With
the help of the renormalization group approach, we
can treat all kinds of fermionic interactions on the the same footing
and then establish the coupled energy-dependent flows of fermionic interaction
parameters via collecting one-loop corrections, from which a number of
interesting results are extracted in the low-energy regime.
At first, various sorts of fermion-fermion interactions furiously
compete with each other and are inevitably attracted by certain fixed point in
the parameter space, which clusters into three qualitatively distinct regions
relying heavily upon the structure parameters of materials. In addition, we notice that
an instability accompanied by some symmetry breaking is triggered around
different sorts of fixed points. Computing and comparing susceptibilities
of twelve potential candidates indicates that charge density wave always
dominates over all other instabilities. Incidently, there exist
several subleading ones including the $x$-current, bond density, and chiral
plus s-wave superconductors. Finally, we realize that strong fluctuations nearby the leading
instability prefer to suppress density of states and specific heat as well compressibility of
quasiparticles in the lowest-energy limit.
\end{abstract}


\maketitle


\section{Introduction}

Past two decades have witnessed a phenomenally rapid
development of semimetal materials~\cite{Lee2005Nature,
Neto2009RMP,Fu2007PRL,Roy2009PRB,Moore2010Nature,Hasan2010RMP,Qi2011RMP,
Sheng2012book,Bernevig2013book,Herbut2018Science,Armitage2018RMP,Roy2018RRX}
that feature well-known discrete Dirac points accompanied by gapless
quasi-particle excitations and linear energy
dispersions along two or three directions~\cite{Lee2005Nature,
Neto2009RMP,Fu2007PRL,Roy2009PRB,Moore2010Nature,Hasan2010RMP,
Qi2011RMP,Sheng2012book,Bernevig2013book,Korshunov2014PRB,Hung2016PRB,
Nandkishore2013PRB,Potirniche2014PRB,Nandkishore2017PRB,Sarma2016PRB,
Herbut2018Science,Alavirad2016,Slager2016,Roy2016SR}. The list consists of Dirac
semimetals~\cite{Wang2012PRB,Young2012PRL,Steinberg2014PRL,Liu2014NM,Liu2014Science,
Xiong2015Science} and Weyl semimetals~\cite{Neto2009RMP,Burkov2011PRL,Yang2011PRB,
Wan2011PRB,Huang2015PRX,Xu2015Science,Xu2015NP,Lv2015NP,Weng2015PRX}
as well as their semi-Dirac cousins~\cite{Hasegawa2006RRB,Katayama2006JPSJ,Dietl2008PRL,Pardo2009PRL, Delplace2010PRB,Wu2014OE}. In recent years, interest has gradually shifted from linear-dispersion toward
quadratic-dispersion fermi materials with up and down bands parabolically touching at
certain quadratic band crossing point (QBCP) for both two~\cite{Chong2008PRB,Fradkin2008PRB,
Fradkin2009PRL,Cvetkovic2012PRB,Murray2014PRB,Herbut2012PRB,Mandal2019CMP,Zhu2016PRL,
Vafek2010PRB,Yang2010PRB,Wang2017PRB,Wang2020-arxiv,Roy2020-arxiv,Janssen2020PRB,Shah2011.00249} and
three dimensions~\cite{Nandkishore2017PRB,Luttinger1956PR,
Murakami2004PRB,Janssen2015PRB,Boettcher2016PRB,Janssen2017PRB,Boettcher2017PRB,
Mandal2018PRB,Lin2018PRB,Savary2014PRX,Savary2017PRB,Vojta1810.07695,
Lai2014arXiv,Goswami2017PRB,Szabo2018arXiv,Foster2019PRB,Wang1911.09654}.
Compared to their Dirac/Weyl counterparts with the vanishment of density of state (DOS)
at Dirac points, 
two-dimensional (2D) QBCP materials
attract more attention and become one of the most active subjects~\cite{Chong2008PRB,Fradkin2008PRB,
Fradkin2009PRL,Cvetkovic2012PRB,Murray2014PRB,Herbut2012PRB,Mandal2019CMP,Janssen2020PRB}. The main
reasons are ascribed to the finite density of states at the Fermi surface together with its unique
gapless quasiparticles (QPs) from discrete QBCPs developed by the crossings of up
and down parabolical bands, leading to the possibility of weak
coupling interaction-driven instability~\cite{Fradkin2009PRL,Murray2014PRB,
Wang2017PRB,Wang2020-arxiv}. These 2D QBCP materials are suggested to be realized on
some collinear spin density wave state~\cite{Chern2012PRL}, Lieb lattice~\cite{Tsai2015NJP}, checkerboard~\cite{Fradkin2008PRB,Murray2014PRB}
and kagom\'{e} lattices~\cite{Huse2003PRB,Fradkin2009PRL,Janssen2020PRB}
with distinct kinds of symmetries under
point group consideration~\cite{Fradkin2009PRL,Cvetkovic2012PRB,Murray2014PRB}.

What is more, 2D QBCP systems are allowed to either host the
time-reversal symmetry (TRS) or present
the TRS breaking depending upon concrete lattices. On one side, the free Hamiltonian
of 2D QBCP semimetals rooted in the checkerboard lattice is protected by TRS~\cite{Fradkin2009PRL,Cvetkovic2012PRB,Murray2014PRB}. With respect to these materials,
unconventional band structures and gapless QPs in tandem with weak couplings yield to
many interesting behaviors in the low-energy regime~\cite{Fradkin2009PRL,
Vafek2010PRB,Yang2010PRB,Murray2014PRB,Venderbos2016PRB,Wu2016PRL,
Zhu2016PRL,Wang2017PRB}. In particular, quantum anomalous Hall (QAH)
and quantum spin Hall (QSH) can be generated by fermion-fermion repulsive
interactions on the checkerboard lattice~\cite{Fradkin2009PRL,Murray2014PRB} or two-valley
bilayer graphene with QBCPs~\cite{Vafek2010PRB,Yang2010PRB}, as well as their low-energy stabilities
under the impacts of impurity scatterings are also examined~\cite{Wang2017PRB,Wang2020-arxiv}.
On the other side, the 2D noninteracting QBCP model that originates from the kagom\'{e} lattice
might be TRS breaking~\cite{Huse2003PRB}. Although this case is equipped with the similar QBCP and quadratic
dispersion, its low-energy physics has hitherto been insufficiently
explored. Given the basic structure of free Hamiltonian is quite far away from
its checkerboard counterpart, a number of tempting questions concerning the kagom\'{e}-lattice
version naturally arise: whether and how the low-energy physical properties are influenced by
fermion-fermion interaction? Whether they can activate the instabilities? Which states
are the suitable candidates and what are the critical behaviors in the vicinity of
potential instabilities? It would be instructive to deepen our understanding
of the 2D QBCP materials once these inquiries are properly answered.

Inspired by these, we within this work put our focus on a 2D QBCP spin-$1/2$ fermionic system
on the kagom\'{e} lattice and investigate its low-energy fate in the presence of
16 types of marginal fermion-fermion interactions. In order to treat all these
physical ingredients on an equal footing, we employ the momentum-shell renormalization-group (RG)
approach~\cite{Shankar1994RMP,Wilson1975RMP,Polchinski9210046},
which is a powerful tool to refine and characterize the hierarchical
physics in the simultaneous presence of various types of interactions.
Practicing the standard procedures of RG framework gives rise to the
one-loop energy-dependent evolutions of all fermion-fermion interaction parameters.
With the help of these RG flows that encode the energy-dependent physics,
several intriguing critical behaviors are extracted in the low-energy regime.

At first, we, with the help of the numerical analysis of RG equations, are
aware that the fermion-fermion interactions are of close relevance to
each other and evolve towards strong couplings due to their intimate interplay
at certain energy scale. Considering the degenerate trajectories of several kinds of
interactions, we only need to put our focus on six nontrivial
fermionic couplings that flow independently. To overcome the strong
couplings and make our study perturbative, we follow the strategy put forward in Refs.~\cite{Vafek2010PRB,Murray2014PRB} and then rescale these six nontrivial parameters
by a non-sign changed coupling to obtain their relative evolutions together
with relatively fixed points which are conventionally in charge of
low-energy properties.

Next, we figure out that the concrete values of relatively fixed points are insusceptible to initial
conditions of fermion-fermion interactions but instead primarily determined by
two structure parameters $d_1$ and $d_3$ in the Hamiltonian. Adjusting
the ratio between these two quantities yields to three qualitative different
regions at which the relatively fixed points exhibit diverse traits.
Since the relatively fixed point is related to some instability that is
always accompanied by certain symmetry breaking and thus some phase
transition~\cite{Cvetkovic2012PRB,Murray2014PRB,Wang2017PRB, Maiti2010PRB, Altland2006Book,Vojta2003RPP,Halboth2000RPL,Halboth2000RPB, Eberlein2014PRB,Chubukov2012ARCMP,Chubukov2016PRX}, it is therefore of intense interest to
identify which is the leading instability for the relatively
fixed point residing in different regions. To this end, we introduce
the source terms of twelve kinds of potential candidates and evaluate
their related susceptibilities approaching a relatively fixed point~\cite{Cvetkovic2012PRB,Murray2014PRB,Wang2017PRB}.
Carrying out both the theoretical and numerical analysis indicates
that the charge density wave always takes a leading
role in the whole region. In addition, four subleading ones involving
the $x$-current, bond density, and chiral or s-wave superconductors largely
hinge upon the relatively fixed points.

Moreover, the critical properties of physical implications are briefly studied around the leading
instability. It is worth pointing out that these quantities are sensitive to
the fluctuation of order parameter triggered by the dominant instability. We notice that ferocious fluctuations induced by the development of
charge density wave are generally detrimental to the density of states and specific heat
as well compressibility of quasiparticles.  Especially, they are all substantially reduced
and even drive to zero as the leading instability is
accessed~\cite{Wang2012PRD, Altland2006Book, Altland2002PR,
Ludwig1994PRB,Xu2008PRB,Zhang2016NJP,Zhao2017PRB}.

The rest of paper is organized as follows. In Sec~\ref{Sec_model}, we present
the microscopic model for a 2D QBCP spin-$1/2$ electronic system on the kagom\'{e} lattice
and construct our effective action consisting of both free terms and all marginal
fermion-fermion interactions. Starting from this effective theory, we within
Sec.~\ref{Sec_RG_analysis} carry out one-loop momentum-shell
RG analysis and derive the coupled RG equations of all fermionic interaction parameters.
By virtue of numerical analysis of RG evolutions,
Sec.~\ref{Sec_RFP} is followed to seek and classify the underlying fixed points
in the low-energy regime. In addition, we bring out the source terms in Sec.~\ref{Sec_instab}
and pinpoint the dominant and subleading instabilities nearby distinct types of
relatively fixed points. Furthermore, several critical behaviors of physical quantities
activated by ferocious fluctuations are concisely investigated in Sec.~\ref{Sec_implication}.
Finally, Sec.~\ref{Sec_summary} briefly summarizes our central points.

\vspace{0.8cm}

\section{Effective theory and RG analysis}

At the outset, we are going to present the microscopic model and build
the effective theory in the low-energy regime as well as
establish the coupled energy-dependent
flow equations of all marginal fermion-fermion couplings by carrying out
the standard momentum-shell RG framework~\cite{Shankar1994RMP,Wilson1975RMP,Polchinski9210046}.

\subsection{Effective theory}\label{Sec_model}

\begin{figure}
\hspace{-1.5cm}
\epsfig{file=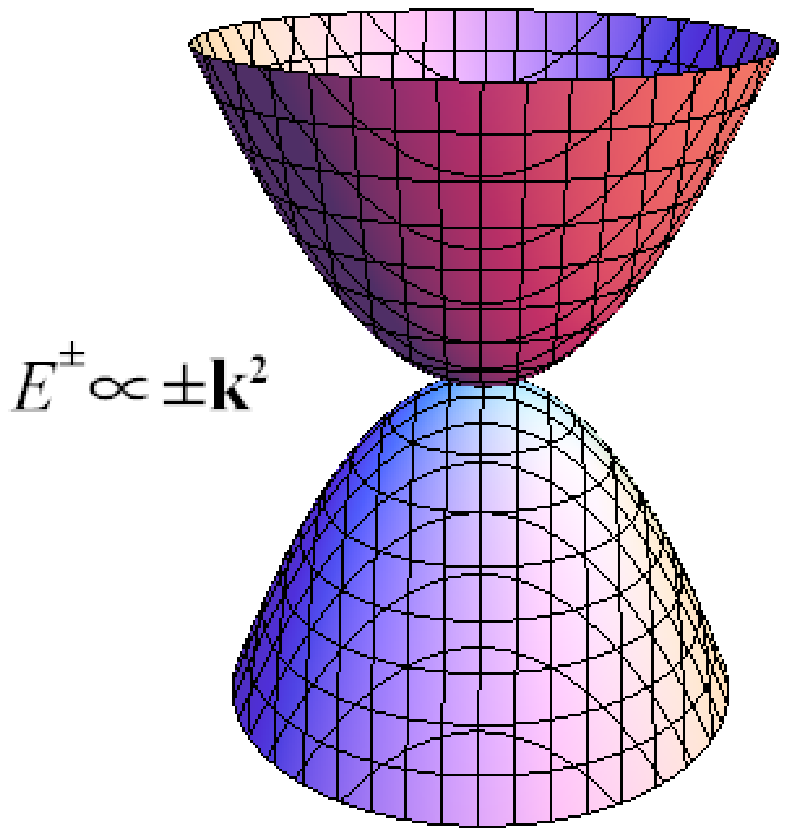,height=5.3cm,width=6.35cm}\\\vspace{-4.36cm}
\hspace{4.61cm}
\includegraphics[width=0.8in]{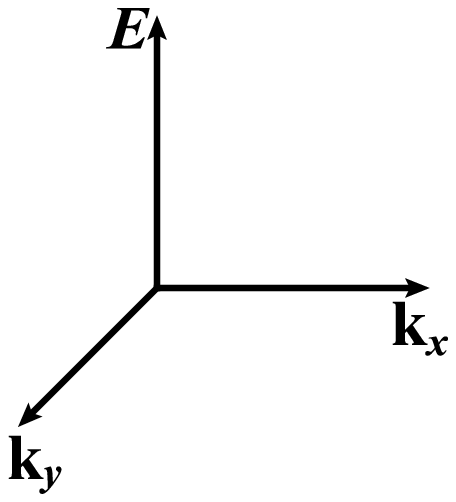}\\
\vspace{1.9cm}
\caption{(Color online) Schematic illustration of the 2D QBCP band
structure with a quadratic band crossing point parabolically touched
by the up and down energy bands.}\label{The_dispersion_of_QBCP_system}
\end{figure}

We hereafter concentrate on the 2D spin-$1/2$ electronic system stemming from
the kagom\'{e} lattice that is characterized by a QBCP at which
up and down energy bands parabolically meet~\cite{Fradkin2009PRL,Huse2003PRB}.
Accordingly, its non-interacting Hamiltonian that captures the low-energy
fermionic excitations nearby the QBCP can be written as~\cite{Fradkin2009PRL,Huse2003PRB},
\begin{eqnarray}
\mathrm{H}_{0}=\sum_{|\mathbf{k}|<\Lambda}\Psi^{\dag}_{\mathbf{k}}
\mathcal{H}_{0}(\mathbf{k})\Psi_{\mathbf{k}},\label{Eq_H_0}
\end{eqnarray}
where $\Lambda$ serves as the momentum cutoff
and the Hamiltonian density is cast as~\cite{Huse2003PRB}
\begin{eqnarray}
\mathcal{H}_{0}(\mathbf{k})=d_{3}\mathbf{k}^{2}\Sigma_{03}
+d_{1}(k_{x}^{2}-k_{y}^{2})\Sigma_{01}+d_{2}k_{x}k_{y}\Sigma_{02},
\label{Eq_h_0}
\end{eqnarray}
with $d_1$, $d_2$, and $d_3$ being microscopic structure parameters of continuum Hamiltonian.
Hereby, $\Psi^\dagger_{\mathbf{k}}=(c^{\dag}_{1\uparrow},
c^{\dag}_{1\downarrow}, c^{\dag}_{2\uparrow},c^{\dag}_{2\downarrow})$
is designated as a four-component spinor to specify the low-energy quasiparticles coming from
two energy bands and unequal spins~\cite{Fradkin2009PRL,Huse2003PRB}.
In addition, the $4\times4$ matrix $\Sigma_{\mu\nu}\equiv\tau_\mu\otimes\sigma_\nu$,
where $\tau_\mu$ and $\sigma_\nu$ with $\mu,\nu=0,1,2,3$ represent
Pauli matrices $\tau_{1,2,3}$, $\sigma_{1,2,3}$ and identity
matrix $\tau_0$, $\sigma_0$, is employed to act on both the spin space
and lattice space. Diagonalizing the free Hamiltonian~(\ref{Eq_h_0})
straightforwardly gives rise to the parabolical energy eigenvalues,
\begin{eqnarray}
E\!=\!\left\{\pm d_{3}\mathbf{k}^{2}, \pm\mathbf{k}^{2}\sqrt{d_{3}^{2}
+d_{1}^{2}\cos^{2}2\theta_{\mathbf{k}}
+\frac{1}{4}d_{2}^{2}\sin^{2}2\theta_{\mathbf{k}}}\right\}\!,
\label{Eigenvalues}
\end{eqnarray}
where $\pm$ specify the upward and downward energy bands
that quadratically touch at $\mathbf{k}=0$ as schematically shown
in Fig.~\ref{The_dispersion_of_QBCP_system} and
$\theta_{\mathbf{k}}\equiv\arctan (k_{y}/k_{x})$ measures the direction
of momentum $\mathbf{k}$. It is of particular interest to point out
that the presence of $\theta_{\mathbf{k}}$ points to rotational asymmetry of the
QBCP system, and its counteraction with the restriction of $d_{2}=2d_{1}$
signals an indication of rotational invariance. Within this work, we put our
focus on the 2D QBCP semimtal owning the rotational symmetry. To this end,
taking $d_{2}=2d_{1}$ yields the reduced energy eigenvalues as
\begin{eqnarray}
E^{\pm}(\mathbf{k})=\pm\mathbf{k}^{2}\sqrt{d_{1}^{2}+d_{3}^{2}}.
\end{eqnarray}

As aforementioned, it has attracted much interest to
investigate the impacts of fermion-fermion interactions on the low-energy
fates of 2D QBCP systems locating at checkerboard lattices~\cite{Fradkin2009PRL,Vafek2010PRB,Murray2014PRB,Wang2017PRB,Wang2018JPCM, Wang2019JPCM,Wang2020-arxiv}. However, compared to its checkerboard counterpart~\cite{Fradkin2008PRB,Murray2014PRB},
we need to bear in mind that the 2D QBCP model~(\ref{Eq_h_0}) with the TRS breaking
is still inadequately explored despite of holding the particle-hole symmetry and sixfold
rotational symmetry~\cite{Fradkin2009PRL,Huse2003PRB}.
%
Due to the qualitative difference of microscopic structures between kagom\'{e}
and checkerboard lattices, we therefore within this work endeavor to verify
how the fermion-fermion interactions impact the low-energy
properties of 2D kagom\'{e}-version QBCP systems.

To be concrete, we herein take into
account all potential marginal short-range four-fermion interactions on an equal footing~\cite{Huse2003PRB,Fradkin2009PRL,Murray2014PRB,Wang2017PRB}
\begin{eqnarray}
S_{\mathrm{int}}\!\!&=&\!\!\!\!\sum_{\mu,\nu=0}^{3}\lambda_{\mu\nu}
\prod^3_{i=1}\int\frac{d\omega_{i}d^{2}\mathbf{k}_{i}}{(2\pi)^{3}}\Psi^{\dag}
(\omega_{1},\mathbf{k}_{1})\Sigma_{\mu\nu}\Psi(\omega_{2},\mathbf{k}_{2})
\nonumber\\
&&\times\Psi^{\dag}(\omega_{3},\mathbf{k}_{3})\Sigma_{\mu\nu}
\Psi(\omega_{1}+\omega_{2}-\omega_{3},
\mathbf{k}_{1}+\mathbf{k}_{2}-\mathbf{k}_{3}),\,\,\,\,\,\,\, \label{S_int}
\end{eqnarray}
where $\lambda_{\mu\nu}$ with $\mu,\nu=0,1,2,3$ are utilized to measure the
coupling strengths of different types of fermion-fermion interactions that
are distinguished by vertex matrixes $\Sigma_{\mu\nu}$
acting on both lattice and spin spaces.

Consequently, we are left with the effective action in the momentum space
after combining the free Hamiltonian~(\ref{Eq_H_0}) and fermion-fermion
interactions~(\ref{S_int}) as follows~\cite{Fradkin2009PRL,Murray2014PRB,Wang2017PRB}
\begin{widetext}
\begin{eqnarray}
S_{\mathrm{eff}}\!\!&=&\!\!\!\!\int_{-\infty}^{\infty}\!\!\frac{d\omega}{2\pi}\!\!
\int^{\Lambda}\!\!\!\frac{d^{2}\mathbf{k}}
{(2\pi)^{2}}\Psi^{\dag}(\omega,\mathbf{k})
\left[-i\omega\Sigma_{00}+d_{3}\mathbf{k}^{2}\Sigma_{03}
+d_{1}(k_{x}^{2}-k_{y}^{2})\Sigma_{01}+d_{2}k_{x}k_{y}\Sigma_{02}\right]
\Psi(\omega,\mathbf{k})+\!\!\!\sum_{\mu,\nu=0}^{3}\!\!\lambda_{\mu\nu}\!\!
\int_{-\infty}^{\infty}\!\!\frac{d\omega_{1}
d\omega_{2}d\omega_{3}}{(2\pi)^{3}}\nonumber\\
\!\!&\times&\!\!\!\!\int^{\Lambda}\!\!\frac{d^{2}\mathbf{k}_{1}
d^{2}\mathbf{k}_{2}d^{2}\mathbf{k}_{3}}{(2\pi)^{6}}\Psi^{\dag}
(\omega_{1},\mathbf{k}_{1})\Sigma_{\mu\nu}\Psi(\omega_{2},\mathbf{k}_{2})
\Psi^{\dag}(\omega_{3},\mathbf{k}_{3})\Sigma_{\mu\nu}
\Psi(\omega_{1}+\omega_{2}-\omega_{3}, \mathbf{k}_{1}
+\mathbf{k}_{2}-\mathbf{k}_{3}).\label{S_eff}
\end{eqnarray}
\end{widetext}
On the basis of quantum field theory,
its noninteracting part directly gives rise to the free fermionic
propagator~\cite{Shankar1994RMP}
\begin{eqnarray}
G_{0}(i\omega, \mathbf{k})\!=\!\frac{1}{-i\omega\!+\!d_{3}\mathbf{k}^{2}\Sigma_{03}
\!+\!d_{1}(k_{x}^{2}\!-\!k_{y}^{2})\Sigma_{01}\!+\!d_{2}k_{x}k_{y}\Sigma_{02}},\label{G_0}
\end{eqnarray}
which plays a crucial role in constructing the RG equations. 

We hereafter adopt the $S_{\mathrm{eff}}$ as our starting point and utilize
the RG approach to inspect the low-energy behaviors of 2D spin-$1/2$ QBCP
electric systems sitting on the kagom\'{e} lattice in the presence of
these marginal fermion-fermion interactions.


\subsection{RG analysis}\label{Sec_RG_analysis}

We within this subsection implement the one-loop momentum-shell RG
analysis~\cite{Shankar1994RMP,Wilson1975RMP,Polchinski9210046}
to construct the entangled energy-dependent evolutions of
all interaction parameters appearing in Eq.~(\ref{S_eff}) that
are of close relevance to the low-energy properties.

Prior to deriving the one-loop RG equations, we are forced to determine
the rescaling transformations of momentum, energy, and fields~\cite{Shankar1994RMP}.
To this end, we can select the non-interacting parts of effective action~(\ref{S_eff})
as an original fixed point at which they are invariant during RG processes. As a result,
we obtain the following RG rescaling transformations
~\cite{Murray2014PRB,Wang2011PRB,Huh2008PRB,Wang2019JPCM}
\begin{eqnarray}
k_x&\longrightarrow&k'_xe^{-l},\label{Eq_rescale-k_x}\\
k_y&\longrightarrow&k'_ye^{-l},\label{Eq_rescale-k_y}\\
\omega&\longrightarrow&\omega'e^{-2l},\label{Eq_rescale-omega}\\
\Psi(i\omega,\mathbf{k})&\longrightarrow&\Psi'(i\omega',\mathbf{k}')e^{3l}.
\label{Eq_rescale-Psi}
\end{eqnarray}

\begin{figure}
\centering
\epsfig{file=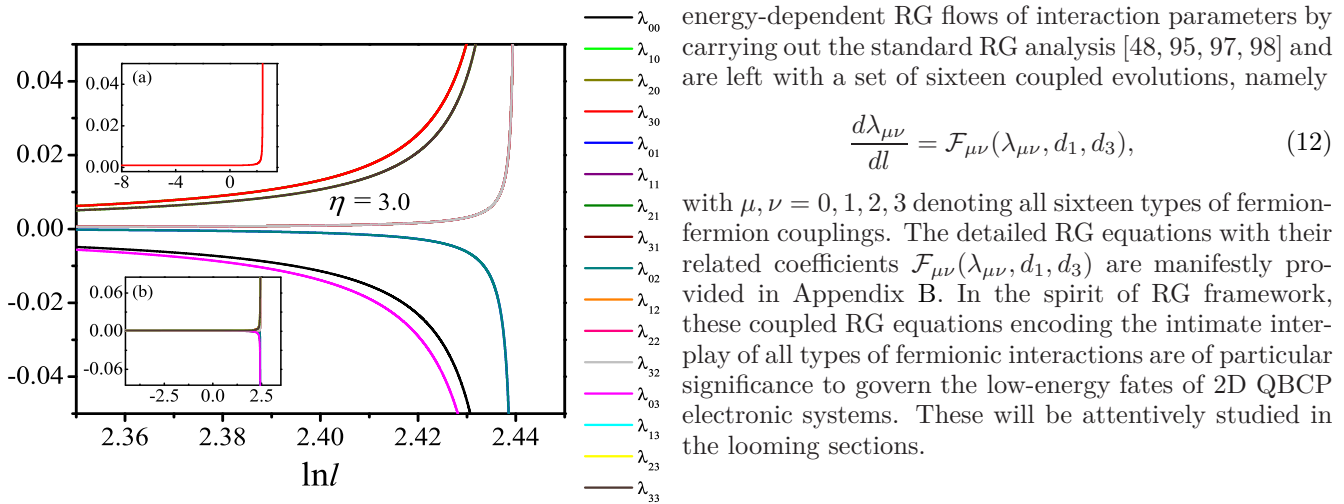,height=6.85cm,width=9.39cm}
\vspace{-0.63cm}
\caption{(Color online) Evolutions of fermion-fermion couplings $\lambda_{\mu\nu}$
at the ratio of structure parameters $\eta=3.0$ (the basic results
are insusceptible to concrete value of $\eta$). Insets: (a) the flows of $\lambda_{\mu\nu}$ in the whole energy region, and (b) the flow of $\lambda_{30}$ without
sign change under the influence of fermion-fermion interactions.
}\label{Evolutions_of_fermion-fermion_interaction}
\end{figure}

\begin{figure*}
\centering
\epsfig{file=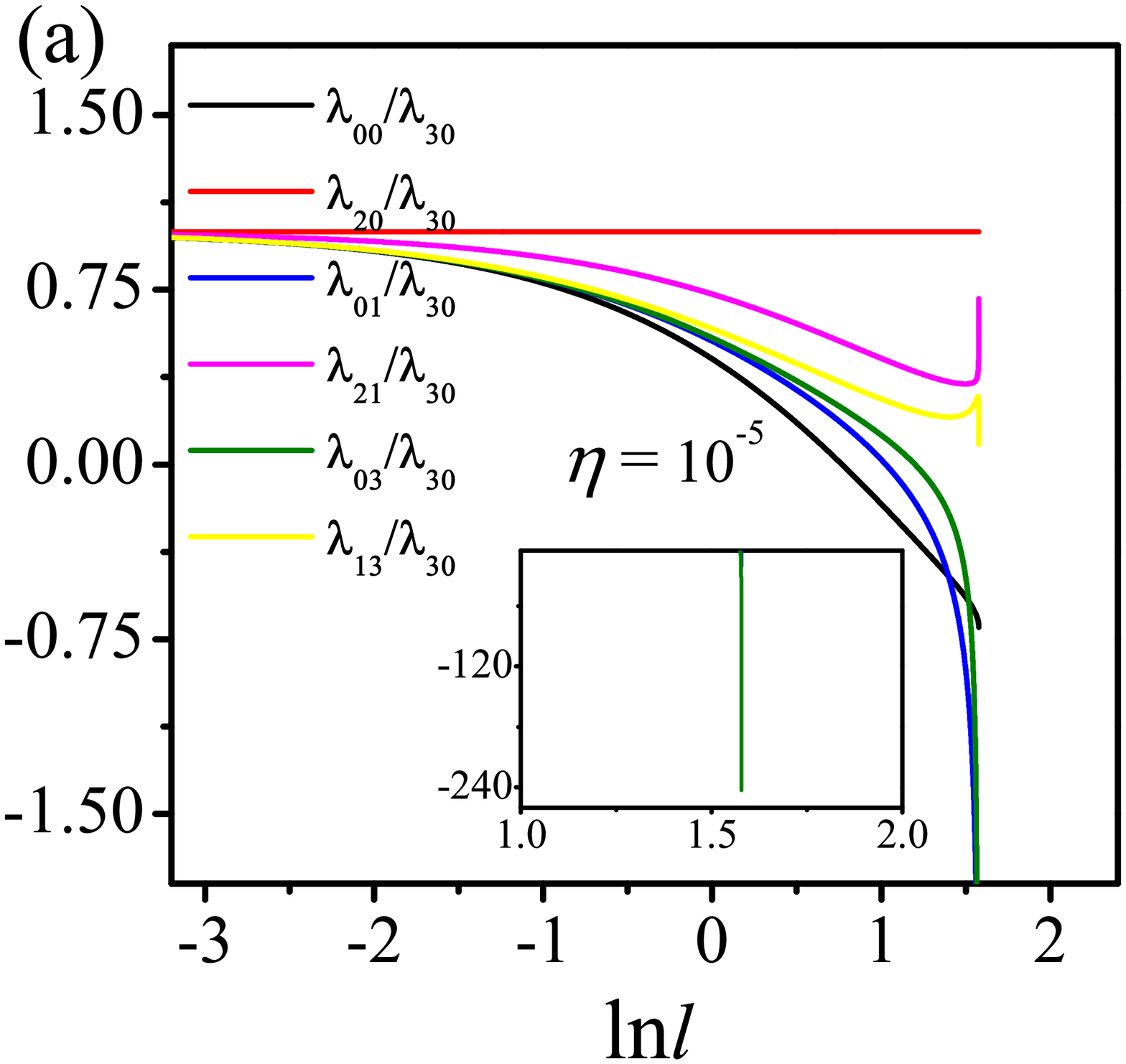,height=5.39cm,width=7.12cm}\hspace{-0.5cm}
\epsfig{file=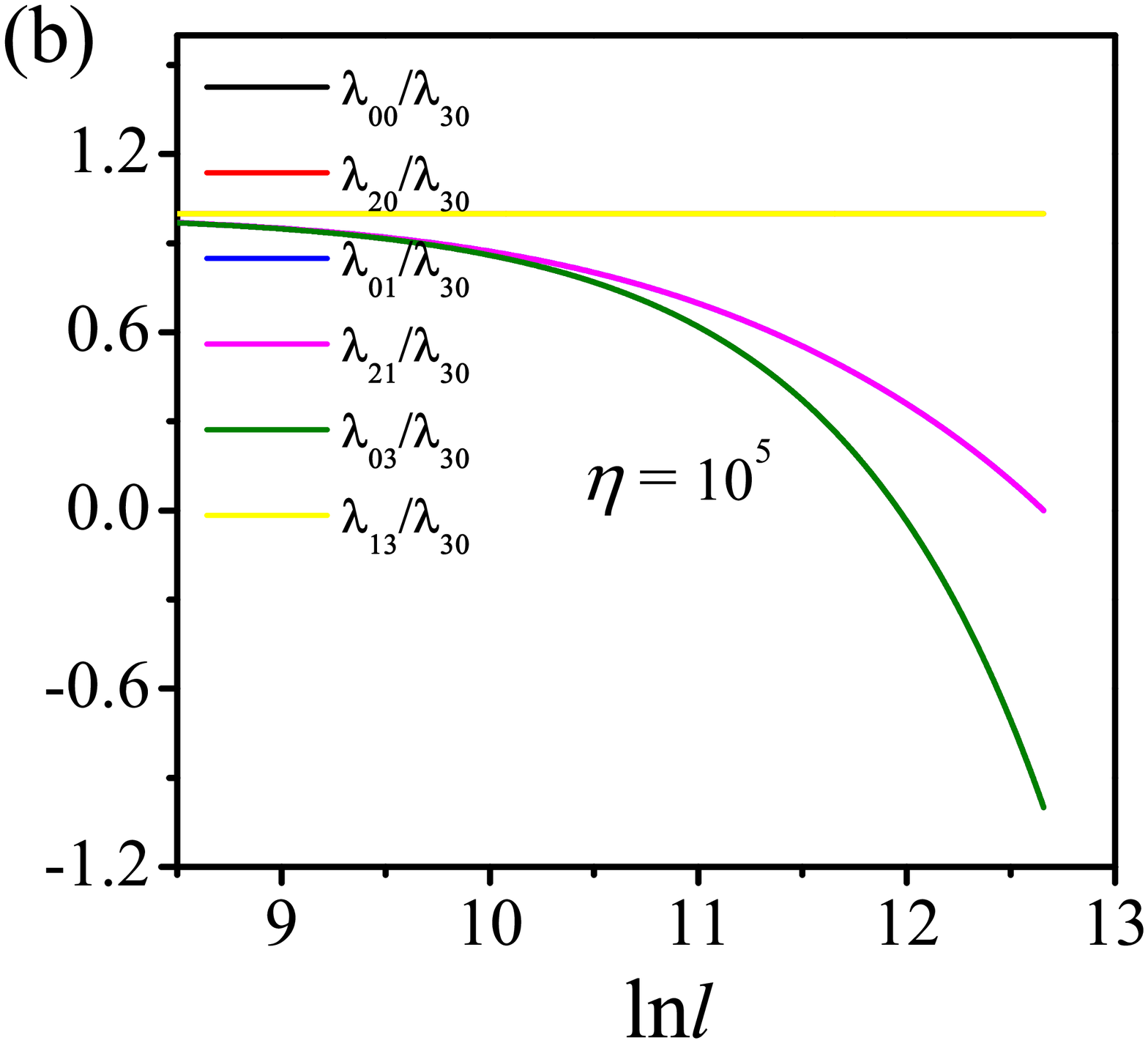,height=5.39cm,width=7.12cm}\\ \vspace{-0.1cm}
\epsfig{file=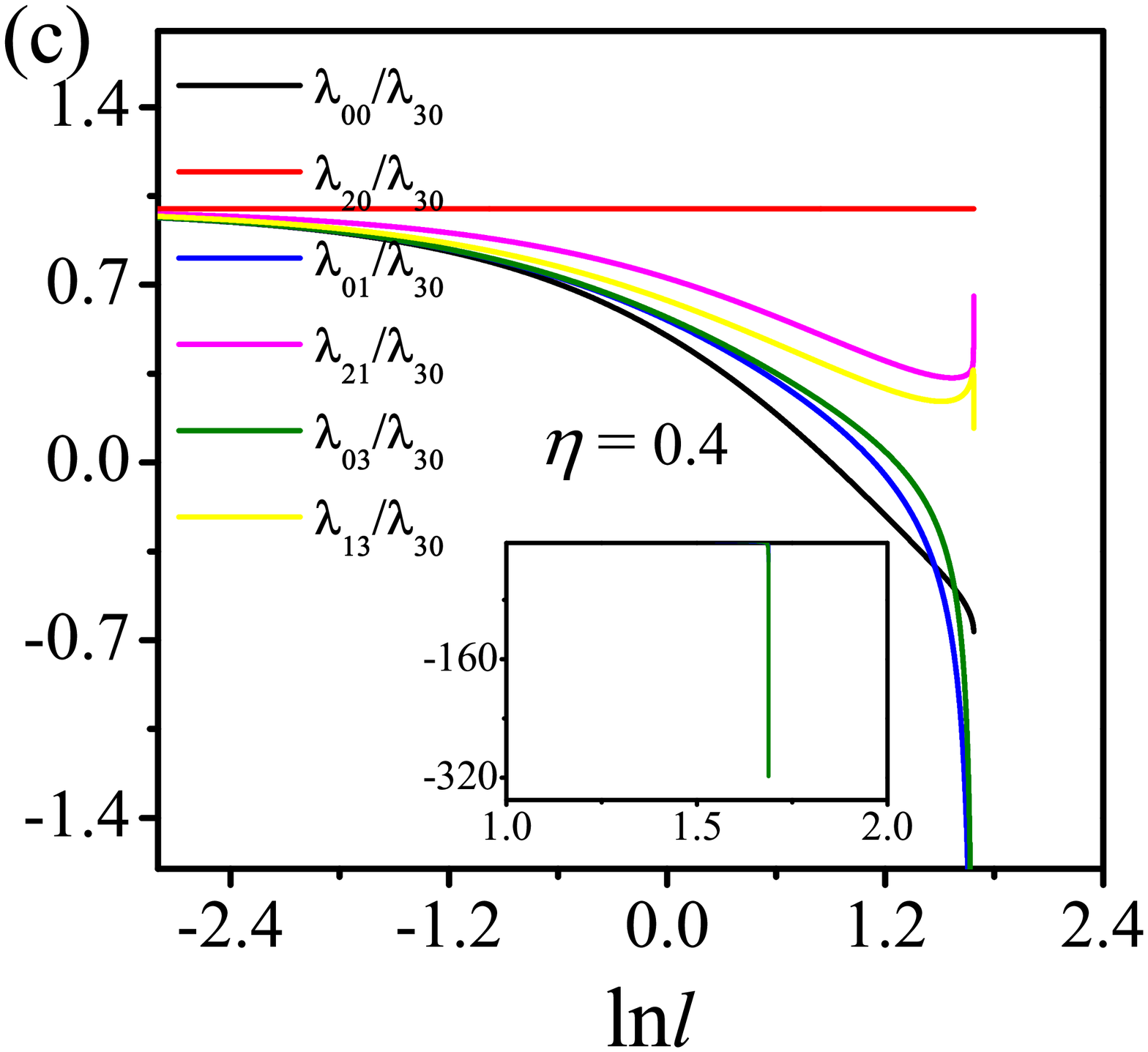,height=5.39cm,width=7.12cm}\hspace{-1.82cm}
\epsfig{file=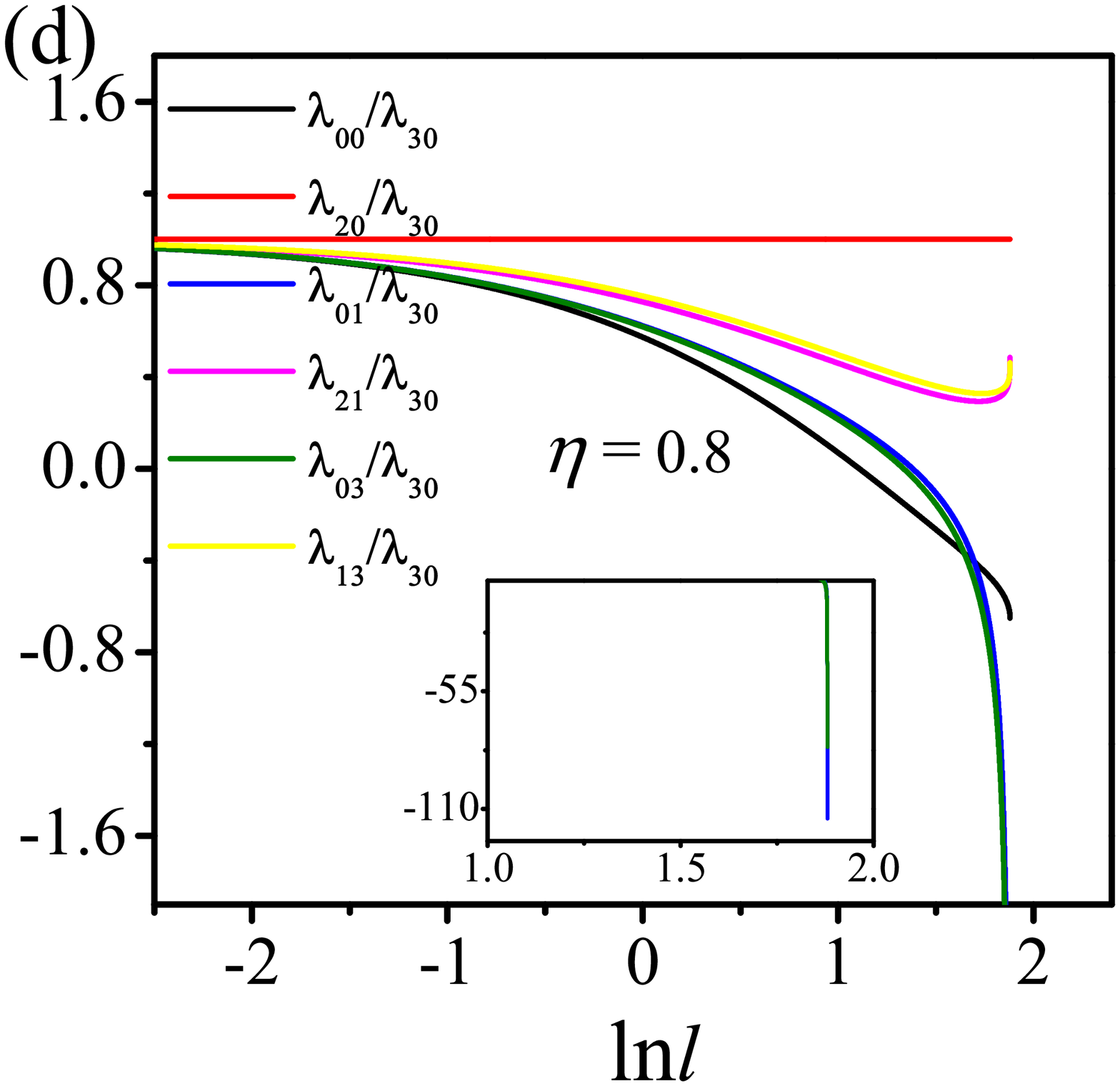,height=5.39cm,width=7.12cm}\hspace{-1.82cm}
\epsfig{file=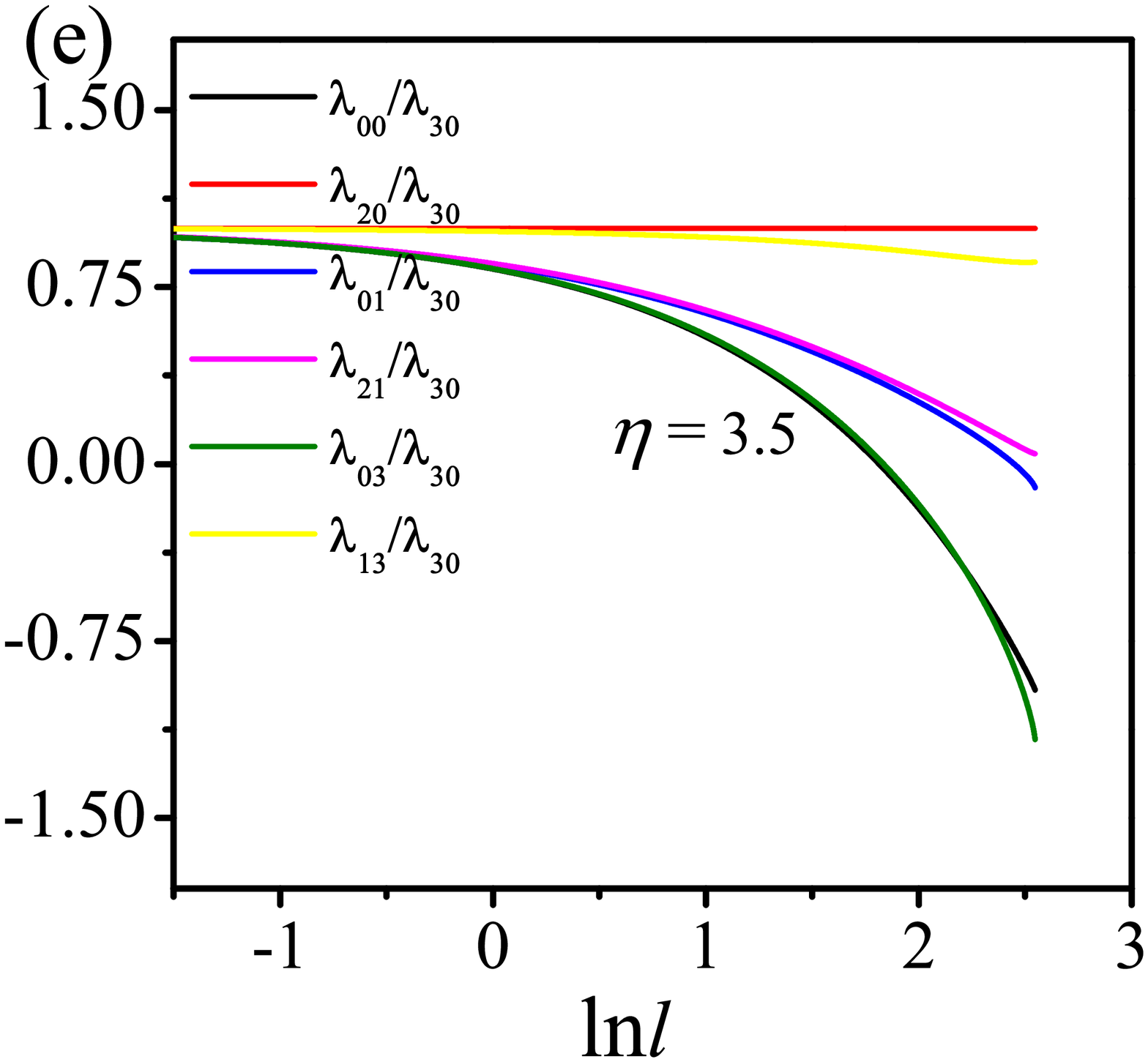,height=5.39cm,width=7.12cm}\\
\vspace{-0.1cm}
\caption{(Color online) Evolutions for six representative sorts of fermion-fermion interaction
parameters $\lambda_{\mu\nu}/\lambda_{30}$ with approaching: (a) Type-I-RFP at $\eta=10^{-5}$;
(b) Type-III-RFP at $\eta=10^{5}$; and (c)-(e) Type-II-RFP at $\eta=0.4$, $\eta=0.8$,
and $\eta=3.5$, respectively.}\label{Fig_relative-flows}
\end{figure*}

In order to grasp the contributions from
fermion-fermion interactions, we have to go beyond this original fixed point and
take into account all one-loop corrections. In the spirit of
momentum-shell RG theory~\cite{Shankar1994RMP}, the ``fast modes" of fermionic fields within
the momentum shell $b\Lambda<k<\Lambda$ are integrated out at first, where $\Lambda$
characterizes the energy scale and variable $b$ is
designated as $b=e^{-l}<1$. It is worth highlighting that $l$
serves as a running length scale and thus its increase is equivalent
to the decrease of energy scale. Next, we insert the corrections within the momentum shell into
the ``slow modes" to yield the new ``slow modes" and then rescale these
``slow modes" to new ``fast modes"~\cite{Cvetkovic2012PRB,Murray2014PRB,Wang2017PRB, Wang2011PRB,Wang2013PRB,She2010PRB,Huh2008PRB,Kim2008PRB,Maiti2010PRB,
She2015PRB,Roy2016PRB,Wang2019JPCM}. In principle, it is convenient to measure the momenta and
energy with cutoff $\Lambda_0$ which is related
to the lattice constant (i.e., $k\rightarrow k/\Lambda_0$
and $\omega\rightarrow\omega/\Lambda_0$) to simply our calculations and write
the results more compactly~\cite{Murray2014PRB,Stauber2005PRB, Wang2011PRB,She2010PRB,Huh2008PRB,Wang2019JPCM}.
To be specific, we unbiasedly take into account all one-loop Feynman diagrams
depicted in Fig.~\ref{Fig_fermion_propagator_and_fermion_interaction_correction}
to capture the one-loop information of ``fast modes". After paralleling the
well-trodden procedures and performing long but straightforward
algebra~\cite{Murray2014PRB, Wang2017PRB, Wang2018JPCM, Wang2019JPCM, Wang2019JPCM},
all one-loop corrections are obtained and presented detailedly
in Appendix~\ref{Appendix_one-loop_corrections}.

\begin{figure}
\hspace{-0.8cm}
\epsfig{file=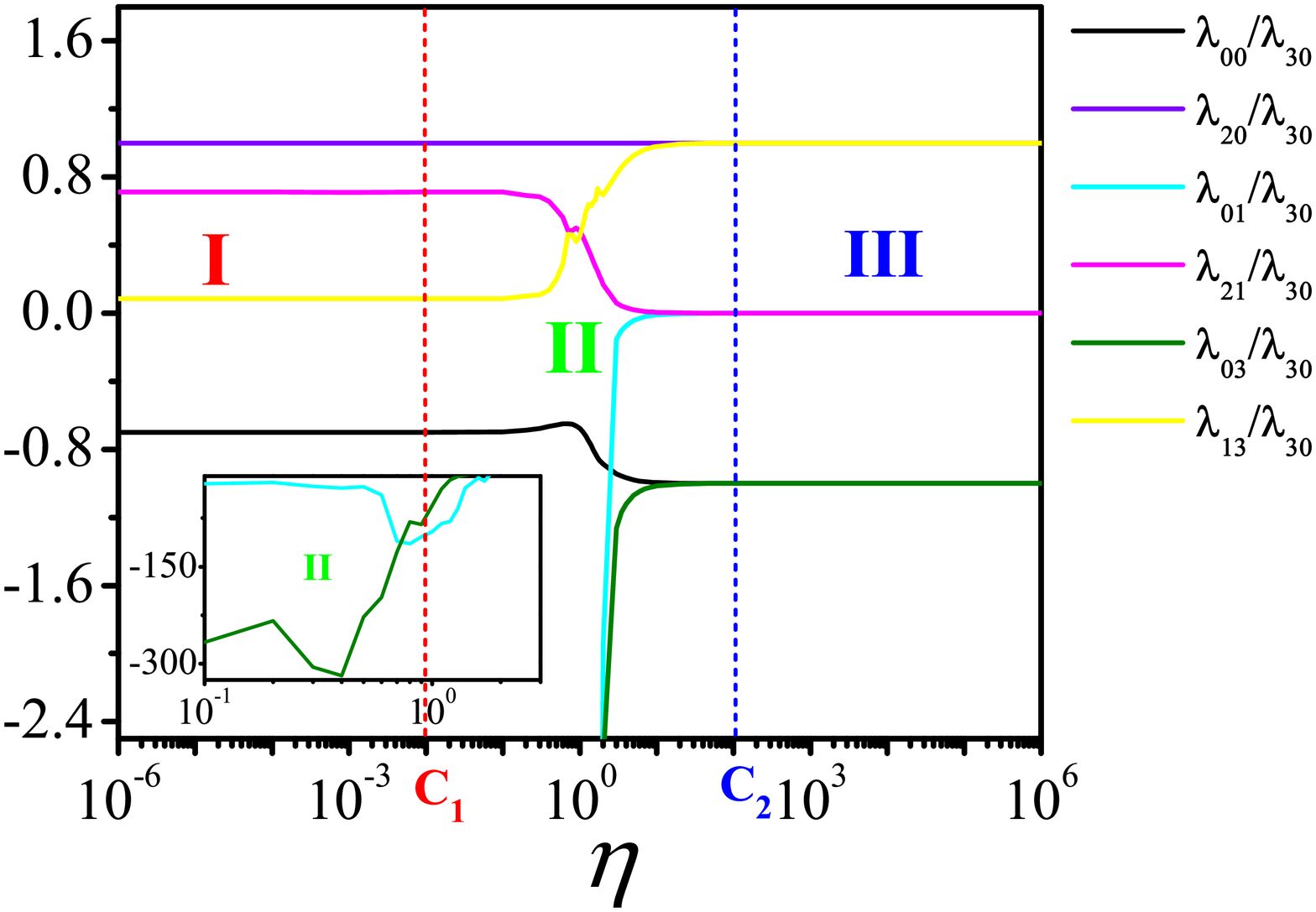,height=7cm,width=9.191cm}
\vspace{-0.3cm}
\caption{(Color online) The $\eta$-dependent evolutions of fermion-fermion
interaction parameters $\lambda_{\mu\nu}/\lambda_{30}$ (i.e., the concrete
values of RFPs). Three distinct regions are separated by $\eta=C_1\approx 10^{-2}$
and $\eta=C_2\approx 10^2$, which are denominated
as Type-I-Region, Type-II-Region, Type-III-Region
(namely, I, II, and III), respectively.}\label{Three_regions}
\end{figure}

Combining RG rescalings~(\ref{Eq_rescale-k_x})-(\ref{Eq_rescale-Psi}) and one-loop corrections~(\ref{One_loop_Eq_S_00})-(\ref{One_loop_Eq_S_33}), we are capable of
deriving the coupled energy-dependent RG flows of interaction parameters by carrying out the
standard RG analysis~\cite{Murray2014PRB,Wang2011PRB,Huh2008PRB,Wang2019JPCM} and are left with a set of
sixteen coupled evolutions, namely
\begin{eqnarray}
\frac{d\lambda_{\mu\nu}}{dl}=\mathcal{F}_{\mu\nu}
(\lambda_{\mu\nu},d_{1},d_{3}),\label{Eq_RG}
\end{eqnarray}
with $\mu,\nu=0,1,2,3$ denoting all sixteen types of fermion-fermion couplings.
The detailed RG equations with their related coefficients
$\mathcal{F}_{\mu\nu}(\lambda_{\mu\nu},d_{1},d_{3})$ are manifestly provided in Appendix~\ref{Appendix_RG_flow_Eqs_of_lambda}.
In the spirit of RG framework, these coupled RG equations encoding the intimate
interplay of all types of fermionic interactions are of particular significance to
govern the low-energy fates of 2D QBCP electronic systems. These will be attentively
studied in the looming sections.

\vspace{0.9cm}

\section{Relative fixed points}\label{Sec_RFP}

With the coupled RG flow equations in hand, we are able to seek the underlying fixed points
with lowering the energy scales, which are always assumed to dictate the critical behaviors.

\subsection{Evolutions of interaction parameters}

We start out by investigating the energy-dependent flows of four-fermion couplings,
which are determined by RG equations and assumed to overarch
the low-energy properties~\cite{Cvetkovic2012PRB,Murray2014PRB,Wang2017PRB, Wang2011PRB,Wang2013PRB,She2010PRB,Huh2008PRB,Kim2008PRB,Maiti2010PRB,
She2015PRB,Roy2016PRB,Wang2019JPCM}. Before proceeding, it is necessary to present several comments
on the initial condition. On one hand, we, without loss of generality, treat all sixteen types
of marginal fermion-fermion interactions unbiasedly and assign them an equal
beginning value. On the other hand, the RG coupled equations also rely upon
two structural parameters $d_1$ and $d_3$. It is of particular importance to
highlight that the flows of interaction parameters are insensitive to their concrete values
but instead the ratio between $d_1$ and $d_3$. To facilitate our analysis, we hereafter
designate $\eta\equiv d_3/d_1$ to capture the basic influence of these
two parameters.

To proceed, taking a concrete value $\eta=3.0$ plus a representative
starting value for $\lambda_{\mu\nu}$ and performing numerical
analysis of Eq.~(\ref{Eq_RG}), we are left with the results shown in Fig.~\ref{Evolutions_of_fermion-fermion_interaction}. Reading off this figure,
we notice that fermion-fermion interactions are strongly energy-dependent
and driven to be divergent at some critical
energy scale denoted by $l_{c}$ in the low-energy regime, which
always is an unambiguous signature for the emergence
of phase transitions~\cite{Cvetkovic2012PRB,Murray2014PRB,Wang2017PRB, Maiti2010PRB, Altland2006Book,Vojta2003RPP,Halboth2000RPL,Halboth2000RPB, Eberlein2014PRB,Chubukov2012ARCMP,Chubukov2016PRX}. In addition,
after completing numerical calculations of coupled RG evolutions of interactions parameters ~(\ref{Eq_RG_lambda00})-(\ref{Eq_RG_lambda33}), we figure out that six groups of interaction parameters
are not coincident and their trajectories do not overlap with decreasing the energy scale
as depicted in Fig.~\ref{Evolutions_of_fermion-fermion_interaction}.
To be concrete, $\lambda_{01}$ is degenerate with $\lambda_{02}$ and
$\lambda_{13}$ is coincident to $\lambda_{23}$ plus $\lambda_{33}$, respectively.
In particular, the tendencies of parameters $\lambda_{10}$,
$\lambda_{20}$, and $\lambda_{30}$ are exactly overlapped. Analogously,
$\lambda_{11}$, $\lambda_{21}$, $\lambda_{31}$,
$\lambda_{12}$, $\lambda_{22}$, and $\lambda_{32}$ share the same evolution.
These imply that some of these parameters may
be not independent. As a result, all of these 16 kinds
of interaction parameters can be clustered into
six groups. In order to compactly exhibit the flow trajectories of
interaction parameters and simplify our discussions, we therefore from
now on choose four representative parameters $\lambda_{01}$, $\lambda_{20}$,
$\lambda_{21}$, and $\lambda_{13}$ to denote their
degenerate counterparts. In other words,
there only exist six independent flows
$\lambda_{00}$, $\lambda_{20}$, $\lambda_{01}$,
$\lambda_{21}$, $\lambda_{03}$, and $\lambda_{13}$ that are employed
to characterize all types of fermion-fermion interactions.

\subsection{Three $\eta$-dependent distinct regions}

In light of the divergence of fermion-fermion couplings in the low-energy regime, we are
conventionally suggested to rescale all parameters with a non-sign changed
parameter~\cite{Vafek2010PRB,Murray2014PRB,Wang2017PRB}.
Based on the relative interaction parameters, one henceforth can safely work
within the perturbative RG framework before the divergence~\cite{Vafek2010PRB,Murray2014PRB,
Wang2017PRB, Maiti2010PRB, Altland2006Book,Vojta2003RPP,Halboth2000RPL,Halboth2000RPB, Eberlein2014PRB,Chubukov2012ARCMP,Chubukov2016PRX}.

\renewcommand\arraystretch{1.5}
\begin{table*}[htbp] 
\centering
\caption{Specific values of fermion-fermion interactions
with several representative ratios of structure parameters
in the vicinity of three different RFPs. The corresponding energy-dependent
evolutions of these couplings are manifestly displayed in Fig.~\ref{Fig_relative-flows}.}\label{Table}
\vspace{0.3cm}
\begin{tabular}{p{3.3cm}<{\centering}|p{2cm}<{\centering}|
p{2cm}<{\centering}|
p{2cm}<{\centering}|p{1.8cm}<{\centering}|
p{2cm}<{\centering}|p{1.8cm}<{\centering}|
p{1.8cm}<{\centering}}
    \hline
    \hline
    RFPs&$\lambda_{00}/\lambda_{30}$ &$\lambda_{20}/\lambda_{30}$ &$\lambda_{01}/\lambda_{30}$ &$\lambda_{21}/\lambda_{30}$ &$\lambda_{03}/\lambda_{30}$ &$\lambda_{13}/\lambda_{30}$ &$\eta$\\
    \hline
    Type-I-RFP &-0.7 &1 &-20.3 &0.7 &-242.8 &0.1&$10^{-5}$\\
    \hline
    \multirow{3}{3cm}&-0.7&1 &-28.4 &0.7 &-318.1 & 0.1&$0.4$\\
    \cline{2-8}
   Type-II-RFP&-0.7 &1 &-114.5 &0.5 &-81&0.5&$0.8$\\
    \cline{2-8}
    &-1 & 1 &-0.1 &0.04 &-1.2 &0.9 &$3.5$\\
    \hline
    Type-III-RFP&-1 &1 &0 &0 &-1&1&$10^{5}$\\
    \hline
    \hline
\end{tabular}
\end{table*}

As explicitly illustrated by Fig.~\ref{Evolutions_of_fermion-fermion_interaction}(a), the
parameter $\lambda_{30}$ flows monotonously and thus does not change sign during
the entire RG flow. It is therefore convenient to measure all interaction parameters with $\lambda_{30}$.
Accordingly, we from now on shift our attention to the low-energy behaviors
of these relative parameters, i.e., $\lambda_{00}/\lambda_{30}$,
$\lambda_{20}/\lambda_{30}$, $\lambda_{01}/\lambda_{30}$,
$\lambda_{21}/\lambda_{30}$, $\lambda_{03}/\lambda_{30}$, and
$\lambda_{13}/\lambda_{30}$. Besides their energy-dependent trajectories,
we are of particular interest to determine the final fates of these parameters
at the low-energy limit, namely, the potential fixed point (FP) which is
expected to govern the physical properties and accompanied by critical behaviors.
Given these parameters are rescaled by $\lambda_{30}$,
we hereafter dub them relatively fixed points (RFPs)~\cite{Vafek2010PRB,Murray2014PRB,Wang2017PRB}.

\begin{figure}
\centering
\includegraphics[width=3.08in]{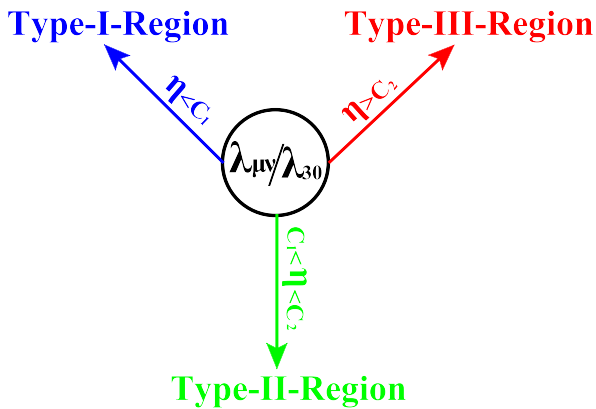}
\vspace{-0.0cm}
\caption{(Color online) Schematic illustration of three different $\eta$-tuned
regions shown in Fig.~\ref{Three_regions} with $C_1\approx 10^{-2}$
and $C_2\approx 10^2$.  Several representative values of RFPs are
presented in Table~\ref{Table} for Type-I-Region,
Type-II-Region, and Type-III-Region,
respectively.}\label{Schematic diagram}
\end{figure}

Although the basic results are hardly susceptible to starting
values of fermion-fermion strengths, the flows of interaction parameters and corresponding RFPs
are heavily $\eta$-dependent. Fig.~\ref{Fig_relative-flows}
manifestly shows the evolving tendencies of fermion-fermion
interaction parameters strongly hinge upon the specific values of $\eta$.
What is more, we learning from Fig.~\ref{Three_regions} notice that concrete
values of RFPs are closely sensitive to $\eta$ as well.
On one side, as long as $\eta$ is either tuned less than certain
small value nominated as $C_{1}$ or adjusted to exceed some critical
value denoted as $C_2$, the final values of parameters
$\lambda_{\mu\nu}/\lambda_{30}$ (i.e., RFPs)
arrive at some constants and then are considerably robust
with lowering or increasing value of $\eta$.
On the other side, the concrete values of RFPs spanning from $C_{1}$ to $C_{2}$
are no longer invariant but instead fairly rely upon $\eta$.
For physical consideration, we infer that $d_1$ is dominant over $d_3$ at $\eta<C_1$
and thus the contribution from $d_3$ is negligible resulting in the stable values of RFPs
and vice versa for $\eta>C_2$. On the contrary, neither $d_1$ nor $d_3$ can completely
win its opponent within $C_1<\eta<C_2$. It is thus the intimate competition between
$d_1$ and $d_3$ that plays a pivotal role in pinning down RFPs
as unambiguously characterized in Fig.~\ref{Three_regions}.

To be specific, we directly notice that the absolute values of negative-divergent
parameters $\lambda_{01}/\lambda_{30}$ and $\lambda_{03}/\lambda_{30}$ as well as
the positive-divergent parameter $\lambda_{21}/\lambda_{30}$ present a clear
downward trend with increasing the values of $\eta$.
Conversely, the increase of $\eta$ is favorable to raise
the absolute value of negative-divergent parameter $\lambda_{00}/\lambda_{30}$
together with positive-divergent parameters $\lambda_{13}/\lambda_{30}$.
In comparison, one can readily figure out $\lambda_{20}/\lambda_{30}=1$ is
hardly susceptible to the modulation of $\eta$. In order to facilitate our
studies under such circumstances, it is profitable to divide $\eta\in(0,\infty)$
into three distinct regions owing to the robustness of RFPs against the change of $\eta$.
As manifestly designated in Fig.~\ref{Three_regions} and schematically illustrated
in Fig.~\ref{Schematic diagram}, they correspond to Type-I-Region ($\eta<C_{1}$), Type-II-Region
($C_{1}<\eta<C_{2}$), and Type-III-Region ($\eta>C_{2}$), respectively.

\subsection{Three types of relatively fixed points}

Concerning the discrepancies of RFPs in these three regions,
we from now on nominate the RFPs locating at
Type-I-Region, Type-II-Region, and Type-III-Region as
Type-I-RFP, Type-II-RFP, Type-III-RFP, respectively.
In order to remedy the insufficiency of qualitative illustrations
in Fig.~\ref{Schematic diagram},
we hereby select some representative values of $\eta$
with three different types of RFPs and present the specific values
of the interaction parameters as collected in Table~\ref{Table}
apparently indicating their individual features.
As aforementioned, the final values of interaction parameters
manifested in Fig.~\ref{Three_regions} are stable in Type-I-Region
and Type-III-Region under the variation of $\eta$. On the contrary, they are rather sensitive to
$\eta$ in the Type-II-Region. Under these respects, Table~\ref{Table} consists of
only one point for both Type-I-RFP and Type-III-RFP as well as
three typical points for Type-II-RFP with considering the tendency of
Type-II-Region in Fig.~\ref{Three_regions}.

To be concrete, Fig.~\ref{Fig_relative-flows}(a) indicates that, at Type-I-RFP,
the parameters $\lambda_{00}/\lambda_{30}$, $\lambda_{01}/\lambda_{30}$, and
$\lambda_{03}/\lambda_{30}$ flow divergently along the negative direction.
In particular, the absolute strengths of $\lambda_{03}/\lambda_{30}$ and
$\lambda_{00}/\lambda_{30}$ are the strongest and weakest.
However, $\lambda_{13}/\lambda_{30}$, $\lambda_{20}/\lambda_{30}$, and $\lambda_{21}/\lambda_{30}$
are sign-unchanged during the whole RG process.
Compared to Type-I-RFP, $\lambda_{01}/\lambda_{30}$ and
$\lambda_{21}/\lambda_{30}$ at Type-III-RFP evolve towards
zero. Additionally, Fig.~\ref{Fig_relative-flows}(b) shows that
$\lambda_{00}/\lambda_{30}$ and $\lambda_{03}/\lambda_{30}$
are driven to be equal but opposite to $\lambda_{20}/\lambda_{30}$
and $\lambda_{13}/\lambda_{30}$ in the lowest-energy limit. As for Type-II-RFP, interaction parameters
delineated in Fig.~\ref{Fig_relative-flows}(c)-(e) share the same
sign-change (unchange) information with Type-I-RFP. In a sharp contrast to
the other two types, it is of remarkable importance to emphasize that interaction
parameters in the Type-II-Region approximately increase or decrease monotonically
with the increase of $\eta$ except some critical point.
In addition to the difference of their concrete values, it is worth pointing out
that the critical energy scales, at which RFPs
are accessed, are nearly constants in Type-I-Region but instead
proportional to $\eta$ in both Type-II-Region and Type-III-Region.

\renewcommand\arraystretch{1.8}
\begin{table}
\setlength{\abovecaptionskip}{0pt}%
\setlength{\belowcaptionskip}{14pt}
\centering
\caption{Twelve different kinds of potential phases triggered by
fermion-fermion interactions, which are associated with
source-term bilinears appearing in Eq.~(\ref{Eq_source-term})~\cite{Roy2018RRX}.
Hereby, SC and AFM denote superconductivity and antiferromagnetism, respectively.
In addition, chiral SC1 and chiral SC2 are adopted to characterize two
distinct sorts of chiral superconducting states.}\label{Twelve_kinds_of_potential_phases}
\vspace{0.0cm}
\begin{tabular}{p{1.7cm}<{\centering}|p{3cm}<{\centering}|
p{3.2cm}<{\centering}}
\hline 
\hline
Order parameters & Vertex matrixes of fermionic bilinears & Potential phases\\
\hline
$\Delta^{c}_{1}$&
$\mathcal{M}^{c}_{1}=\tau_{0}\otimes\sigma_{0}$ & charge instability\\
\hline
$\Delta^{c}_{2}$&$\mathcal{M}^{c}_{2}=\tau_{0}\otimes\sigma_{1}$ &$x$-current \\
\hline
$\Delta^{c}_{3}$&$\mathcal{M}^{c}_{3}=\tau_{0}\otimes\sigma_{2}$ &bond density \\
\hline
$\Delta^{c}_{4}$&$\mathcal{M}^{c}_{4}=\tau_{0}\otimes\sigma_{3}$ &charge density wave \\
\hline
$\vec{\Delta}^{s}_{1}$&$\vec{\mathcal{M}}^{s}_{1}=\vec{\tau}\otimes\sigma_{0}$ &Ferromagnet \\
\hline
$\vec{\Delta}^{s}_{2}$&$\vec{\mathcal{M}}^{s}_{2}=\vec{\tau}\otimes\sigma_{1}$ &$x$-spin-current \\
\hline
$\vec{\Delta}^{s}_{3}$&$\vec{\mathcal{M}}^{s}_{3}=\vec{\tau}\otimes\sigma_{2}$ &spin bond density \\
\hline
$\vec{\Delta}^{s}_{4}$&$\vec{\mathcal{M}}^{s}_{4}=\vec{\tau}\otimes\sigma_{3}$ &AFM \\
\hline
$\Delta^{\mathrm{PP}}_{1}$&
$\mathcal{M}^{\mathrm{PP}}_{1}=\tau_{2}\otimes\sigma_{3}$ &s-wave SC\\
\hline
$\Delta^{\mathrm{PP}}_{2}$&$\mathcal{M}^{\mathrm{PP}}_{2}=\tau_{2}\otimes\sigma_{1}$ &chiral SC1 \\
\hline
$\Delta^{\mathrm{PP}}_{3}$&$\mathcal{M}^{\mathrm{PP}}_{3}=\tau_{2}\otimes\sigma_{0}$ &chiral SC2 \\
\hline
$\Delta^{\mathrm{PP}}_{4}$&$\mathcal{M}^{\mathrm{PP}}_{4}=\tau_{0,1,3}\otimes\sigma_{2}$ &triplet SC\\
\hline
\hline
\end{tabular}
\end{table}

Before closing this section, it is necessary to present several comments on the
flow tendencies of Fig.~\ref{Evolutions_of_fermion-fermion_interaction}
and Fig.~\ref{Fig_relative-flows}. Roughly speaking, there are two facets
responsible for this issue. On one hand, the correlations among different sorts of
fermionic interactions are less important and even negligible
once the system is far enough away from the instability point.
In comparison, both interactions and fluctuations play more important
roles with accessing the instability point, yielding sensitive changes of
interaction parameters. This indicates that the competition among all these distinct types of
fermion-fermion interactions becomes stronger and stronger as
the instability is approached, but instead weaker and weaker with deviating from that point.
As a consequence, the RFP with strong couplings is a significant impetus to trigger
a multitude of unusual phenomena including instabilities and critical physical
behaviors~\cite{Vafek2010PRB,Murray2014PRB,Wang2017PRB,Maiti2010PRB, Altland2006Book,Vojta2003RPP,Halboth2000RPL,Halboth2000RPB, Eberlein2014PRB,Chubukov2012ARCMP,Chubukov2016PRX}. Stimulated by these,
we are going to investigate the potential instabilities around
three distinct types of RFPs in the upcoming section
and defer the underlying critical physical behaviors
to Sec.~\ref{Sec_implication}.

\vspace{0.25cm}

\section{Instabilities induced by fermion-fermion interactions}\label{Sec_instab}

With the variation of structural coefficient $\eta$, we have confirmed in previous section
there exist three distinct types (regions) of RFPs attesting to the subtle fermion-fermion interactions.
As mentioned previously, instabilities are tightly linked to these RFPs,
which are well-known signatures of symmetry breaking~\cite{Cvetkovic2012PRB,Murray2014PRB,Wang2017PRB, Maiti2010PRB, Altland2006Book,Vojta2003RPP,Halboth2000RPL,Halboth2000RPB, Eberlein2014PRB,Chubukov2012ARCMP,Nandkishore2012NP,Chubukov2016PRX,Roy2018RRX}.
In this respect, it is of enormous interest to seek and identify the leading
instability and its related phase transition. 

\begin{figure}
\centering
\includegraphics[width=3.2in]{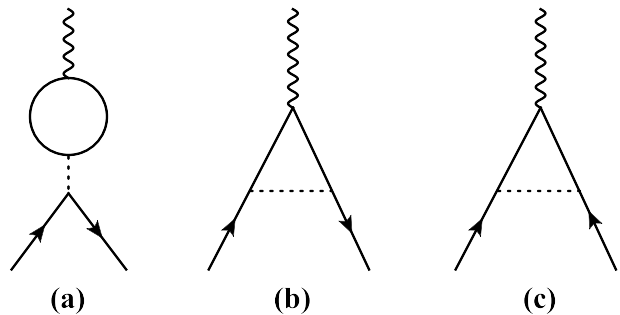}
\vspace{-0.0cm}
\caption{One-loop corrections to the bilinear fermion-source
terms~\cite{Murray2014PRB,Wang2017PRB}: (a) and (b) represent the particle-hole channel and (c) specifies
particle-particle channel. The solid, dash, and wave lines correspond to the fermion,
fermion-fermion interaction and source term fields,
respectively.}\label{One_loop_corrections_to_source term}
\end{figure}

\begin{figure*}
\centering
\epsfig{file=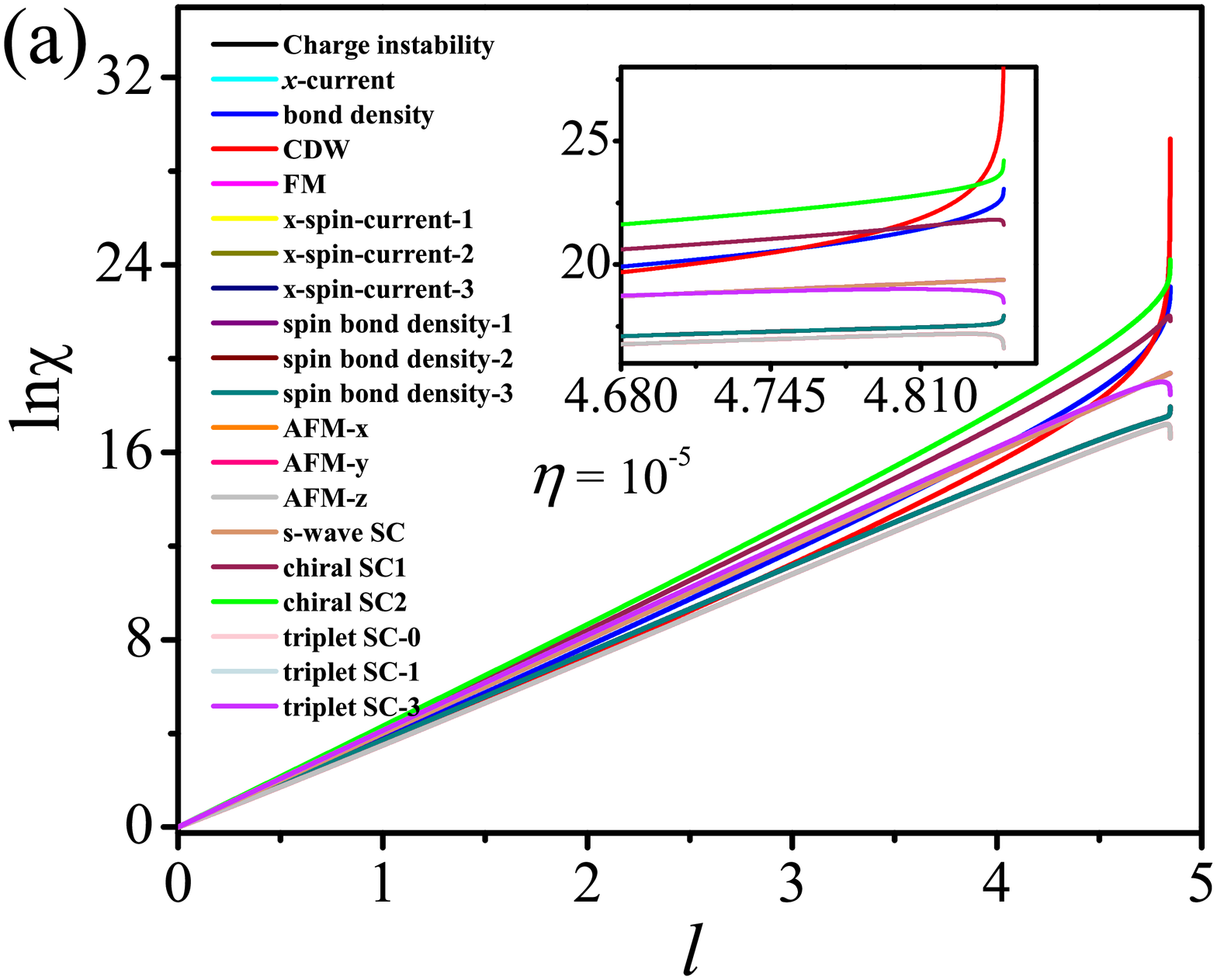,height=7.1cm,width=9.31cm}\hspace{-0.8cm}
\epsfig{file=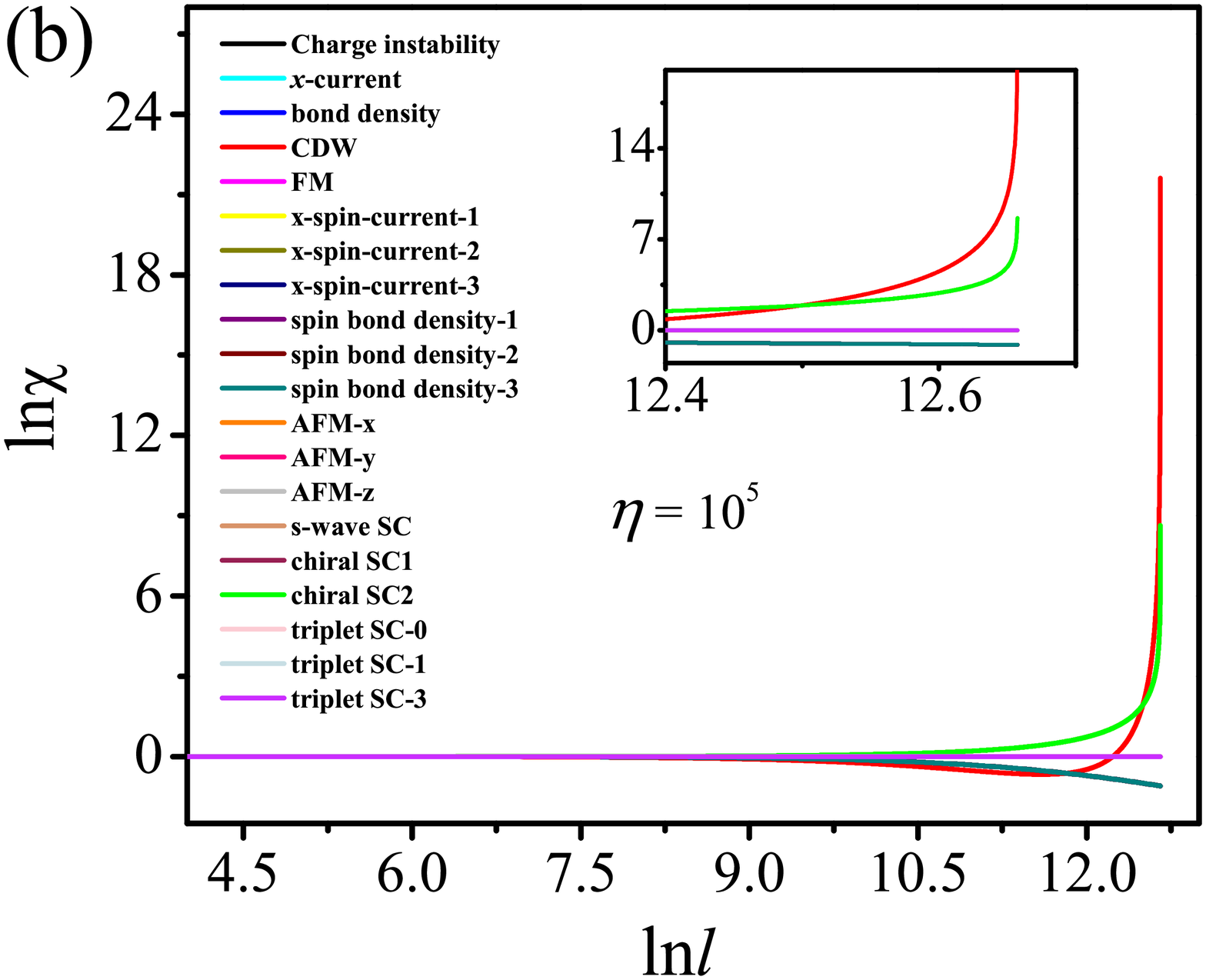,height=7.1cm,width=9.31cm}\\ \vspace{0.1cm}
\epsfig{file=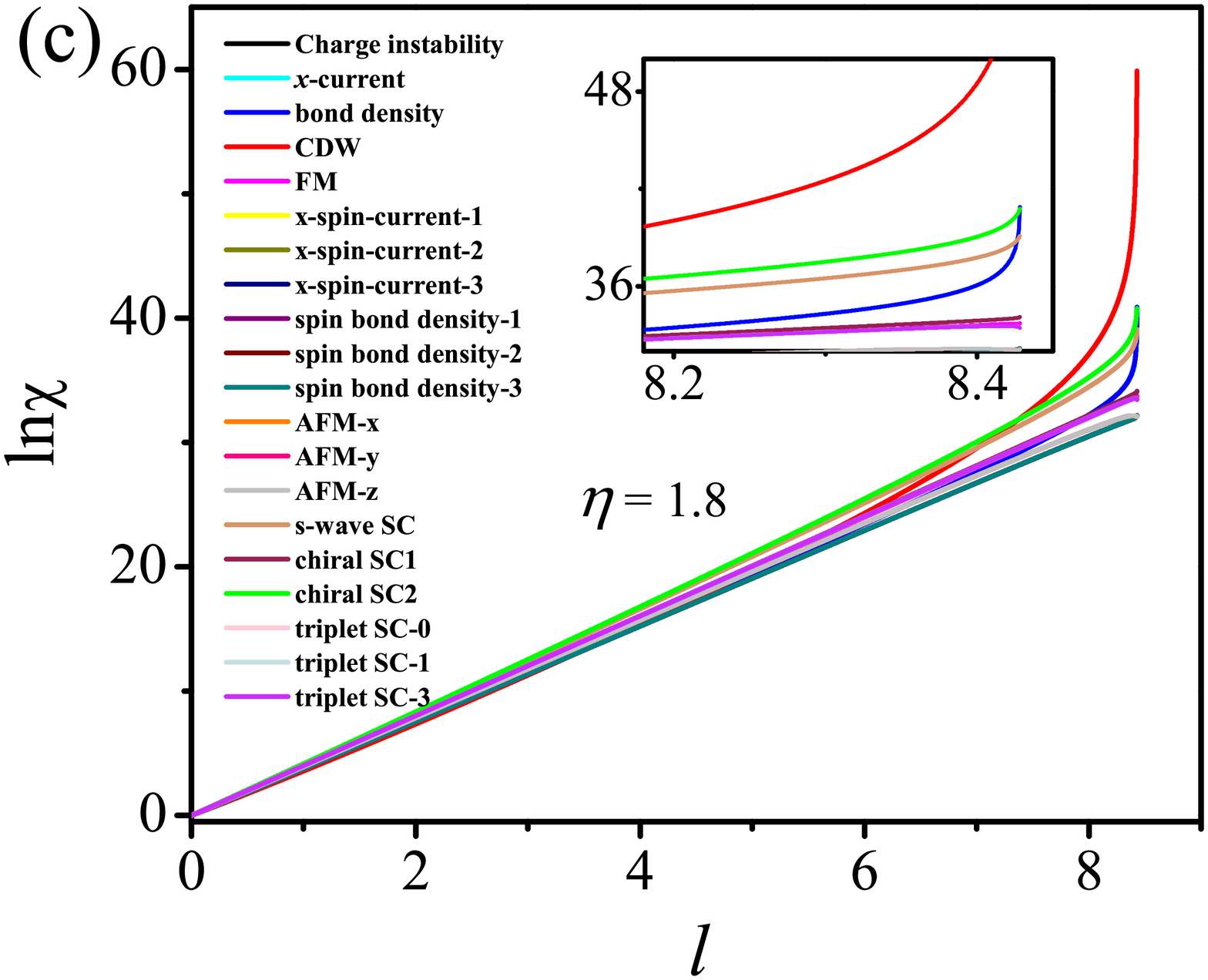,height=7.1cm,width=9.31cm}\hspace{-0.8cm}
\epsfig{file=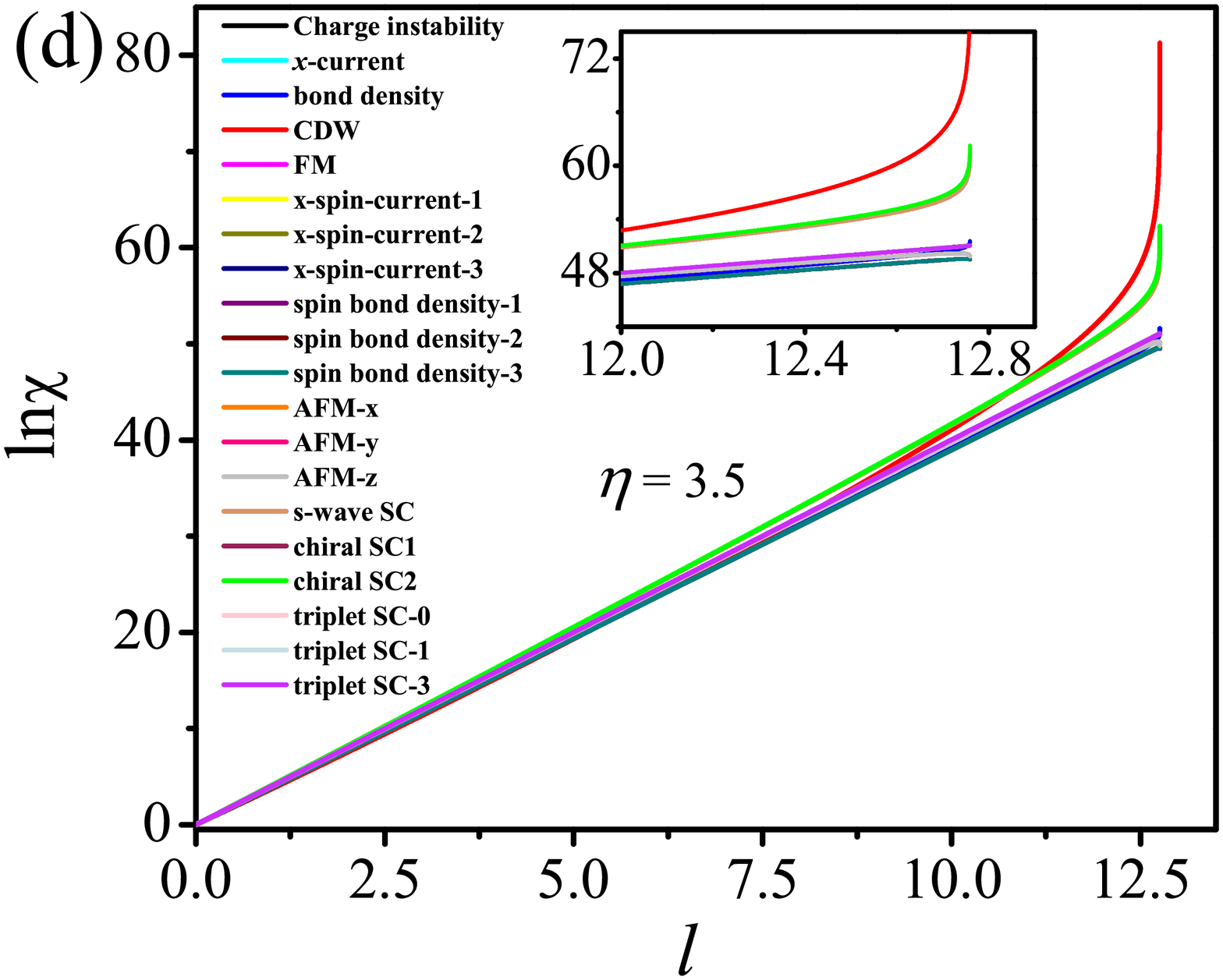,height=7.1cm,width=9.31cm}\\
\vspace{0.15cm}
\caption{(Color online) Flows of all particle-hole and particle-particle susceptibilities
catalogued in Table~\ref{Twelve_kinds_of_potential_phases} as functions of the RG evolution
parameter $l$ by approaching distinct types of RFPs classified and illustrated
in Fig.~\ref{Three_regions} and Fig.~\ref{Schematic diagram}.
The $x$-spin-current-$i$ and spin bond density-$i$ with $i=1,2,3$ as well as triplet SC-$j$ with
$j=0,1,3$ and AFM-$\zeta$ with $\zeta=x,y,z$ are employed to specify distinct components of corresponding
states as detailedly shown in Table~\ref{Twelve_kinds_of_potential_phases}.}\label{Phase_transitions}
\end{figure*}

In order to justify potential types of symmetry breaking, we are suggested to bring out the
following source terms that collect both the charge and spin channels~\cite{Vafek2010PRB,Murray2014PRB,Roy2018RRX,Roy2009.05055}
\begin{eqnarray}
S_{\mathrm{sou}}\!\!&=&\!\!\!\int \!d\tau\!\!\int \!d^{2}\mathbf{x}
\sum^4_{i=1}(\Delta^{c}_{i}\Psi^{\dag}
\mathcal{M}_{i}^{c}\Psi+\vec{\Delta}^{s}_{i}\cdot\Psi^{\dag}
\vec{\mathcal{M}}_{i}^{s}\Psi)\nonumber\\
&&+\int \!d\tau\!\!\int \!d^{2}\mathbf{x}\sum^4_{i=1}(\Delta^{\mathrm{PP}}_{i}\Psi^{\dag}
\mathcal{M}_{i}^{\mathrm{PP}}\Psi^*+\mathrm{H.c}).\label{Eq_source-term}
\end{eqnarray}
Hereby, the vertex matrixes $\mathcal{M}^{c/s}_{i}$ with $i=1,2,3,4$
denote various sorts of fermionic bilinears in the particle-hole part consisting
of both charge and spin channels. In comparison,
$\mathcal{M}_{i}^{\mathrm{PP}}$ correspond to possible fermionic bilinears
in the particle-particle situation~\cite{Vafek2010PRB,Murray2014PRB}. In addition, the couplings
$\Delta^{c/s}_{i}$ and $\Delta^{\mathrm{PP}}_{i}$ serve as
the strengths of associated fermion-source terms, which can be regarded as
order parameters accompanied by corresponding symmetry breakings. In principle,
the onset of fermionic bilinear is a manifest signal for certain instability
and hence implies a phase transition tied to some symmetry
breaking~\cite{Cvetkovic2012PRB,Maiti2010PRB, Halboth2000RPL,Halboth2000RPB,Wang2020NPB,Nandkishore2012NP}. Table~\ref{Twelve_kinds_of_potential_phases} 
catalog the primary candidates of fermion bilinears and related phase transitions for our
system.

A question is then naturally raised, which instability is the dominant one around
three different types of RFPs. To elucidate this, we need to add the source terms~(\ref{Eq_source-term})
into our effective model. In this sense, the parameters $\Delta_i$ are entangled with the
fermion-fermion interactions after taking into account one-loop fermion-fermion
corrections as diagrammatically
illustrated in Fig.~\ref{One_loop_corrections_to_source term}.
After carrying out analogous procedures in Sec.~\ref{Sec_RG_analysis}~\cite{Vafek2010PRB,Murray2014PRB},
we notice that the strength of source term would be sensitive to energy scales and subject to
the following set of RG evolutions
\begin{eqnarray}
\frac{d\Delta_{i}^{c/s,\mathrm{PP}}}{dl}=\mathcal{G}^{c/s,\mathrm{PP}}_i\Delta_{i}^{c/s,\mathrm{PP}},
\label{RG_Eqs_source}
\end{eqnarray}
where the index $i$ runs from $1$ to $4$ and
the coefficients $\mathcal{G}^{c/s,\mathrm{PP}}_i$ are closely dependent
upon the fermion-fermion interactions plus structural parameters $d_1$ and $d_3$.
The details for the flows of source terms and
coefficients $\mathcal{G}^{c/s,\mathrm{PP}}_i$
are stored in Appendix~\ref{Appendix_source}.

To proceed, we are capable of capturing the corresponding susceptibilities
around the RFPs by adopting the relationship~\cite{Vafek2010PRB,Murray2014PRB}
\begin{eqnarray}
\delta\chi=-\frac{\partial^{2}\delta f}{\partial\Delta(0)
\partial\Delta^{*}(0)},\label{Suscep}
\end{eqnarray}
with $f$ being the free energy density. Concerning the ground state
can be characterized by the susceptibility with the strongest divergence~\cite{Cvetkovic2012PRB,Murray2014PRB,Wang2017PRB, Maiti2010PRB, Altland2006Book,Vojta2003RPP,Halboth2000RPL,Halboth2000RPB, Eberlein2014PRB,Chubukov2012ARCMP,Nandkishore2012NP,Chubukov2016PRX,Roy2018RRX,Nandkishore2008.05485},
we are now in a suitable position to identify the very dominant instability and
the associated phase transition nearby the corresponding RFP by computing and comparing
the susceptibilities of all underlying instabilities listed in Table~\ref{Twelve_kinds_of_potential_phases}.

To this end, we are forced to combine the RG evolutions of fermion-fermion interactions~(\ref{Eq_RG})
and energy-dependent strengths of source terms~(\ref{RG_Eqs_source}) in conjunction with
the connections between susceptibilities and source-term couplings~(\ref{Suscep}).
After implementing long but straightforward numerical analysis,
we are eventually left with the energy-dependent susceptibilities
of all potential instabilities, 
which carry the low-energy physical information and determine the fates of all sorts of
possible instabilities nearby distinct types of RFPs as
apparently displayed in Fig.~\ref{Phase_transitions}.

Subsequently, we deliver the primary results covered in Fig.~\ref{Phase_transitions}.
At the first sight, we figure out that all kinds of susceptibilities are
fairly susceptible to energy scales and climb up quickly with accessing
any sorts of the potential RFPs. Especially, it is
of peculiar interest to address that fermion-fermion interactions
are rather in favor of charge density wave (CDW). On one hand, the CDW susceptibility
manifestly dominates over all other types once the system is tuned towards the expected RFP.
On the other, this result is qualitatively insensitive to $\eta$. In other words,
this kind of susceptibility is inevitable to be the strongest one no matter which type (region) of RFP
is approached. Consequently, we come to a conclusion schematically illustrated
in Fig.~\ref{Schematic diagram for phase transition} that the leading instability is directly
associated with a phase transition from QBCP semimetal to CDW state under the influence of
fermion-fermion couplings in the 2D QBCP materials sited on the kagom\'{e} lattice
without time-reversal symmetry. Compared to the traditional Peierls
instability activated by the spontaneous symmetry breaking of
ground state~\cite{Peierls1995Book}, this CDW instability essentially originates from
the dynamical spontaneous symmetry breaking which is generated by
the particle-hole condensations owing to marginally relevant fermion-fermion
interactions in the low-energy regime. It is worth pointing out that this result
is basically in agreement with recent works on
analogous compounds~\cite{Saran2019PRB,Tada2020arXiv,Trott2020arXiv,Sehayek2020arXiv}.
This suggests that the CDW state is another
winner driven by fermion-fermion interactions in the 2D QBCP materials besides quantum anomalous
Hall and quantum spin Hall sates in its checkerboard-lattice counterpart with time-reversal
symmetry~\cite{Murray2014PRB,Wang2017PRB}.

Next, we move to consider the subleading instabilities. Unlike the leading instability, we
learning from Fig.~\ref{Phase_transitions} find that the subleading instabilities
are of close relevance to the structural coefficient $\eta$ and
exhibit diverse fates in the vicinity of three different types of RFPs.
To be concrete, as depicted in Fig.~\ref{Phase_transitions}(a),
susceptibilities of $x$-current, bond density, and chiral SC-2 are
increased and become subdominant nearby Type-I-RFP.
With respect to Type-II-RFP, the tendencies of all instabilities share the basic fates with
Type-I-RFP once $\eta$ is small. However, while $\eta$ is increased to $1.8$,
the susceptibility of s-wave SC illustrated in Fig.~\ref{Phase_transitions}(c)
becomes comparable with that of $x$-current, bond density, and chiral SC-2 and thus can be regarded
as another subleading instability. Further, s-wave SC and chiral SC-2 are gradually enhanced with the
continue increase of $\eta$ as delineated in Fig.~\ref{Phase_transitions}(d) and become
only two subdominant ones via eliminating $x$-current and bond density at $\eta=3.5$.
As for Type-III-RFP, Fig.~\ref{Phase_transitions}(b) indicates that the qualitative results
are analogous to Type-II-RFP's at $\eta=3.5$, namely s-wave SC and chiral SC-2 are
subleading phases. At last, it is notable to highlight that the
critical energy scale ($E_c$) is rapidly decreased (i.e. $l_c$ is rapidly increased)
with tuning up the value of $\eta$. This means that an enhancement of $\eta$ is
harmful to the emergence of instability.

To be brief, the dominant instability driven
by fermion-fermion interactions is always tied to the CDW state irrespective
of the value of $\eta$. In contrast, there exist four
$\eta$-dependent candidates including the $x$-current and bond density
as well as chiral SC-2 plus s-wave SC that are subordinate to the CDW state.
Especially, we realize that the chiral SC-2 state is always subleading irrespective
of concrete value of $\eta$. Rather, both the $x$-current and bond density can only
play a subdominant role at a weak $\eta$. As the $\eta$ is progressively increased,
they share the positions with and eventually are replaced by the s-wave SC. Albeit the dominant instability
generally takes a major responsibility for the low-energy physics, these subdominant
ones might be in charge of related phenomena while the system is impacted by unexpected
facets. For the sake of completeness, we are about to investigate how the behaviors
of physical quantities are affected by the leading instability in the forthcoming section.

\vspace{0.35cm}

\section{Critical physical implications}\label{Sec_implication}

On the basis of intimate fermion-fermion interactions
in the low-energy sector, we have presented potential instabilities induced
by fermionic interactions and corroborated in Sec.~\ref{Sec_instab} a direct
connection between the overarching instability and CDW phase transition at certain RFP.
At this stage, the RFP is tantamount to a phase transition point,
at which the fluctuations are always so ferocious that usually render
a plenty of singular critical behaviors~\cite{Altland2002PR,Lee2006RMP,Fradkin2010ARCMP,Sarma2011RMP,Sachdev2011Book,
Kotov2012PMP,Mahan1990Book}. Accordingly, it is interesting to investigate possible physical implications triggered by the onset of CDW state.



\begin{figure}
\includegraphics[width=3.3in]{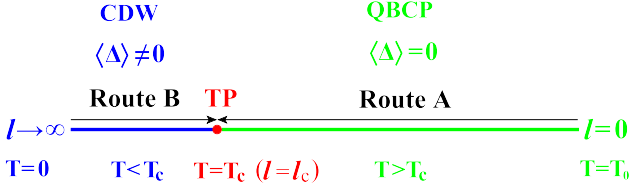}\\
\vspace{0.0cm}
\caption{(Color online) Schematic diagram for temperature-tuned phase transition
from a 2D QBCP semimetal to a CDW state. The $T_c$ and ``TP" designate the critical
temperature and the transition point (or critical point) associated with the CDW instability,
respectively. Route A and Route B exhibit
two distinct paths to access the critical point (despite $T_c$
($T_c\gtrsim 0$) is very small as $l_ {c}$ is a relatively bigger value as shown
in Fig.~\ref{Evolutions_of_fermion-fermion_interaction},
it hereby is enlarged to make a sharp contrast).}\label{Schematic diagram for phase transition}
\end{figure}

For this purpose, it is necessary to clarify the overall scenario of our basic results
sketched in Fig.~\ref{Schematic diagram for phase transition}. It consists of two subspaces separated
by the critical point with $l=l_c$ ($T=T_c$) corresponding to a disordered QBCP semimetal state
located at the region $(T_{c},T_0]$ and an ordered CDW state at $T<T_{c}$, respectively.
Obviously, there exist two routes that are Route A starting from the
disordered phase (QBCP) and Route B from the ordered phase (CDW) to
access the phase transition point (TP). In principle, one can either go along
Route A or Route B to track the effects of phase transition on the
physical observables. However, it is insensible to probe into the physical behaviors
as it approaches the critical point through Route B since the RG approach
is strictly based upon the perturbative theory.
In this respect, we within this section go along with Route A to tentatively investigate the
physical observables on the right side of critical point with $T>T_{c}$.
Given that the band structure and dispersion of the QBCP system are stable
in region $T>T_{c}$, but destroyed at $T<T_{c}$, this strategy at
least can provide a preliminary understanding
of criticality nearby CDW instability.

Technically, it seems inappropriate to capture the effects of
CDW instability for the region  $T>T_{c}$ since the
average value of order parameter $\Delta$ vanishes
at $T>T_c$ as illustrated in Fig.~\ref{Schematic diagram for phase transition}.
However, it is worth pointing out
the fluctuation of order parameter is nonzero albeit $\langle\Delta\rangle=0$, and
becomes stronger and stronger as the critical point
is approached along with Route A, eventually diverges at the critical
point. In this sense, we are suggested to regard $\Delta$ as a fluctuation of CDW order
parameter, which is generated by the fermion-fermion interactions, to collect
the influences of CDW instability, and then examine how the physical quantities
vary by adjusting the value of $\Delta$ (it is equivalent to
approaching the TP along with Route A). To this end, we introduce
a constant $\Delta$ to represent the fluctuation associated with the leading stability
and add it by hand into the fermionic free propagator of the 2D QBCP system~\cite{Wang1911.09654}.
Then the free fermionic propagator~(\ref{G_0}) dressed by one-loop
corrections is thus recast as
\begin{eqnarray}
G_{0}(i\omega_{n},\mathbf{k})
\!\!&=&\!\!\Bigl(-i\omega_{n}+(d_{3}\mathbf{k}^{2}+\Delta)\Sigma_{03}
+d_{1}\Sigma_{01}(k_{x}^{2}-k_{y}^{2})\nonumber\\
&&+d_{2}\Sigma_{02}k_{x}k_{y}\Bigr)^{-1}\!\!\!,
\end{eqnarray}
where $\omega_{n}=(2n+1)\pi T$ with $n$ being an integer
stands for the Matsubara frequency. To proceed, performing
analytical continuation $i\omega_{n}\rightarrow\omega_{n}+i\delta$,
we are subsequently left with the retarded fermion propagator as follows~\cite{Mahan1990Book},
\begin{widetext}
\begin{eqnarray}
G_{0}^{\mathrm{ret}}(\omega_{n},\mathbf{k})
&=&\frac{(\omega_{n}+i\delta)+(d_{3}k^{2}
+\Delta)\Sigma_{03}+d_{1}k^{2}(\cos2\theta\Sigma_{01}
+\sin2\theta\Sigma_{02})}{(-\omega_{n}^{2}-2i\omega_{n}\delta+
d_{3}^{2}k^{4}+2d_{3}k^{2}\Delta
+\Delta^{2}\!+\!d_{1}^{2}k^{4})}.\label{Eq_G_0_ret}
\end{eqnarray}
\end{widetext}
It is now in a suitable position to extract the qualitative effects sparked by the
formation of order parameter on the physical quantities.
Without loss of generality, we within this section put our focus on the density of states (DOS)
as well as specific heat and compressibility of quasiparticles.

\begin{figure}
\centering
\epsfig{file=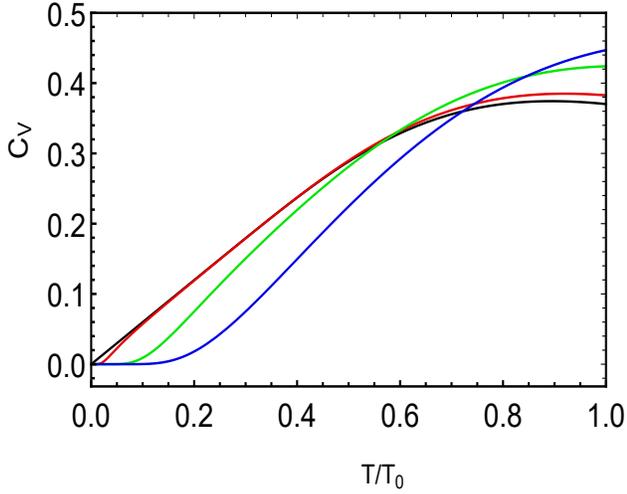,height=6.5cm,width=8.2cm}\vspace{0.55cm}\\
\vspace{-0.55cm}
\caption{(Color online) Temperature dependence of the specific heat $C_{V}$ accessing the
Type-II-RFP under the CDW fluctuation. The black, red, green, and blue lines
correspond to $\Delta^{\prime}=0$, $\Delta^{\prime}=0.1$, $\Delta^{\prime}=0.5$,
and $\Delta^{\prime}=1$, respectively (the qualitative results for Type-I-RFP
are analogous and hence not shown here).}\label{CV-RFP-II}
\end{figure}


\subsection{DOS}

At first, let us concentrate on the DOS under the influence of
CDW fluctuation with approaching the critical point through Route A.
To achieve this goal, one necessitates the corresponding spectral function, which
is directly connected to the retarded fermion
propagator~(\ref{Eq_G_0_ret})~\cite{Wang2012PRD,Mahan1990Book} and
of the following form
\begin{eqnarray}
&&\mathcal{A}(\omega_{n},\mathbf{k})\nonumber\\
\!&=&\!-\frac{1}{\pi}\mathrm{Tr}\left(\mathrm{Im}G_{0}^{\mathrm{ret}}
(\omega_{n},\mathbf{k})\right)\nonumber\\
\!&=&\!4N|\omega_{n}|\delta(d_{3}^{2}k^{4}\!+\!2d_{3}
k^{2}\Delta\!+\!\Delta^{2}\!+\!d_{1}^{2}k^{4}\!-\!\omega_{n}^{2}),\label{Spectral_function_of_DOS}
\end{eqnarray}
where the number $N$ characterizes the fermion flavor. With this respect, the DOS of quasiparticles consequently can be cast as
\begin{eqnarray}
\rho(\omega_{n})
&=&N\int_{0}^{\Lambda_{0}}\int_{0}^{2\pi}\frac{kdkd\theta}
{(2\pi)^{2}}\mathcal{A}
(\omega_{n},\mathbf{k}).\label{The_density_of_state}
\end{eqnarray}

Prior to inspecting the impact of order parameter, it is necessary to briefly
discuss the circumstance in the absence of instability. Removing the fluctuation via
taking $\Delta\rightarrow0$ limit in Eq.~(\ref{The_density_of_state}) directly gives rise to
\begin{eqnarray}
\rho(\omega_{n})=\frac{N}{2\pi\sqrt{d_{3}^{2}+d_{1}^{2}}}.\label{DOS_(i)}
\end{eqnarray}
This is in reminiscence of the fact that the DOS of 2D QBCP system is
a finite constant at the Fermi surface~\cite{Fradkin2009PRL,Murray2014PRB}.

Subsequently, we go to examine the very influence of CDW fluctuation. Based upon the general
result~(\ref{The_density_of_state}) in tandem with the essential properties of $\delta$ function,
we notice that the DOS would be broken down into two distinct situations depending upon $\Delta$'s magnitude.
On one hand, one can find the $\omega_n$ or temperature ($T$) dependence of DOS at $|\Delta|<T$ is
rewritten as follows
\begin{eqnarray}
\rho(\omega_{n})
=\frac{N}{2\pi\sqrt{d_{3}^{2}
+d_{1}^{2}(1-\frac{\Delta^{2}}{\omega_{n}^{2}})}}.\label{DOS_(ii)}
\end{eqnarray}
On the other hand, a large order parameter with $|\Delta|>T$ is of particular detriment to
DOS which disappears exactly at the instability, namely
\begin{eqnarray}
\rho(\omega_{n})=0.\label{DOS_(iii)}
\end{eqnarray}

These manifestly shed light on the important role of CDW fluctuation in the proximity of instability.
Compared to a finite DOS for 2D QBCP materials with $\Delta=0$, we figure out that it nearby
Fermi surface (QBCP) is slightly enhanced while the system is a little far away from instability
with $|\Delta|<T$. In addition, its concrete values as clearly designated in Eqs.~(\ref{DOS_(i)})-(\ref{DOS_(ii)}) are heavily dependent upon two microscopic
parameters $d_{3}$ and $d_{1}$, which are closely associated with different types
of RFPs. However, once the RFP is sufficiently approached with $|\Delta|>T$,
the onset of large order parameter substantially suppresses the DOS due to ferocious fluctuations.
In other words, the band structure of 2D QBCP systems would be
completely sabotaged~\cite{Wang2020-arxiv,Altland2006Book}. It is
worth emphasizing that we have checked all sorts of RFPs share the analogous qualitative results.


\subsection{Specific heat}\label{Sec_C_v}

Next, we are going to shift our target to the specific heat of quasiparticles.
For the sake of completeness, we hereby bring out an
infinitesimal chemical potential $\mu$ into our effective theory~\cite{Wang2012PRD,Kapusta1994Book}.
As a result, the corresponding free fermionic propagator in the Matsubara
formalism is reformulated as
\begin{widetext}
\begin{eqnarray}
G_{0}(i\omega_{n}, \mathbf{k})=\frac{1}{i\omega_{n}+\mu-\mathcal{H}_{0}(k)
-\Delta\Sigma_{03}}
=-\frac{i\omega_{n}+\mu+(d_{3}k^{2}+\Delta)\Sigma_{03}
+d_{1}\Sigma_{01}k^{2}\cos2\theta
+d_{1}\Sigma_{02}k^{2}\sin2\theta}{(\omega_{n}-i\mu)^{2}+
(d_{3}k^{2}+\Delta)^{2}+d_{1}^{2}k^{4}}.
\end{eqnarray}
\end{widetext}

Following the tactic in Ref.~\cite{Wang2012PRD}, we integrate over all frequencies
and then write the free energy of the fermions as
\begin{eqnarray}
f(T,\mu)\!\!=\!\!-2N\!\sum_{\alpha\pm1}\!\int\!\!\!\frac{d^{2}\mathbf{k}}
{(2\pi)^{2}}\!\!\left[\!\epsilon(\mathbf{k})\!+\!T\!\ln\!\left(\!1\!+\!
e^{-\frac{\epsilon(\mathbf{k})+\alpha\mu}{T}}\!\right)\!\right],
\end{eqnarray}
where the energy is designated as
\begin{eqnarray}
\epsilon(\mathbf{k})\equiv\sqrt{(d_{3}\mathbf{k}^{2}+\Delta)^{2}+d_{1}^{2}\mathbf{k}^{4}}.\label{Eq_epsilon}
\end{eqnarray}
To simplify our study, we take advantage of transformation $f(T)-f(0)\rightarrow f(T)$ to
eliminate the zero-point energy and obtain a compact free energy as follows
\begin{eqnarray}
f(T,\mu)\!\!=\!\!-2NT\!\sum_{\alpha\pm1}\!\int_{0}^{\Lambda_{0}}
\!\!\!\!\!\int_{0}^{2\pi}\!\!\frac{kdkd\theta}{(2\pi)^{2}}\ln\left
(1\!+\!e^{-\frac{\epsilon(\mathbf{k})+\alpha\mu}{T}}\right).\,\,\,
\label{The_free_energy_with_finite_chemical_potential}
\end{eqnarray}

In this sense, we hereafter only pour our attention into $\mu=0$
situation for specific heat, which is based upon two points. On one side, the
starting point~(\ref{S_eff}) is restricted to zero chemical potential
at the QBCP. On the other side, this work only concerns qualitative phenomena of physical
implications triggered by an instability irrespective of the value of $\mu$. To proceed,
supposing $\mu=0$ in Eq.~(\ref{The_free_energy_with_finite_chemical_potential}) gives
rise to
\begin{eqnarray}
f(T)\!=\!-4NT\int_{0}^{\Lambda_{0}}\!\int_{0}^{2\pi}\frac{kdkd\theta}{(2\pi)^{2}}
\ln\left(1+e^{-\frac{\epsilon(\mathbf{k})}{T}}\right).
\label{The_free_energy_with_zero_chemical_potential}
\end{eqnarray}
Taking the derivatives of free energy with respect to temperature forthrightly yields to the
specific heat ($C_V$)~\cite{Wang2012PRD,Kapusta1994Book}
\begin{eqnarray}
C_{V}(T)=-T\frac{\partial^{2}f(T)}{\partial T^{2}}
=\frac{2N}{\pi T^{2}}\int_{0}^{\Lambda_{0}} \frac{kdk\epsilon^2(\mathbf{k})e^{\frac{\epsilon(\mathbf{k})}
{T}}}{\left(e^{\frac{\epsilon(\mathbf{k})}
{T}}+1\right)^{2}}.
\end{eqnarray}

In order to facilitate our calculations, it is convenient to rescale the momentum and order parameter
with the cutoff temperature $T_0$ that is related to the cutoff $\Lambda_0$ by
$T_{0}\equiv\Lambda_{0}^{2}$, namely $k^{\prime}\equiv k/\sqrt{T_{0}}$ and $\Delta^{\prime}\equiv\Delta/T_{0}$~\cite{Wang2012PRD}.
As a corollary, we can convert the $C_V(T)$ into the following form
\begin{eqnarray}
C_{V}(T)&=&\frac{2NT_{0}^{3}}{\pi T^{2}}\int_{0}^{1} \frac{k^{\prime}dk^{\prime}\epsilon'^2(\mathbf{k}',\Delta')
e^{\frac{\epsilon'(\mathbf{k}',\Delta')}{T/T_{0}}}}
{\left(e^{\frac{\epsilon'(\mathbf{k}',\Delta')}
{T/T_{0}}}+1\right)^{2}},\label{General_specific_heat}
\end{eqnarray}
where the $\epsilon'$ is designated as
\begin{eqnarray}
\epsilon'(\mathbf{k}',\Delta')\equiv\sqrt{(d_{3}k^{\prime2}
+\Delta^{\prime})^{2}+d_{1}^{2}k^{\prime4}}.\label{Eq_epsilon-2}
\end{eqnarray}

On the basis of the general expression for $C_V(T)$~(\ref{General_specific_heat}), 
a few comments on the specific heat are addressed under the fluctuation of an
order parameter kindled by the CDW instability. We at first tackle the limit case
with $\Delta^{\prime}=0$ (i.e., the 2D QBCP state). In this circumstance, it is fortunate that
the analytical result~(\ref{General_specific_heat}) can be obtained by integrating out the momenta
\begin{widetext}
\begin{eqnarray}
C_{V}(T)\!\!&=&\!\!\frac{2NT_{0}^{\frac{3}{2}}}{\pi \sqrt{T}}\!\frac{\left[6\frac{(d_{3}^{2}+d_{1}^{2})}
{(T/T_{0})^2}\frac{e^{\frac{\sqrt{d_{3}^{2}+d_{1}^{2}}}{T/T_{0}}}
}{1+e^{\frac{\sqrt{d_{3}^{2}+d_{1}^{2}}}{T/T_{0}}}}-12\frac{\sqrt{d_{3}^{2}
+d_{1}^{2}}}{T/T_{0}}\log\left(1+e^{\frac{\sqrt{d_{3}^{2}+d_{1}^{2}}}
{T/T_{0}}}\right)-12\mathrm{Li}_{2}
\left(-e^{\frac{\sqrt{d_{3}^{2}+d_{1}^{2}}}{T/T_{0}}}\right)-\pi^{2}\right]}
{12(d_{3}^{2}+d_{1}^{2})^{\frac{1}{2}}}
\!,\label{specific_heat-limit}
\end{eqnarray}
\end{widetext}
with $\mathrm{Li}_{2}(z)$ corresponding to a polylogarithmic function~\cite{Gradshteyn2007Book}.
This will be utilized to compare with its $\Delta'\neq0$ counterparts.

\begin{figure}
\centering
\epsfig{file=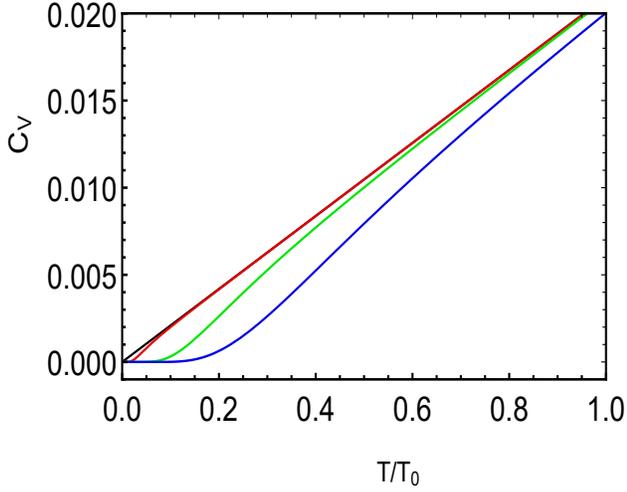,height=6.5cm,width=8.2cm}\\
\vspace{-0.05cm}
\caption{(Color online) Temperature dependence of the specific heat $C_{V}$ approaching the
Type-III-RFP under the CDW fluctuation. The black, red, green, and blue lines
correspond to $\Delta^{\prime}=0$, $\Delta^{\prime}=0.1$, $\Delta^{\prime}=0.5$,
and $\Delta^{\prime}=1$, respectively.}\label{CV-RFP-III}
\end{figure}

Subsequently, we endeavor to investigate the nontrivial situation with a moderate order parameter once
the system is adjacent to RFP (not exactly accessed). Carrying out the numerical analysis
of Eqs.~(\ref{General_specific_heat}) and (\ref{specific_heat-limit}) yields several
interesting features shown in Fig.~\ref{CV-RFP-II} and Fig.~\ref{CV-RFP-III}.
Studying from Fig.~\ref{CV-RFP-II}, one broadly realizes that the fates of $C_{V}(T)$
around Type-II-RFP are fairly dependent
upon $\Delta^{\prime}$ (the qualitative result for Type-I-RFP is analogous and
thus not shown here). In the low-temperature region,
the order parameter apparently hampers the specific heat. On the contrary, $C_V$ gains a slight
lift in the high-temperature region. In comparison, Fig.~\ref{CV-RFP-III}
displays that $\Delta^{\prime}$ always brings some detriments to
$C_V$ as the system is close to the Type-III-RFP. In particular,
CDW instability drives the specific heat $C_V(T)\rightarrow0$ as $T\rightarrow T_c$,
which is a finite constant in the absence of fluctuation.
Eventually, we examine the limit $\Delta^{\prime} \gg1$ (namely $\Delta\gg T_{0}$), which
would be ignited once the RFP is approached. In this respect, combining Eq.~(\ref{General_specific_heat})
with this limit is expected to grasp the central point of specific heat
\begin{eqnarray}
C_{V}(T)&\approx&\frac{NT_{0}^{3}\Delta^{\prime2}}{\pi T^{2}e^{\frac{\Delta^{\prime}}{T/T_{0}}}},
\end{eqnarray}
which signals $\lim_{\Delta^{\prime}\rightarrow\infty}C_V(T)\rightarrow0$. It manifestly indicates that $C_V$ is profoundly reduced by the strong fluctuation of
order parameter at the CDW instability.

\subsection{Compressibility}\label{Sec_compress}

\begin{figure}
\centering
\epsfig{file=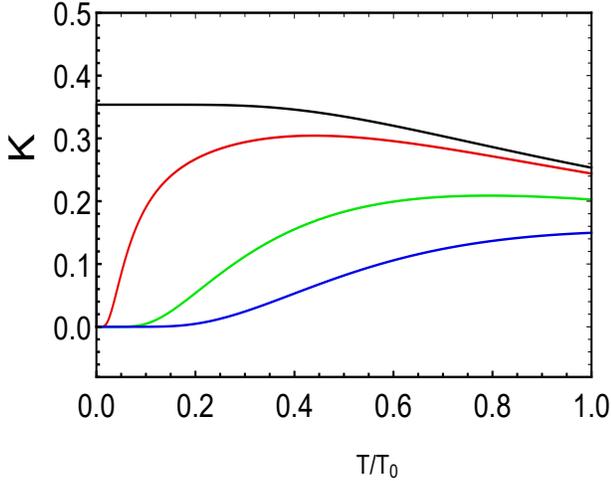,height=6.39cm,width=8.0cm}
\vspace{-0.05cm}
\caption{(Color online) Temperature dependence of the compressibility $\kappa$
approaching the Type-II-RFP under the CDW fluctuation. The black, red, green, and blue lines
correspond to $\Delta^{\prime}=0$, $\Delta^{\prime}=0.1$, $\Delta^{\prime}=0.5$,
and $\Delta^{\prime}=1$, respectively (the basic conclusions for both
Type-I-RFP and Type-III-RFP are similar and hence not shown here).}\label{The_diagram_of_compressibility_around_Type-III-RFP}
\end{figure}

At last, with the help of temperature- and $\mu$-dependent free energy derived
in Eq.~(\ref{The_free_energy_with_finite_chemical_potential}),
we are allowed to verify how the compressibility of quasiparticles
labeled by $\kappa$ behaves in the proximity of the leading instability.
Concretely, the temperature dependence of $\kappa$ reads~\cite{Wang2012PRD,Sheehy2007PRL,Schwabl2006Book,Hwang2007PRL}
\begin{eqnarray}
\kappa(T)=\!-\!\left.\frac{\partial^{2}f(T,\mu)}{\partial\mu^{2}}\right|_{\mu=0}
\!\!\!=\frac{2N}{\pi T}\!\!\int_{0}^{\Lambda_{0}}\!\!\!kdk \frac{e^{\frac{\epsilon(\mathbf{k})
}{T}}}{\left(e^{\frac{\epsilon(\mathbf{k})}{T}}+1\right)^{2}},
\end{eqnarray}
with $\epsilon(\mathbf{k})$ being denominated in Eq.~(\ref{Eq_epsilon}) for the absence of
chemical potential. We then adopt the same rescalings employed in Sec.~\ref{Sec_C_v}
and are left with
\begin{eqnarray}
\kappa(T)
\!\!=\!\!\frac{2N}{\pi(T/T_{0})}\!\int_{0}^{1}\!\!k^{\prime}dk^{\prime}\! \frac{e^{\frac{\epsilon'(\mathbf{k}',\Delta')}{T/T_{0}}}}
{\left(e^{\frac{\epsilon'(\mathbf{k}',\Delta')}{T/T_{0}}}+1\right)^{2}},\,\,\,\label{The_compressibility}
\end{eqnarray}
where $\epsilon'(\mathbf{k}',\Delta')$ is designated in Eq.~(\ref{Eq_epsilon-2}).

Accordingly, the analytical expressions for $\Delta^{\prime}=0$ and
$\Delta^{\prime}\gg 1$ can also be easily obtained,
\begin{eqnarray}
\kappa(T,\Delta^{\prime}=0)
&=&\frac{N\tanh\left(\frac{\sqrt{d_{3}^{2}+d_{1}^{2}}}
{2(T/T_{0})}\right)}{2\pi\sqrt{d_{3}^{2}+d_{1}^{2}}},\\
\kappa(T,\Delta^{\prime}\gg 1)
&\approx&\frac{N}{\pi(T/T_{0})} e^{-\frac{\Delta^{\prime}}{T/T_{0}}}.\label{Eq_kappa_large}
\end{eqnarray}
Numerically implementing Eq.~(\ref{The_compressibility}) leads to Fig.~\ref{The_diagram_of_compressibility_around_Type-III-RFP}, which implies that
the compressibility is severely suppressed by a finite order parameter.
Additionally, Eq.~(\ref{Eq_kappa_large}) proposes that the compressibility goes
toward vanishment because of the divergent fluctuation of order parameter exactly at the RFP.
It is worth emphasizing that all these basic conclusions
are insensitive to concrete values of RFP. In such circumstance,
this phenomenon can be regarded as the third critical behavior driven by fermion-fermion
interactions~\cite{Wang2012PRD, Altland2006Book}.

To recapitulate, the primary task of this work is to judge whether and
which kind of phase transition can be induced by the fermion-fermion interactions,
which, to a large extent, is basically finished in Sec.~\ref{Sec_RFP} and
Sec.~\ref{Sec_instab}. In comparison, the physical behaviors around the
critical point are only our secondary concerns. Although
the strategy employed in this section may be not the best one, we are able to
qualitatively capture the behaviors of these
physical quantities via regarding the parameter $\Delta$ as a fluctuation and tuning the variation
of $\Delta$ to simulate the access of critical point along with Route A illustrated in Fig.~\ref{Schematic diagram for phase transition}.

\vspace{0.8cm}

\section{Summary}\label{Sec_summary}

In summary, we attentively verify how the low-energy properties of 2D spin-$1/2$ QBCP fermionic
systems on the kagom\'{e} lattice are impacted by all sixteen sorts of marginal
fermion-fermion interactions. For the purpose of treating these degrees of freedom
on the same footing, we resort to the momentum-shell RG method~\cite{Shankar1994RMP,Wilson1975RMP,Polchinski9210046},
which is a well-trodden strategy for the description of hierarchical
physics under the coexistence of multiple sorts of interactions.
In the spirit of standard RG analysis, a set of coupled flow equations of all
interaction strengths are derived by taking into account one-loop corrections
of the correlated Feynman diagrams. After vigilantly analyzing these RG evolutions,
several physical properties ignited by marginally relevant fermion-fermion interactions are addressed
in the low-energy sector.

At first, we notice that some sorts of fermion-fermion couplings
coalesce with the decrease of energy scale due to their intimate correlations.
This tenders just six of them can flow independently and evolve divergently
with lowering energy scales. In this respect, we only need to contemplate the energy-dependent
trajectories of six nontrivial fermionic couplings.
In order to work in the perturbative theory, it is convenient to rescale these strong-coupling
interactions with a non-sign changed parameters (such as $\lambda_{30}$) and obtain the
relative flows of interaction parameters as well as their RFPs that directly govern the critical
physics~\cite{Vafek2010PRB,Murray2014PRB,Wang2017PRB}.
In particular, these RFPs are of close association with
the coefficient composed by two structure parameters (i.e., $\eta\equiv d_{3}/d_{1}$).
To be concrete, Fig.~\ref{Three_regions} and Fig.~\ref{Schematic diagram} unambiguously
manifest three qualitatively distinct $\eta-$dependent regions,
which are named as Type-I-Region, Type-II-Region, and Type-III-Region residing in $\eta<C_{1}$,
$C_{1}<\eta<C_{2}$, and $\eta>C_{2}$, respectively. Focusing on the vicinity of RFPs
in these three different regions, we then carefully investigate the underlying instabilities, which
are accompanied by corresponding symmetry breakings and tied to related phase
transitions~\cite{Cvetkovic2012PRB,Murray2014PRB,Wang2017PRB, Maiti2010PRB, Altland2006Book,Vojta2003RPP,Halboth2000RPL,Halboth2000RPB, Eberlein2014PRB,Chubukov2012ARCMP,Chubukov2016PRX}. On the basis of numerical RG studies together
with comparisons of susceptibilities, we, reading off Fig.~\ref{Phase_transitions},
draw a conclusion that the CDW instability is always dominant over all other
candidates irrespective of the specific value of $\eta$.
In contrast, four subleading ones are also manifestly proposed, which
include $x$-current, bond density, chiral SC-2, and s-wave SC depending upon the
variation of $\eta$.
Furthermore, as the leading instability is accessed along with Route A shown
in Fig.~\ref{Schematic diagram for phase transition}, we tentatively examine the effects of
fluctuation induced by the phase transition from a QBCP semimetal to CDW state
on physical implications consisting of DOS and specific heat as well as
compressibility. To be specific, the development of CDW state is very harmful to
these three kinds of physical observables. Especially, they are all
considerably suppressed and even vanish once the fluctuation of
order parameter is strong enough in the vicinity of
CDW instability. These results are reminiscent of quantum critical behaviors in the low-energy regime~\cite{Altland2002PR,Lee2006RMP,Fradkin2010ARCMP,Sarma2011RMP,Sachdev2011Book,
Kotov2012PMP,Mahan1990Book}.

We wish these studies would supplement current understandings of 2D QBCP semimetals and
open helpful routes to promote further research of 2D QBCP materials as well as
explore their cousin materials in the future.

\section*{ACKNOWLEDGEMENTS}

J.W. is partially supported by the National Natural
Science Foundation of China under Grant No. 11504360.
The authors would like to thank Jie-Qiong Li, Xiao-Yue Ren,
and Yao-Ming Dong for helpful discussions.

\vspace{0.5cm}

\section*{AUTHOR CONTRIBUTIONS}

J. W. initiated and supervised the project as well as performed the numerical analysis and wrote the manuscript. Y.H.Z carried out the analytical calculations and plotted figures.

\section*{ADDITIONAL INFORMATION}

\textbf{Competing interests:} The authors declare no Competing Financial or Non-Financial Interests.


\vspace{0.5cm}
\appendix

\section{One-loop corrections}\label{Appendix_one-loop_corrections}

The one-loop corrections to self-energy and fermion-fermion interactions are depicted in Fig.~\ref{Fig_fermion_propagator_and_fermion_interaction_correction}. After
long but straightforward calculations, we obtain~\cite{Murray2014PRB,Wang2017PRB}
\begin{widetext}
\begin{small}
\begin{eqnarray}
S_{00}&=&-\int_{-\infty}^{\infty}\frac
{d\omega_{1}
d\omega_{2}d\omega_{3}}{(2\pi)^{3}}\int\frac{d^{2}\mathbf{k}_{1}
d^{2}\mathbf{k}_{2}d^{2}\mathbf{k}_{3}}{(2\pi)^{6}}\Psi^{\dag}
(\omega_{1},\mathbf{k}_{1})\Sigma_{00}\Psi
(\omega_{2},\mathbf{k}_{2})
\Psi^{\dag}(\omega_{3},\mathbf{k}_{3})
\Sigma_{00}\Psi(\omega_{1}+\omega_{2}-\omega_{3}, \mathbf{k}_{1}+\mathbf{k}_{2}-\mathbf{k}_{3})\nonumber\\
&&\times[2d_{3}^{2}(\lambda_{00}\lambda_{03}\!+\!\lambda_{10}
\lambda_{13}\!+\!\lambda_{20}\lambda_{23}\!+\!\lambda_{30}\lambda_{33})
+d_{1}^{2}(\lambda_{00}\lambda_{01}\!+\!\lambda_{10}
\lambda_{11}+\lambda_{20}\lambda_{21}+\lambda_{30}\lambda_{31}
+\lambda_{00}\lambda_{02}+\lambda_{10}
\lambda_{12}+\lambda_{20}\lambda_{22}\nonumber\\
&&+\lambda_{30}\lambda_{32})]
\frac{l}{4\pi(d_{3}^{2}+d_{1}^{2})^{\frac{3}{2}}},\label{One_loop_Eq_S_00}\\
S_{10}&=&\int_{-\infty}^{\infty}\frac
{d\omega_{1}
d\omega_{2}d\omega_{3}}{(2\pi)^{3}}\int\frac{d^{2}\mathbf{k}_{1}
d^{2}\mathbf{k}_{2}d^{2}\mathbf{k}_{3}}{(2\pi)^{6}}\Psi^{\dag}
(\omega_{1},\mathbf{k}_{1})\Sigma_{10}\Psi
(\omega_{2},\mathbf{k}_{2})
\Psi^{\dag}(\omega_{3},\mathbf{k}_{3})
\Sigma_{10}\Psi(\omega_{1}+\omega_{2}-\omega_{3}, \mathbf{k}_{1}+\mathbf{k}_{2}-\mathbf{k}_{3})\nonumber\\
&&\times\{2d_{3}^{2}[\lambda_{20}\lambda_{30}
+\lambda_{21}\lambda_{31}+\lambda_{22}
\lambda_{32}+\lambda_{23}\lambda_{33}
-(\lambda_{13}\lambda_{00}+\lambda_{10}
\lambda_{03}+\lambda_{21}\lambda_{32}+\lambda_{22}
\lambda_{31})]
+d_{1}^{2}[2(\lambda_{20}\lambda_{30}
+\lambda_{21}\lambda_{31}\nonumber\\
&&\!+\!\lambda_{22}
\lambda_{32}\!+\!\lambda_{23}\lambda_{33})
\!-(\lambda_{01}\lambda_{10}+\lambda_{00}
\lambda_{11}+\lambda_{22}\lambda_{33}
+\lambda_{23}\lambda_{32}+\lambda_{12}\lambda_{00}+\lambda_{10}
\lambda_{02}+\lambda_{23}\lambda_{31}+\lambda_{21}
\lambda_{33})]\}\frac{l}
{4\pi(d_{3}^{2}+d_{1}^{2})^{\frac{3}{2}}},\label{One_loop_Eq_S_10}\\
S_{20}&=&\int_{-\infty}^{\infty}\frac
{d\omega_{1}
d\omega_{2}d\omega_{3}}{(2\pi)^{3}}\int\frac{d^{2}\mathbf{k}_{1}
d^{2}\mathbf{k}_{2}d^{2}\mathbf{k}_{3}}{(2\pi)^{6}}\Psi^{\dag}
(\omega_{1},\mathbf{k}_{1})\Sigma_{20}\Psi
(\omega_{2},\mathbf{k}_{2})
\Psi^{\dag}(\omega_{3},\mathbf{k}_{3})
\Sigma_{20}\Psi(\omega_{1}+\omega_{2}-\omega_{3}, \mathbf{k}_{1}+\mathbf{k}_{2}-\mathbf{k}_{3})\nonumber\\
&&\times\{2d_{3}^{2}[\lambda_{30}\lambda_{10}+\lambda_{31}\lambda_{11}
+\lambda_{32}\lambda_{12}+\lambda_{33}\lambda_{13}
-(\lambda_{23}\lambda_{00}+\lambda_{20}\lambda_{03}
+\lambda_{31}\lambda_{12}+\lambda_{32}\lambda_{11})]
+d_{1}^{2}[2(\lambda_{30}\lambda_{10}+\lambda_{31}\lambda_{11}\nonumber\\
&&\!+\!\lambda_{32}\lambda_{12}\!+\!\lambda_{33}\lambda_{13})
\!-(\lambda_{21}\lambda_{00}+\lambda_{20}\lambda_{01}
+\lambda_{32}\lambda_{13}+\lambda_{33}\lambda_{12}
+\lambda_{22}\lambda_{00}+\lambda_{20}\lambda_{02}
+\lambda_{33}\lambda_{11}+\lambda_{31}\lambda_{13})]\}
\frac{l}{4\pi(d_{3}^{2}+d_{1}^{2})^{\frac{3}{2}}},\label{One_loop_Eq_S_20}\\
S_{30}&=&\int_{-\infty}^{\infty}\frac
{d\omega_{1}
d\omega_{2}d\omega_{3}}{(2\pi)^{3}}\int\frac{d^{2}\mathbf{k}_{1}
d^{2}\mathbf{k}_{2}d^{2}\mathbf{k}_{3}}{(2\pi)^{6}}\Psi^{\dag}
(\omega_{1},\mathbf{k}_{1})\Sigma_{30}\Psi
(\omega_{2},\mathbf{k}_{2})
\Psi^{\dag}(\omega_{3},\mathbf{k}_{3})
\Sigma_{30}\Psi(\omega_{1}+\omega_{2}-\omega_{3}, \mathbf{k}_{1}+\mathbf{k}_{2}-\mathbf{k}_{3})\nonumber\\
&&\times\{2d_{3}^{2}[\lambda_{10}\lambda_{20}+\lambda_{11}\lambda_{21}
+\lambda_{12}\lambda_{22}+\lambda_{13}\lambda_{23}
-(\lambda_{33}\lambda_{00}+\lambda_{30}\lambda_{03}
+\lambda_{11}\lambda_{22}+\lambda_{12}\lambda_{21})]
+d_{1}^{2}[2(\lambda_{10}\lambda_{20}+\lambda_{11}\lambda_{21}\nonumber\\
&&\!+\!\lambda_{12}\lambda_{22}\!+\!\lambda_{13}\lambda_{23})
\!-(\lambda_{31}\lambda_{00}+\lambda_{30}\lambda_{01}
+\lambda_{12}\lambda_{23}+\lambda_{13}\lambda_{22}
+\lambda_{32}\lambda_{00}+\lambda_{30}\lambda_{02}
+\lambda_{13}\lambda_{21}+\lambda_{11}\lambda_{23})]\}
\frac{l}{4\pi(d_{3}^{2}+d_{1}^{2})^{\frac{3}{2}}},\label{One_loop_Eq_S_30}\\
S_{01}&=&\int_{-\infty}^{\infty}\frac
{d\omega_{1}
d\omega_{2}d\omega_{3}}{(2\pi)^{3}}\int\frac{d^{2}\mathbf{k}_{1}
d^{2}\mathbf{k}_{2}d^{2}\mathbf{k}_{3}}{(2\pi)^{6}}\Psi^{\dag}
(\omega_{1},\mathbf{k}_{1})\Sigma_{01}\Psi
(\omega_{2},\mathbf{k}_{2})
\Psi^{\dag}(\omega_{3},\mathbf{k}_{3})
\Sigma_{01}\Psi(\omega_{1}+\omega_{2}-\omega_{3},
\mathbf{k}_{1}+\mathbf{k}_{2}-\mathbf{k}_{3})\nonumber\\
&&\times\{4d_{3}^{2}[\lambda_{01}\lambda_{00}
+\lambda_{01}\lambda_{10}+\lambda_{01}\lambda_{11}
+\lambda_{01}\lambda_{20}+\lambda_{01}\lambda_{21}
+\lambda_{01}\lambda_{30}+\lambda_{01}\lambda_{31}+\lambda_{03}\lambda_{02}
+\lambda_{13}\lambda_{12}
+\lambda_{23}\lambda_{22}+\lambda_{33}\lambda_{32}\nonumber\\
&&
-(\lambda_{01}\lambda_{02}+\lambda_{01}\lambda_{03}
+\lambda_{01}\lambda_{12}+\lambda_{01}\lambda_{13}
+\lambda_{01}\lambda_{22}+\lambda_{01}\lambda_{23}
+\lambda_{01}\lambda_{32}+\lambda_{01}\lambda_{33}
+3\lambda_{01}\lambda_{01}+\lambda_{02}\lambda_{00}
+\lambda_{12}\lambda_{10}\nonumber\\
&&+\lambda_{22}\lambda_{20}
+\lambda_{32}\lambda_{30})]+d_{1}^{2}[2(\lambda_{01}\lambda_{00}
+\lambda_{01}\lambda_{10}+\lambda_{01}\lambda_{11}
+\lambda_{01}\lambda_{20}+\lambda_{01}\lambda_{21}+\lambda_{01}\lambda_{30}
+\lambda_{01}\lambda_{31})+4(\lambda_{03}\lambda_{02}\nonumber\\
&&+\lambda_{13}\lambda_{12}
+\lambda_{23}\lambda_{22}+\lambda_{33}\lambda_{32})
-(\lambda_{00}\lambda_{00}+\lambda_{10}\lambda_{10}
+\lambda_{20}\lambda_{20}+\lambda_{30}\lambda_{30}
+\lambda_{01}\lambda_{01}+\lambda_{11}\lambda_{11}
+\lambda_{21}\lambda_{21}+\lambda_{31}\lambda_{31}\nonumber\\
&&
+\lambda_{02}\lambda_{02}+\lambda_{12}\lambda_{12}+\lambda_{22}\lambda_{22}
+\lambda_{32}\lambda_{32}+\lambda_{03}\lambda_{03}+\lambda_{13}\lambda_{13}
+\lambda_{23}\lambda_{23}+\lambda_{33}\lambda_{33})
-2(\lambda_{01}\lambda_{02}+\lambda_{01}\lambda_{03}
+\lambda_{01}\lambda_{12}\nonumber\\
&&\!+\!\lambda_{01}\lambda_{13}
\!+\!\lambda_{01}\lambda_{22}+\lambda_{01}\lambda_{23}
+\lambda_{01}\lambda_{32}+\lambda_{01}\lambda_{33}
+3\lambda_{01}\lambda_{01}+
\lambda_{03}\lambda_{00}+\lambda_{13}\lambda_{10}
+\lambda_{23}\lambda_{20}+\lambda_{33}\lambda_{30})]\}
\frac{l}{8\pi(d_{3}^{2}+d_{1}^{2})^{\frac{3}{2}}},\label{One_loop_Eq_S_01}\\
S_{11}&=&\int_{-\infty}^{\infty}\frac
{d\omega_{1}
d\omega_{2}d\omega_{3}}{(2\pi)^{3}}\int\frac{d^{2}\mathbf{k}_{1}
d^{2}\mathbf{k}_{2}d^{2}\mathbf{k}_{3}}{(2\pi)^{6}}\Psi^{\dag}
(\omega_{1},\mathbf{k}_{1})\Sigma_{11}\Psi
(\omega_{2},\mathbf{k}_{2})
\Psi^{\dag}(\omega_{3},\mathbf{k}_{3})
\Sigma_{11}\Psi(\omega_{1}+\omega_{2}-\omega_{3}, \mathbf{k}_{1}+\mathbf{k}_{2}-\mathbf{k}_{3})\nonumber\\
&&\times\{2d_{3}^{2}[\lambda_{12}\lambda_{03}+\lambda_{13}\lambda_{02}
+\lambda_{31}\lambda_{20}+\lambda_{30}\lambda_{21}
+\lambda_{11}\lambda_{00}+\lambda_{11}\lambda_{01}
+\lambda_{11}\lambda_{10}+\lambda_{11}\lambda_{22}
+\lambda_{11}\lambda_{23}
+\lambda_{11}\lambda_{32}+\lambda_{11}\lambda_{33}\nonumber\\
&&
-(\lambda_{12}\lambda_{00}+\lambda_{10}\lambda_{02}
+\lambda_{33}\lambda_{21}+\lambda_{31}\lambda_{23}
+\lambda_{11}\lambda_{02}+\lambda_{11}\lambda_{03}
+\lambda_{11}\lambda_{12}+\lambda_{11}\lambda_{13}
+\lambda_{11}\lambda_{20}+\lambda_{11}\lambda_{21}
+\lambda_{11}\lambda_{30}\nonumber\\
&&+\lambda_{11}\lambda_{31}+3\lambda_{11}\lambda_{11})]
+d_{1}^{2}[\lambda_{11}\lambda_{00}+\lambda_{11}\lambda_{01}
+\lambda_{11}\lambda_{10}+\lambda_{11}\lambda_{22}
+\lambda_{11}\lambda_{23}+\lambda_{11}\lambda_{32}+\lambda_{11}\lambda_{33}
+2(\lambda_{12}\lambda_{03}+\lambda_{13}\lambda_{02}\nonumber\\
&&+\lambda_{31}\lambda_{20}+\lambda_{30}\lambda_{21})
-(\lambda_{10}\lambda_{00}+\lambda_{11}\lambda_{01}
+\lambda_{12}\lambda_{02}+\lambda_{13}\lambda_{03}
+\lambda_{13}\lambda_{00}+\lambda_{10}\lambda_{03}
+\lambda_{31}\lambda_{22}+\lambda_{32}\lambda_{21}
+\lambda_{11}\lambda_{02}\nonumber\\
&&+\lambda_{11}\lambda_{03}
+\lambda_{11}\lambda_{12}+\lambda_{11}\lambda_{13}
+\lambda_{11}\lambda_{20}+\lambda_{11}\lambda_{21}
+\lambda_{11}\lambda_{30}
+\lambda_{11}\lambda_{31}+3\lambda_{11}\lambda_{11})]\}\frac{l}
{4\pi(d_{3}^{2}+d_{1}^{2})^{\frac{3}{2}}},\label{One_loop_Eq_S_11}\\
S_{21}&=&\int_{-\infty}^{\infty}\frac
{d\omega_{1}
d\omega_{2}d\omega_{3}}{(2\pi)^{3}}\int\frac{d^{2}\mathbf{k}_{1}
d^{2}\mathbf{k}_{2}d^{2}\mathbf{k}_{3}}{(2\pi)^{6}}\Psi^{\dag}
(\omega_{1},\mathbf{k}_{1})\Sigma_{21}\Psi
(\omega_{2},\mathbf{k}_{2})
\Psi^{\dag}(\omega_{3},\mathbf{k}_{3})
\Sigma_{21}\Psi(\omega_{1}+\omega_{2}-\omega_{3}, \mathbf{k}_{1}+\mathbf{k}_{2}-\mathbf{k}_{3})\nonumber\\
&&\times\{2d_{3}^{2}[\lambda_{22}\lambda_{03}+\lambda_{23}\lambda_{02}
+\lambda_{11}\lambda_{30}+\lambda_{10}\lambda_{31}
+\lambda_{21}\lambda_{00}+\lambda_{21}\lambda_{01}
+\lambda_{21}\lambda_{12}+\lambda_{21}\lambda_{13}+\lambda_{21}\lambda_{20}
+\lambda_{21}\lambda_{32}+\lambda_{21}\lambda_{33}\nonumber\\
&&
-(\lambda_{22}\lambda_{00}+\lambda_{20}\lambda_{02}
+\lambda_{13}\lambda_{31}+\lambda_{11}\lambda_{33}
+\lambda_{21}\lambda_{02}+\lambda_{21}\lambda_{03}
+\lambda_{21}\lambda_{10}+\lambda_{21}\lambda_{11}
+\lambda_{21}\lambda_{22}+\lambda_{21}\lambda_{23}
+\lambda_{21}\lambda_{30}\nonumber\\
&&+\lambda_{21}\lambda_{31}
+3\lambda_{21}\lambda_{21})]
+d_{1}^{2}[2(\lambda_{22}\lambda_{03}+\lambda_{23}\lambda_{02}
+\lambda_{11}\lambda_{30}+\lambda_{10}\lambda_{31})
+\lambda_{21}\lambda_{00}+\lambda_{21}\lambda_{01}
+\lambda_{21}\lambda_{12}+\lambda_{21}\lambda_{13}\nonumber\\
&&+\lambda_{21}\lambda_{20}
+\lambda_{21}\lambda_{32}+\lambda_{21}\lambda_{33}
-(\lambda_{20}\lambda_{00}+\lambda_{21}\lambda_{01}
+\lambda_{22}\lambda_{02}+\lambda_{23}\lambda_{03}
+\lambda_{23}\lambda_{00}+\lambda_{20}\lambda_{03}
+\lambda_{11}\lambda_{32}+\lambda_{12}\lambda_{31}\nonumber\\
&&+\lambda_{21}\lambda_{02}+\lambda_{21}\lambda_{03}
+\lambda_{21}\lambda_{10}+\lambda_{21}\lambda_{11}
+\lambda_{21}\lambda_{22}+\lambda_{21}\lambda_{23}
+\lambda_{21}\lambda_{30}+\lambda_{21}\lambda_{31}
+3\lambda_{21}\lambda_{21})]\}
\frac{l}{4\pi(d_{3}^{2}+d_{1}^{2})^{\frac{3}{2}}},\label{One_loop_Eq_S_21}\\
S_{31}&=&\int_{-\infty}^{\infty}\frac
{d\omega_{1}
d\omega_{2}d\omega_{3}}{(2\pi)^{3}}\int\frac{d^{2}\mathbf{k}_{1}
d^{2}\mathbf{k}_{2}d^{2}\mathbf{k}_{3}}{(2\pi)^{6}}\Psi^{\dag}
(\omega_{1},\mathbf{k}_{1})\Sigma_{31}\Psi
(\omega_{2},\mathbf{k}_{2})
\Psi^{\dag}(\omega_{3},\mathbf{k}_{3})
\Sigma_{31}\Psi(\omega_{1}+\omega_{2}-\omega_{3}, \mathbf{k}_{1}+\mathbf{k}_{2}-\mathbf{k}_{3})\nonumber\\
&&\times\{2d_{3}^{2}[\lambda_{32}\lambda_{03}+\lambda_{33}\lambda_{02}
+\lambda_{21}\lambda_{10}+\lambda_{20}\lambda_{11}
+\lambda_{31}\lambda_{00}+\lambda_{31}\lambda_{01}
+\lambda_{31}\lambda_{12}+\lambda_{31}\lambda_{13}
+\lambda_{31}\lambda_{22}+\lambda_{31}\lambda_{23}
+\lambda_{31}\lambda_{30}\nonumber\\
&&-(\lambda_{32}\lambda_{00}+\lambda_{30}\lambda_{02}
+\lambda_{23}\lambda_{11}+\lambda_{21}\lambda_{13}
+\lambda_{31}\lambda_{02}+\lambda_{31}\lambda_{03}
+\lambda_{31}\lambda_{10}+\lambda_{31}\lambda_{11}
+\lambda_{31}\lambda_{20}+\lambda_{31}\lambda_{21}
+\lambda_{31}\lambda_{32}\nonumber\\
&&+\lambda_{31}\lambda_{33}
+3\lambda_{31}\lambda_{31})]+d_{1}^{2}[2(\lambda_{32}\lambda_{03}
+\lambda_{33}\lambda_{02}
+\lambda_{21}\lambda_{10}+\lambda_{20}\lambda_{11})
+\lambda_{31}\lambda_{00}+\lambda_{31}\lambda_{01}
+\lambda_{31}\lambda_{12}+\lambda_{31}\lambda_{13}\nonumber\\
&&+\lambda_{31}\lambda_{22}+\lambda_{31}\lambda_{23}
+\lambda_{31}\lambda_{30}
-(\lambda_{30}\lambda_{00}+\lambda_{31}\lambda_{01}
+\lambda_{32}\lambda_{02}+\lambda_{33}\lambda_{03}
+\lambda_{33}\lambda_{00}+\lambda_{30}\lambda_{03}
+\lambda_{21}\lambda_{12}+\lambda_{22}\lambda_{11}\nonumber\\
&&+\lambda_{31}\lambda_{02}+\lambda_{31}\lambda_{03}
+\lambda_{31}\lambda_{10}+\lambda_{31}\lambda_{11}
+\lambda_{31}\lambda_{20}+\lambda_{31}\lambda_{21}
+\lambda_{31}\lambda_{32}+\lambda_{31}\lambda_{33}
+3\lambda_{31}\lambda_{31})]\}
\frac{l}{4\pi(d_{3}^{2}+d_{1}^{2})^{\frac{3}{2}}},\label{One_loop_Eq_S_31}\\
S_{02}&=&\int_{-\infty}^{\infty}\frac
{d\omega_{1}
d\omega_{2}d\omega_{3}}{(2\pi)^{3}}\int\frac{d^{2}\mathbf{k}_{1}
d^{2}\mathbf{k}_{2}d^{2}\mathbf{k}_{3}}{(2\pi)^{6}}\Psi^{\dag}
(\omega_{1},\mathbf{k}_{1})\Sigma_{02}\Psi
(\omega_{2},\mathbf{k}_{2})
\Psi^{\dag}(\omega_{3},\mathbf{k}_{3})
\Sigma_{02}\Psi(\omega_{1}+\omega_{2}-\omega_{3}, \mathbf{k}_{1}+\mathbf{k}_{2}-\mathbf{k}_{3})\nonumber\\
&&\times\{4d_{3}^{2}[\lambda_{01}\lambda_{03}+\lambda_{11}\lambda_{13}
+\lambda_{21}\lambda_{23}+\lambda_{31}\lambda_{33}
+\lambda_{02}\lambda_{00}
+\lambda_{02}\lambda_{10}+\lambda_{02}\lambda_{12}
+\lambda_{02}\lambda_{20}+\lambda_{02}\lambda_{22}
+\lambda_{02}\lambda_{30}+\lambda_{02}\lambda_{32}\nonumber\\
&&-(\lambda_{01}\lambda_{00}+\lambda_{11}\lambda_{10}
+\lambda_{21}\lambda_{20}+\lambda_{31}\lambda_{30}
+\lambda_{02}\lambda_{01}+\lambda_{02}\lambda_{03}
+\lambda_{02}\lambda_{11}+\lambda_{02}\lambda_{13}
+\lambda_{02}\lambda_{21}+\lambda_{02}\lambda_{23}
+\lambda_{02}\lambda_{31}\nonumber\\
&&+\lambda_{02}\lambda_{33}
+3\lambda_{02}\lambda_{02})]
+d_{1}^{2}[ 4(\lambda_{01}\lambda_{03}
+\lambda_{11}\lambda_{13}+\lambda_{21}\lambda_{23}
+\lambda_{31}\lambda_{33})+\lambda_{02}\lambda_{00}
+\lambda_{02}\lambda_{10}+\lambda_{02}\lambda_{12}
+\lambda_{02}\lambda_{20}\nonumber\\
&&+\lambda_{02}\lambda_{22}
+\lambda_{02}\lambda_{30}+\lambda_{02}\lambda_{32}
-(\lambda_{00}\lambda_{00}+\lambda_{10}\lambda_{10}
+\lambda_{20}\lambda_{20}+\lambda_{30}\lambda_{30}
+\lambda_{01}\lambda_{01}+\lambda_{11}\lambda_{11}
+\lambda_{21}\lambda_{21}+\lambda_{31}\lambda_{31}\nonumber\\
&&+\lambda_{02}\lambda_{02}+\lambda_{12}\lambda_{12}
+\lambda_{22}\lambda_{22}+\lambda_{32}\lambda_{32}
+\lambda_{03}\lambda_{03}+\lambda_{13}\lambda_{13}
+\lambda_{23}\lambda_{23}+\lambda_{33}\lambda_{33})
-2(\lambda_{03}\lambda_{00}+\lambda_{13}\lambda_{10}
+\lambda_{23}\lambda_{20}\nonumber\\
&&\!+\!\lambda_{33}\lambda_{30}
\!+\!\lambda_{02}\lambda_{01}\!+\!\lambda_{02}\lambda_{03}
\!+\lambda_{02}\lambda_{11}+\lambda_{02}\lambda_{13}
+\lambda_{02}\lambda_{21}+\lambda_{02}\lambda_{23}
+\lambda_{02}\lambda_{31}+\lambda_{02}\lambda_{33}
+3\lambda_{02}\lambda_{02})]\}
\frac{l}{8\pi(d_{3}^{2}+d_{1}^{2})^{\frac{3}{2}}},\label{One_loop_Eq_S_02}\\
S_{12}&=&\int_{-\infty}^{\infty}\frac
{d\omega_{1}
d\omega_{2}d\omega_{3}}{(2\pi)^{3}}\int\frac{d^{2}\mathbf{k}_{1}
d^{2}\mathbf{k}_{2}d^{2}\mathbf{k}_{3}}{(2\pi)^{6}}\Psi^{\dag}
(\omega_{1},\mathbf{k}_{1})\Sigma_{12}\Psi
(\omega_{2},\mathbf{k}_{2})
\Psi^{\dag}(\omega_{3},\mathbf{k}_{3})
\Sigma_{12}\Psi(\omega_{1}+\omega_{2}-\omega_{3}, \mathbf{k}_{1}+\mathbf{k}_{2}-\mathbf{k}_{3})\nonumber\\
&&\times\{2d_{3}^{2}[\lambda_{13}\lambda_{01}+\lambda_{11}\lambda_{03}
+\lambda_{32}\lambda_{20}+\lambda_{30}\lambda_{22}
+\lambda_{12}\lambda_{00}+\lambda_{12}\lambda_{02}
+\lambda_{12}\lambda_{10}+\lambda_{12}\lambda_{21}
+\lambda_{12}\lambda_{23}+\lambda_{12}\lambda_{31}
+\lambda_{12}\lambda_{33}\nonumber\\
&&-(\lambda_{11}\lambda_{00}
+\lambda_{10}\lambda_{01}
+\lambda_{32}\lambda_{23}+\lambda_{33}\lambda_{22}
+\lambda_{12}\lambda_{01}+\lambda_{12}\lambda_{03}
+\lambda_{12}\lambda_{11}+\lambda_{12}\lambda_{13}
+\lambda_{12}\lambda_{20}+\lambda_{12}\lambda_{22}
+\lambda_{12}\lambda_{30}\nonumber\\
&&+\lambda_{12}\lambda_{32}
+3\lambda_{12}\lambda_{12})]
+d_{1}^{2}[2(\lambda_{13}\lambda_{01}+\lambda_{11}\lambda_{03}
+\lambda_{32}\lambda_{20}+\lambda_{30}\lambda_{22})
+\lambda_{12}\lambda_{00}+\lambda_{12}\lambda_{02}
+\lambda_{12}\lambda_{10}+\lambda_{12}\lambda_{21}\nonumber\\
&&+\lambda_{12}\lambda_{23}
+\lambda_{12}\lambda_{31}+\lambda_{12}\lambda_{33}
-(\lambda_{13}\lambda_{00}+\lambda_{10}\lambda_{03}
+\lambda_{31}\lambda_{22}+\lambda_{32}\lambda_{21}
+\lambda_{10}\lambda_{00}+\lambda_{11}\lambda_{01}
+\lambda_{12}\lambda_{02}+\lambda_{13}\lambda_{03}\nonumber\\
&&+\lambda_{12}\lambda_{01}+\lambda_{12}\lambda_{03}
+\lambda_{12}\lambda_{11}+\lambda_{12}\lambda_{13}
+\lambda_{12}\lambda_{20}+\lambda_{12}\lambda_{22}
+\lambda_{12}\lambda_{30}+\lambda_{12}\lambda_{32}
+3\lambda_{12}\lambda_{12})]\}
\frac{l}{4\pi(d_{3}^{2}+d_{1}^{2})^{\frac{3}{2}}},\label{One_loop_Eq_S_12}\\
S_{22}&=&\int_{-\infty}^{\infty}\frac
{d\omega_{1}
d\omega_{2}d\omega_{3}}{(2\pi)^{3}}\int\frac{d^{2}\mathbf{k}_{1}
d^{2}\mathbf{k}_{2}d^{2}\mathbf{k}_{3}}{(2\pi)^{6}}\Psi^{\dag}
(\omega_{1},\mathbf{k}_{1})\Sigma_{22}\Psi
(\omega_{2},\mathbf{k}_{2})
\Psi^{\dag}(\omega_{3},\mathbf{k}_{3})
\Sigma_{22}\Psi(\omega_{1}+\omega_{2}-\omega_{3}, \mathbf{k}_{1}+\mathbf{k}_{2}-\mathbf{k}_{3})\nonumber\\
&&\times\{2d_{3}^{2}[\lambda_{23}\lambda_{01}+\lambda_{21}\lambda_{03}
+\lambda_{12}\lambda_{30}+\lambda_{10}\lambda_{32}
+\lambda_{22}\lambda_{00}+\lambda_{22}\lambda_{02}
+\lambda_{22}\lambda_{11}+\lambda_{22}\lambda_{13}
+\lambda_{22}\lambda_{20}+\lambda_{22}\lambda_{31}
+\lambda_{22}\lambda_{33}\nonumber\\
&&-(\lambda_{21}\lambda_{00}+\lambda_{20}\lambda_{01}
+\lambda_{12}\lambda_{33}+\lambda_{13}\lambda_{32}
+\lambda_{22}\lambda_{01}+\lambda_{22}\lambda_{03}
+\lambda_{22}\lambda_{10}+\lambda_{22}\lambda_{12}
+\lambda_{22}\lambda_{21}+\lambda_{22}\lambda_{23}
+\lambda_{22}\lambda_{30}\nonumber\\
&&+\lambda_{22}\lambda_{32}
+3\lambda_{22}\lambda_{22})]
+d_{1}^{2}[2(\lambda_{23}\lambda_{01}+\lambda_{21}\lambda_{03}
+\lambda_{12}\lambda_{30}+\lambda_{10}\lambda_{32})
+\lambda_{22}\lambda_{00}+\lambda_{22}\lambda_{02}
+\lambda_{22}\lambda_{11}+\lambda_{22}\lambda_{13}\nonumber\\
&&+\lambda_{22}\lambda_{20}+\lambda_{22}\lambda_{31}
+\lambda_{22}\lambda_{33}-(\lambda_{23}\lambda_{00}
+\lambda_{20}\lambda_{03}
+\lambda_{11}\lambda_{32}+\lambda_{12}\lambda_{31}
+\lambda_{20}\lambda_{00}+\lambda_{21}\lambda_{01}
+\lambda_{22}\lambda_{02}+\lambda_{23}\lambda_{03}\nonumber\\
&&+\lambda_{22}\lambda_{01}+\lambda_{22}\lambda_{03}
+\lambda_{22}\lambda_{10}+\lambda_{22}\lambda_{12}
+\lambda_{22}\lambda_{21}+\lambda_{22}\lambda_{23}+\lambda_{22}\lambda_{30}
+\lambda_{22}\lambda_{32}+3\lambda_{22}\lambda_{22})]\}
\frac{l}{4\pi(d_{3}^{2}+d_{1}^{2})^{\frac{3}{2}}},\label{One_loop_Eq_S_22}\\
S_{32}&=&\int_{-\infty}^{\infty}\frac
{d\omega_{1}
d\omega_{2}d\omega_{3}}{(2\pi)^{3}}\int\frac{d^{2}\mathbf{k}_{1}
d^{2}\mathbf{k}_{2}d^{2}\mathbf{k}_{3}}{(2\pi)^{6}}\Psi^{\dag}
(\omega_{1},\mathbf{k}_{1})\Sigma_{32}\Psi
(\omega_{2},\mathbf{k}_{2})
\Psi^{\dag}(\omega_{3},\mathbf{k}_{3})
\Sigma_{32}\Psi(\omega_{1}+\omega_{2}-\omega_{3}, \mathbf{k}_{1}+\mathbf{k}_{2}-\mathbf{k}_{3})\nonumber\\
&&\times\{2d_{3}^{2}[\lambda_{33}\lambda_{01}+\lambda_{31}\lambda_{03}
+\lambda_{22}\lambda_{10}+\lambda_{20}\lambda_{12}
+\lambda_{32}\lambda_{00}+\lambda_{32}\lambda_{02}
+\lambda_{32}\lambda_{11}+\lambda_{32}\lambda_{13}
+\lambda_{32}\lambda_{21}+\lambda_{32}\lambda_{23}
+\lambda_{32}\lambda_{30}\nonumber\\
&&
-(\lambda_{31}\lambda_{00}+\lambda_{30}\lambda_{01}
+\lambda_{22}\lambda_{13}+\lambda_{23}\lambda_{12}
+\lambda_{32}\lambda_{01}+\lambda_{32}\lambda_{03}
+\lambda_{32}\lambda_{10}+\lambda_{32}\lambda_{12}
+\lambda_{32}\lambda_{20}+\lambda_{32}\lambda_{22}
+\lambda_{32}\lambda_{31}\nonumber\\
&&+\lambda_{32}\lambda_{33}
+3\lambda_{32}\lambda_{32})]
+d_{1}^{2}[2(\lambda_{33}\lambda_{01}+\lambda_{31}\lambda_{03}
+\lambda_{22}\lambda_{10}+\lambda_{20}\lambda_{12})
+\lambda_{32}\lambda_{00}
+\lambda_{32}\lambda_{02}+\lambda_{32}\lambda_{11}
+\lambda_{32}\lambda_{13}\nonumber\\
&&+\lambda_{32}\lambda_{21}
+\lambda_{32}\lambda_{23}+\lambda_{32}\lambda_{30}
-(\lambda_{33}\lambda_{00}+\lambda_{30}\lambda_{03}
+\lambda_{21}\lambda_{12}+\lambda_{22}\lambda_{11}
+\lambda_{30}\lambda_{00}+\lambda_{31}\lambda_{01}
+\lambda_{32}\lambda_{02}+\lambda_{33}\lambda_{03}\nonumber\\
&&
+\lambda_{32}\lambda_{01}+\lambda_{32}\lambda_{03}
+\lambda_{32}\lambda_{10}+\lambda_{32}\lambda_{12}
+\lambda_{32}\lambda_{20}+\lambda_{32}\lambda_{22}
+\lambda_{32}\lambda_{31}+\lambda_{32}\lambda_{33}
+3\lambda_{32}\lambda_{32})]\}
\frac{l}{4\pi(d_{3}^{2}+d_{1}^{2})^{\frac{3}{2}}},\label{One_loop_Eq_S_32}\\
S_{03}&=&\int_{-\infty}^{\infty}\frac
{d\omega_{1}
d\omega_{2}d\omega_{3}}{(2\pi)^{3}}\int\frac{d^{2}\mathbf{k}_{1}
d^{2}\mathbf{k}_{2}d^{2}\mathbf{k}_{3}}{(2\pi)^{6}}\Psi^{\dag}
(\omega_{1},\mathbf{k}_{1})\Sigma_{03}\Psi
(\omega_{2},\mathbf{k}_{2})
\Psi^{\dag}(\omega_{3},\mathbf{k}_{3})
\Sigma_{03}\Psi(\omega_{1}+\omega_{2}-\omega_{3}, \mathbf{k}_{1}+\mathbf{k}_{2}-\mathbf{k}_{3})\nonumber\\
&&\times\{d_{3}^{2}[2(\lambda_{02}\lambda_{01}+\lambda_{12}\lambda_{11}
+\lambda_{22}\lambda_{21}+\lambda_{32}\lambda_{31})
-(\lambda_{00}\lambda_{00}+\lambda_{10}\lambda_{10}
+\lambda_{20}\lambda_{20}+\lambda_{30}\lambda_{30}
+\lambda_{01}\lambda_{01}+\lambda_{11}\lambda_{11}\nonumber\\
&&
+\lambda_{21}\lambda_{21}+\lambda_{31}\lambda_{31}
+\lambda_{02}\lambda_{02}+\lambda_{12}\lambda_{12}
+\lambda_{22}\lambda_{22}+\lambda_{32}\lambda_{32}
+\lambda_{03}\lambda_{03}+\lambda_{13}\lambda_{13}
+\lambda_{23}\lambda_{23}+\lambda_{33}\lambda_{33})]
+d_{1}^{2}[2(\lambda_{02}\lambda_{01}\nonumber\\
&&+\lambda_{12}\lambda_{11}
+\lambda_{22}\lambda_{21}+\lambda_{32}\lambda_{31}
+\lambda_{03}\lambda_{00}
+\lambda_{03}\lambda_{10}+\lambda_{03}\lambda_{13}
+\lambda_{03}\lambda_{20}+\lambda_{03}\lambda_{23}
+\lambda_{03}\lambda_{30}+\lambda_{03}\lambda_{33})
-(\lambda_{01}\lambda_{00}\nonumber\\
&&+\lambda_{11}\lambda_{10}
+\lambda_{21}\lambda_{20}+\lambda_{31}\lambda_{30}
+\lambda_{02}\lambda_{00}+\lambda_{12}\lambda_{10}
+\lambda_{22}\lambda_{20}+\lambda_{32}\lambda_{30})
-2(\lambda_{03}\lambda_{01}+\lambda_{03}\lambda_{02}
+\lambda_{03}\lambda_{11}+\lambda_{03}\lambda_{12}\nonumber\\
&&
+\lambda_{03}\lambda_{21}+\lambda_{03}\lambda_{22}
+\lambda_{03}\lambda_{31}+\lambda_{03}\lambda_{32}
+3\lambda_{03}\lambda_{03})]\}
\frac{l}{4\pi(d_{3}^{2}+d_{1}^{2})^{\frac{3}{2}}},\label{One_loop_Eq_S_03}\\
S_{13}&=&\int_{-\infty}^{\infty}\frac
{d\omega_{1}
d\omega_{2}d\omega_{3}}{(2\pi)^{3}}\int\frac{d^{2}\mathbf{k}_{1}
d^{2}\mathbf{k}_{2}d^{2}\mathbf{k}_{3}}{(2\pi)^{6}}\Psi^{\dag}
(\omega_{1},\mathbf{k}_{1})\Sigma_{13}\Psi
(\omega_{2},\mathbf{k}_{2})
\Psi^{\dag}(\omega_{3},\mathbf{k}_{3})
\Sigma_{13}\Psi(\omega_{1}+\omega_{2}-\omega_{3}, \mathbf{k}_{1}+\mathbf{k}_{2}-\mathbf{k}_{3})\nonumber\\
&&\times\{2d_{3}^{2}[\lambda_{11}\lambda_{02}+\lambda_{12}\lambda_{01}
+\lambda_{33}\lambda_{20}+\lambda_{30}\lambda_{23}
-(\lambda_{10}\lambda_{00}+\lambda_{11}\lambda_{01}
+\lambda_{12}\lambda_{02}+\lambda_{13}\lambda_{03})]
+d_{1}^{2}[2(\lambda_{11}\lambda_{02}+\lambda_{12}\lambda_{01}\nonumber\\
&&
+\lambda_{33}\lambda_{20}+\lambda_{30}\lambda_{23}
+\lambda_{13}\lambda_{00}+\lambda_{13}\lambda_{03}
+\lambda_{13}\lambda_{10}+\lambda_{13}\lambda_{21}
+\lambda_{13}\lambda_{22}+\lambda_{13}\lambda_{31}
+\lambda_{13}\lambda_{32})-(\lambda_{12}\lambda_{00}
+\lambda_{10}\lambda_{02}\nonumber\\
&&+\lambda_{33}\lambda_{21}+\lambda_{31}\lambda_{23}
+\lambda_{11}\lambda_{00}+\lambda_{10}\lambda_{01}
+\lambda_{32}\lambda_{23}+\lambda_{33}\lambda_{22})
-2(\lambda_{13}\lambda_{01}+\lambda_{13}\lambda_{02}
+\lambda_{13}\lambda_{11}+\lambda_{13}\lambda_{12}
+\lambda_{13}\lambda_{20}\nonumber\\
&&+\lambda_{13}\lambda_{23}
+\lambda_{13}\lambda_{30}+\lambda_{13}\lambda_{33}
+3\lambda_{13}\lambda_{13})]\}
\frac{l}{4\pi(d_{3}^{2}+d_{1}^{2})^{\frac{3}{2}}},\label{One_loop_Eq_S_13}\\
S_{23}&=&\int_{-\infty}^{\infty}\frac
{d\omega_{1}
d\omega_{2}d\omega_{3}}{(2\pi)^{3}}\int\frac{d^{2}\mathbf{k}_{1}
d^{2}\mathbf{k}_{2}d^{2}\mathbf{k}_{3}}{(2\pi)^{6}}\Psi^{\dag}
(\omega_{1},\mathbf{k}_{1})\Sigma_{23}\Psi
(\omega_{2},\mathbf{k}_{2})
\Psi^{\dag}(\omega_{3},\mathbf{k}_{3})
\Sigma_{23}\Psi(\omega_{1}+\omega_{2}-\omega_{3}, \mathbf{k}_{1}+\mathbf{k}_{2}-\mathbf{k}_{3})\nonumber\\
&&\times\{2d_{3}^{2}[\lambda_{21}\lambda_{02}+\lambda_{22}\lambda_{01}
+\lambda_{13}\lambda_{30}+\lambda_{10}\lambda_{33}
-(\lambda_{20}\lambda_{00}+\lambda_{21}\lambda_{01}
+\lambda_{22}\lambda_{02}+\lambda_{23}\lambda_{03})]
+d_{1}^{2}[2(\lambda_{21}\lambda_{02}+\lambda_{22}\lambda_{01}\nonumber\\
&&
+\lambda_{13}\lambda_{30}+\lambda_{10}\lambda_{33}
+\lambda_{23}\lambda_{00}+\lambda_{23}\lambda_{03}
+\lambda_{23}\lambda_{11}+\lambda_{23}\lambda_{12}+\lambda_{23}\lambda_{20}
+\lambda_{23}\lambda_{31}+\lambda_{23}\lambda_{32})
-(\lambda_{22}\lambda_{00}+\lambda_{20}\lambda_{02}\nonumber\\
&&
+\lambda_{13}\lambda_{31}+\lambda_{11}\lambda_{33}
+\lambda_{21}\lambda_{00}+\lambda_{20}\lambda_{01}
+\lambda_{12}\lambda_{33}+\lambda_{13}\lambda_{32})
-2(\lambda_{23}\lambda_{01}+\lambda_{23}\lambda_{02}
+\lambda_{23}\lambda_{10}+\lambda_{23}\lambda_{13}
+\lambda_{23}\lambda_{21}\nonumber\\
&&+\lambda_{23}\lambda_{22}
+\lambda_{23}\lambda_{30}+\lambda_{23}\lambda_{33}
+3\lambda_{23}\lambda_{23})]\}
\frac{l}{4\pi(d_{3}^{2}+d_{1}^{2})^{\frac{3}{2}}},\label{One_loop_Eq_S_23}\\
S_{33}&=&\int_{-\infty}^{\infty}\frac
{d\omega_{1}
d\omega_{2}d\omega_{3}}{(2\pi)^{3}}\int\frac{d^{2}\mathbf{k}_{1}
d^{2}\mathbf{k}_{2}d^{2}\mathbf{k}_{3}}{(2\pi)^{6}}\Psi^{\dag}
(\omega_{1},\mathbf{k}_{1})\Sigma_{33}\Psi
(\omega_{2},\mathbf{k}_{2})
\Psi^{\dag}(\omega_{3},\mathbf{k}_{3})
\Sigma_{33}\Psi(\omega_{1}+\omega_{2}-\omega_{3}, \mathbf{k}_{1}+\mathbf{k}_{2}-\mathbf{k}_{3})\nonumber\\
&&\times\{2d_{3}^{2}[\lambda_{31}\lambda_{02}+\lambda_{32}\lambda_{01}
+\lambda_{23}\lambda_{10}+\lambda_{20}\lambda_{13}
-(\lambda_{30}\lambda_{00}+\lambda_{31}\lambda_{01}
+\lambda_{32}\lambda_{02}+\lambda_{33}\lambda_{03})]
+d_{1}^{2}[2(\lambda_{31}\lambda_{02}
+\lambda_{32}\lambda_{01}\nonumber\\
&&+\lambda_{23}\lambda_{10}
+\lambda_{20}\lambda_{13}+\lambda_{33}\lambda_{00}
+\lambda_{33}\lambda_{03}+\lambda_{33}\lambda_{11}
+\lambda_{33}\lambda_{12}+\lambda_{33}\lambda_{21}
+\lambda_{33}\lambda_{22}+\lambda_{33}\lambda_{30})
-(\lambda_{32}\lambda_{00}+\lambda_{30}\lambda_{02}\nonumber\\
&&+\lambda_{23}\lambda_{11}+\lambda_{21}\lambda_{13}
+\lambda_{31}\lambda_{00}+\lambda_{30}\lambda_{01}
+\lambda_{22}\lambda_{13}+\lambda_{23}\lambda_{12})
-2(\lambda_{33}\lambda_{01}+\lambda_{33}\lambda_{02}
+\lambda_{33}\lambda_{10}+\lambda_{33}\lambda_{13}
+\lambda_{33}\lambda_{20}\nonumber\\
&&+\lambda_{33}\lambda_{23}
+\lambda_{33}\lambda_{31}+\lambda_{33}\lambda_{32}
+3\lambda_{33}\lambda_{33})]\}
\frac{l}{4\pi(d_{3}^{2}+d_{1}^{2})^{\frac{3}{2}}}.\label{One_loop_Eq_S_33}
\end{eqnarray}
\end{small}
\begin{figure}
\centering
\includegraphics[width=7in]{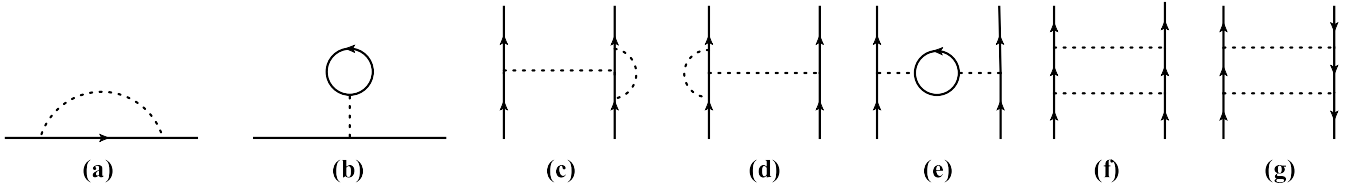}
\vspace{-0.05cm}
\caption{One-loop corrections to (a)-(b): the fermion propagator and
(c)-(g): the fermion-fermion interactions. The solid and dashed lines characterize
the fermion propagator and four-fermion interactions, respectively~\cite{Murray2014PRB,Wang2017PRB}.}
\label{Fig_fermion_propagator_and_fermion_interaction_correction}
\end{figure}

\section{RG flow equations of all interaction parameters}\label{Appendix_RG_flow_Eqs_of_lambda}

Combining our effective action~(\ref{S_eff}) and the
RG rescalings~(\ref{Eq_rescale-k_x})-(\ref{Eq_rescale-Psi}) as well as
all the one-loop corrections presented in Appendix~\ref{Appendix_one-loop_corrections},
we consequently are left with the following coupled RG evolutions of
fermion-fermion interactions after carrying out the standard procedures of RG analysis~\cite{Murray2014PRB,Wang2011PRB,Huh2008PRB,Wang2019JPCM}
\begin{small}
\begin{eqnarray}
\frac{d\lambda_{00}}{dl}
&=&-\frac{1}{4\pi(d_{3}^{2}+d_{1}^{2})^{\frac{3}{2}}}
[2d_{3}^{2}(\lambda_{00}\lambda_{03}+\lambda_{10}
\lambda_{13}+\lambda_{20}\lambda_{23}+\lambda_{30}\lambda_{33})
+d_{1}^{2}(\lambda_{00}\lambda_{01}+\lambda_{10}
\lambda_{11}+\lambda_{20}\lambda_{21}+\lambda_{30}\lambda_{31}\nonumber\\
&&
+\lambda_{00}\lambda_{02}+\lambda_{10}
\lambda_{12}+\lambda_{20}\lambda_{22}
+\lambda_{30}\lambda_{32})],\label{Eq_RG_lambda00}\\
\frac{d\lambda_{10}}{dl}
&=&\frac{1}{4\pi(d_{3}^{2}+d_{1}^{2})^{\frac{3}{2}}}\{2d_{3}^{2}
[\lambda_{20}\lambda_{30}
+\lambda_{21}\lambda_{31}+\lambda_{22}
\lambda_{32}+\lambda_{23}\lambda_{33}
-(\lambda_{13}\lambda_{00}+\lambda_{10}
\lambda_{03}+\lambda_{21}\lambda_{32}+\lambda_{22}
\lambda_{31})]\nonumber\\
&&+d_{1}^{2}[2(\lambda_{20}\lambda_{30}
+\lambda_{21}\lambda_{31}+\lambda_{22}
\lambda_{32}+\lambda_{23}\lambda_{33})
-(\lambda_{01}\lambda_{10}+\lambda_{00}
\lambda_{11}+\lambda_{22}\lambda_{33}+\lambda_{23}\lambda_{32}
+\lambda_{12}\lambda_{00}\nonumber\\
&&+\lambda_{10}\lambda_{02}+\lambda_{23}\lambda_{31}
+\lambda_{21}\lambda_{33})]\},\label{Eq_RG_lambda10}\\
\frac{d\lambda_{20}}{dl}
&=&\frac{1}{4\pi(d_{3}^{2}+d_{1}^{2})^{\frac{3}{2}}}
\{2d_{3}^{2}[\lambda_{30}\lambda_{10}+\lambda_{31}\lambda_{11}
+\lambda_{32}\lambda_{12}+\lambda_{33}\lambda_{13}
-(\lambda_{23}\lambda_{00}+\lambda_{20}\lambda_{03}
+\lambda_{31}\lambda_{12}+\lambda_{32}\lambda_{11})]\nonumber\\
&&+d_{1}^{2}[2(\lambda_{30}\lambda_{10}+\lambda_{31}\lambda_{11}
+\lambda_{32}\lambda_{12}+\lambda_{33}\lambda_{13})
-(\lambda_{21}\lambda_{00}+\lambda_{20}\lambda_{01}
+\lambda_{32}\lambda_{13}+\lambda_{33}\lambda_{12}
+\lambda_{22}\lambda_{00}\nonumber\\
&&+\lambda_{20}\lambda_{02}
+\lambda_{33}\lambda_{11}+\lambda_{31}\lambda_{13})]\}
,\label{Eq_RG_lambda20}\\
\frac{d\lambda_{30}}{dl}
&=&\frac{1}{4\pi(d_{3}^{2}+d_{1}^{2})^{\frac{3}{2}}}
\{2d_{3}^{2}[\lambda_{10}\lambda_{20}+\lambda_{11}\lambda_{21}
+\lambda_{12}\lambda_{22}+\lambda_{13}\lambda_{23}
-(\lambda_{33}\lambda_{00}+\lambda_{30}\lambda_{03}
+\lambda_{11}\lambda_{22}+\lambda_{12}\lambda_{21})]\nonumber\\
&&
+d_{1}^{2}[2(\lambda_{10}\lambda_{20}+\lambda_{11}\lambda_{21}
+\lambda_{12}\lambda_{22}+\lambda_{13}\lambda_{23})
-(\lambda_{31}\lambda_{00}+\lambda_{30}\lambda_{01}
+\lambda_{12}\lambda_{23}+\lambda_{13}\lambda_{22}
+\lambda_{32}\lambda_{00}\nonumber\\
&&+\lambda_{30}\lambda_{02}
+\lambda_{13}\lambda_{21}+\lambda_{11}\lambda_{23})]\}
,\label{Eq_RG_lambda30}\\
\frac{d\lambda_{01}}{dl}
&=&\frac{1}{8\pi(d_{3}^{2}+d_{1}^{2})^{\frac{3}{2}}}
\{4d_{3}^{2}[\lambda_{01}\lambda_{00}
+\lambda_{01}\lambda_{10}+\lambda_{01}\lambda_{11}
+\lambda_{01}\lambda_{20}+\lambda_{01}\lambda_{21}
+\lambda_{01}\lambda_{30}+\lambda_{01}\lambda_{31}
+\lambda_{03}\lambda_{02}\nonumber\\
&&
+\lambda_{13}\lambda_{12}
+\lambda_{23}\lambda_{22}+\lambda_{33}\lambda_{32}
-(\lambda_{01}\lambda_{02}+\lambda_{01}\lambda_{03}
+\lambda_{01}\lambda_{12}+\lambda_{01}\lambda_{13}
+\lambda_{01}\lambda_{22}+\lambda_{01}\lambda_{23}
+\lambda_{01}\lambda_{32}\nonumber\\
&&+\lambda_{01}\lambda_{33}
+3\lambda_{01}\lambda_{01}+\lambda_{02}\lambda_{00}
+\lambda_{12}\lambda_{10}+\lambda_{22}\lambda_{20}
+\lambda_{32}\lambda_{30})]+d_{1}^{2}[2(\lambda_{01}\lambda_{00}
+\lambda_{01}\lambda_{10}+\lambda_{01}\lambda_{11}\nonumber\\
&&+\lambda_{01}\lambda_{20}+\lambda_{01}\lambda_{21}
+\lambda_{01}\lambda_{30}
+\lambda_{01}\lambda_{31})+4(\lambda_{03}\lambda_{02}
+\lambda_{13}\lambda_{12}
+\lambda_{23}\lambda_{22}+\lambda_{33}\lambda_{32})
-(\lambda_{00}\lambda_{00}+\lambda_{10}\lambda_{10}\nonumber\\
&&+\lambda_{20}\lambda_{20}+\lambda_{30}\lambda_{30}
+\lambda_{01}\lambda_{01}+\lambda_{11}\lambda_{11}
+\lambda_{21}\lambda_{21}+\lambda_{31}\lambda_{31}
+\lambda_{02}\lambda_{02}+\lambda_{12}\lambda_{12}+\lambda_{22}\lambda_{22}
+\lambda_{32}\lambda_{32}\nonumber\\
&&+\lambda_{03}\lambda_{03}+\lambda_{13}\lambda_{13}
+\lambda_{23}\lambda_{23}+\lambda_{33}\lambda_{33})
-2(\lambda_{01}\lambda_{02}+\lambda_{01}\lambda_{03}
+\lambda_{01}\lambda_{12}+\lambda_{01}\lambda_{13}
+\lambda_{01}\lambda_{22}+\lambda_{01}\lambda_{23}\nonumber\\
&&+\lambda_{01}\lambda_{32}+\lambda_{01}\lambda_{33}
+3\lambda_{01}\lambda_{01}+
\lambda_{03}\lambda_{00}+\lambda_{13}\lambda_{10}
+\lambda_{23}\lambda_{20}+\lambda_{33}\lambda_{30})
]\},\label{Eq_RG_lambda01}\\
\frac{d\lambda_{11}}{dl}&=&\frac{1}{4\pi(d_{3}^{2}+d_{1}^{2})^{\frac{3}{2}}}
\{2d_{3}^{2}[\lambda_{12}\lambda_{03}+\lambda_{13}\lambda_{02}
+\lambda_{31}\lambda_{20}+\lambda_{30}\lambda_{21}
+\lambda_{11}\lambda_{00}+\lambda_{11}\lambda_{01}
+\lambda_{11}\lambda_{10}+\lambda_{11}\lambda_{22}
+\lambda_{11}\lambda_{23}\nonumber\\
&&
+\lambda_{11}\lambda_{32}+\lambda_{11}\lambda_{33}
-(\lambda_{12}\lambda_{00}+\lambda_{10}\lambda_{02}
+\lambda_{33}\lambda_{21}+\lambda_{31}\lambda_{23}
+\lambda_{11}\lambda_{02}+\lambda_{11}\lambda_{03}
+\lambda_{11}\lambda_{12}+\lambda_{11}\lambda_{13}
+\lambda_{11}\lambda_{20}\nonumber\\
&&+\lambda_{11}\lambda_{21}
+\lambda_{11}\lambda_{30}+\lambda_{11}\lambda_{31}
+3\lambda_{11}\lambda_{11})]
+d_{1}^{2}[\lambda_{11}\lambda_{00}+\lambda_{11}\lambda_{01}
+\lambda_{11}\lambda_{10}+\lambda_{11}\lambda_{22}
+\lambda_{11}\lambda_{23}+\lambda_{11}\lambda_{32}
+\lambda_{11}\lambda_{33}\nonumber\\
&&+2(\lambda_{12}\lambda_{03}+\lambda_{13}\lambda_{02}
+\lambda_{31}\lambda_{20}+\lambda_{30}\lambda_{21})
-(\lambda_{10}\lambda_{00}+\lambda_{11}\lambda_{01}
+\lambda_{12}\lambda_{02}+\lambda_{13}\lambda_{03}
+\lambda_{13}\lambda_{00}+\lambda_{10}\lambda_{03}
+\lambda_{31}\lambda_{22}\nonumber\\
&&+\lambda_{32}\lambda_{21}
+\lambda_{11}\lambda_{02}+\lambda_{11}\lambda_{03}
+\lambda_{11}\lambda_{12}+\lambda_{11}\lambda_{13}
+\lambda_{11}\lambda_{20}+\lambda_{11}\lambda_{21}
+\lambda_{11}\lambda_{30}
+\lambda_{11}\lambda_{31}+3\lambda_{11}\lambda_{11})]\}
,\label{Eq_RG_lambda11}\\
\frac{d\lambda_{21}}{dl}&=&\frac{1}{4\pi(d_{3}^{2}+d_{1}^{2})^{\frac{3}{2}}}
\{2d_{3}^{2}[\lambda_{22}\lambda_{03}+\lambda_{23}\lambda_{02}
+\lambda_{11}\lambda_{30}+\lambda_{10}\lambda_{31}
+\lambda_{21}\lambda_{00}+\lambda_{21}\lambda_{01}
+\lambda_{21}\lambda_{12}+\lambda_{21}\lambda_{13}
+\lambda_{21}\lambda_{20}\nonumber\\
&&+\lambda_{21}\lambda_{32}+\lambda_{21}\lambda_{33}
-(\lambda_{22}\lambda_{00}+\lambda_{20}\lambda_{02}
+\lambda_{13}\lambda_{31}+\lambda_{11}\lambda_{33}
+\lambda_{21}\lambda_{02}+\lambda_{21}\lambda_{03}
+\lambda_{21}\lambda_{10}+\lambda_{21}\lambda_{11}
+\lambda_{21}\lambda_{22}\nonumber\\
&&+\lambda_{21}\lambda_{23}
+\lambda_{21}\lambda_{30}+\lambda_{21}\lambda_{31}
+3\lambda_{21}\lambda_{21})]
+d_{1}^{2}[2(\lambda_{22}\lambda_{03}+\lambda_{23}\lambda_{02}
+\lambda_{11}\lambda_{30}+\lambda_{10}\lambda_{31})
+\lambda_{21}\lambda_{00}+\lambda_{21}\lambda_{01}\nonumber\\
&&+\lambda_{21}\lambda_{12}+\lambda_{21}\lambda_{13}
+\lambda_{21}\lambda_{20}
+\lambda_{21}\lambda_{32}+\lambda_{21}\lambda_{33}
-(\lambda_{20}\lambda_{00}+\lambda_{21}\lambda_{01}
+\lambda_{22}\lambda_{02}+\lambda_{23}\lambda_{03}
+\lambda_{23}\lambda_{00}+\lambda_{20}\lambda_{03}\nonumber\\
&&
+\lambda_{11}\lambda_{32}+\lambda_{12}\lambda_{31}
+\lambda_{21}\lambda_{02}+\lambda_{21}\lambda_{03}
+\lambda_{21}\lambda_{10}+\lambda_{21}\lambda_{11}
+\lambda_{21}\lambda_{22}+\lambda_{21}\lambda_{23}
+\lambda_{21}\lambda_{30}+\lambda_{21}\lambda_{31}
+3\lambda_{21}\lambda_{21})]\},\label{Eq_RG_lambda21}\\
\frac{d\lambda_{31}}{dl}
&=&\frac{1}{4\pi(d_{3}^{2}+d_{1}^{2})^{\frac{3}{2}}}
\{2d_{3}^{2}[\lambda_{32}\lambda_{03}+\lambda_{33}\lambda_{02}
+\lambda_{21}\lambda_{10}+\lambda_{20}\lambda_{11}
+\lambda_{31}\lambda_{00}+\lambda_{31}\lambda_{01}
+\lambda_{31}\lambda_{12}+\lambda_{31}\lambda_{13}
+\lambda_{31}\lambda_{22}\nonumber\\
&&+\lambda_{31}\lambda_{23}+\lambda_{31}\lambda_{30}
-(\lambda_{32}\lambda_{00}+\lambda_{30}\lambda_{02}
+\lambda_{23}\lambda_{11}+\lambda_{21}\lambda_{13}
+\lambda_{31}\lambda_{02}+\lambda_{31}\lambda_{03}
+\lambda_{31}\lambda_{10}+\lambda_{31}\lambda_{11}
+\lambda_{31}\lambda_{20}\nonumber\\
&&+\lambda_{31}\lambda_{21}
+\lambda_{31}\lambda_{32}+\lambda_{31}\lambda_{33}
+3\lambda_{31}\lambda_{31})]+d_{1}^{2}[2(\lambda_{32}\lambda_{03}
+\lambda_{33}\lambda_{02}
+\lambda_{21}\lambda_{10}+\lambda_{20}\lambda_{11})
+\lambda_{31}\lambda_{00}+\lambda_{31}\lambda_{01}\nonumber\\
&&
+\lambda_{31}\lambda_{12}+\lambda_{31}\lambda_{13}
+\lambda_{31}\lambda_{22}+\lambda_{31}\lambda_{23}
+\lambda_{31}\lambda_{30}
-(\lambda_{30}\lambda_{00}+\lambda_{31}\lambda_{01}
+\lambda_{32}\lambda_{02}+\lambda_{33}\lambda_{03}
+\lambda_{33}\lambda_{00}+\lambda_{30}\lambda_{03}\nonumber\\
&&+\lambda_{21}\lambda_{12}+\lambda_{22}\lambda_{11}
+\lambda_{31}\lambda_{02}+\lambda_{31}\lambda_{03}
+\lambda_{31}\lambda_{10}+\lambda_{31}\lambda_{11}
+\lambda_{31}\lambda_{20}+\lambda_{31}\lambda_{21}
+\lambda_{31}\lambda_{32}+\lambda_{31}\lambda_{33}
+3\lambda_{31}\lambda_{31})]\},\label{Eq_RG_lambda31}\\
\frac{\lambda_{02}}{dl}&=&\frac{l}{8\pi(d_{3}^{2}+d_{1}^{2})^{\frac{3}{2}}}
\{4d_{3}^{2}[\lambda_{01}\lambda_{03}+\lambda_{11}\lambda_{13}
+\lambda_{21}\lambda_{23}+\lambda_{31}\lambda_{33}
+\lambda_{02}\lambda_{00}
+\lambda_{02}\lambda_{10}+\lambda_{02}\lambda_{12}
+\lambda_{02}\lambda_{20}+\lambda_{02}\lambda_{22}\nonumber\\
&&
+\lambda_{02}\lambda_{30}+\lambda_{02}\lambda_{32}
-(\lambda_{01}\lambda_{00}+\lambda_{11}\lambda_{10}
+\lambda_{21}\lambda_{20}+\lambda_{31}\lambda_{30}
+\lambda_{02}\lambda_{01}+\lambda_{02}\lambda_{03}
+\lambda_{02}\lambda_{11}+\lambda_{02}\lambda_{13}
+\lambda_{02}\lambda_{21}\nonumber\\
&&\!+\!\lambda_{02}\lambda_{23}
\!+\!\lambda_{02}\lambda_{31}\!+\!\lambda_{02}\lambda_{33}
\!+\!3\lambda_{02}\lambda_{02})]
+d_{1}^{2}[ 4(\lambda_{01}\lambda_{03}
\!+\!\lambda_{11}\lambda_{13}\!+\!\lambda_{21}\lambda_{23}
\!+\!\lambda_{31}\lambda_{33})\!+\!\lambda_{02}\lambda_{00}
\!+\!\lambda_{02}\lambda_{10}+\lambda_{02}\lambda_{12}\nonumber\\
&&+\lambda_{02}\lambda_{20}+\lambda_{02}\lambda_{22}
+\lambda_{02}\lambda_{30}+\lambda_{02}\lambda_{32}
-(\lambda_{00}\lambda_{00}+\lambda_{10}\lambda_{10}
+\lambda_{20}\lambda_{20}+\lambda_{30}\lambda_{30}
+\lambda_{01}\lambda_{01}+\lambda_{11}\lambda_{11}
+\lambda_{21}\lambda_{21}\nonumber\\
&&+\lambda_{31}\lambda_{31}
+\lambda_{02}\lambda_{02}+\lambda_{12}\lambda_{12}+\lambda_{22}\lambda_{22}
+\lambda_{32}\lambda_{32}
+\lambda_{03}\lambda_{03}+\lambda_{13}\lambda_{13}
+\lambda_{23}\lambda_{23}+\lambda_{33}\lambda_{33})
-2(\lambda_{03}\lambda_{00}+\lambda_{13}\lambda_{10}\nonumber\\
&&+\lambda_{23}\lambda_{20}+\lambda_{33}\lambda_{30}
+\lambda_{02}\lambda_{01}+\lambda_{02}\lambda_{03}
+\lambda_{02}\lambda_{11}+\lambda_{02}\lambda_{13}
+\lambda_{02}\lambda_{21}+\lambda_{02}\lambda_{23}
+\lambda_{02}\lambda_{31}+\lambda_{02}\lambda_{33}
+3\lambda_{02}\lambda_{02})]\},\label{Eq_RG_lambda02}\\
\frac{d\lambda_{12}}{dl}&=&\frac{1}{4\pi(d_{3}^{2}+d_{1}^{2})^{\frac{3}{2}}}
\{2d_{3}^{2}[\lambda_{13}\lambda_{01}+\lambda_{11}\lambda_{03}
+\lambda_{32}\lambda_{20}+\lambda_{30}\lambda_{22}
+\lambda_{12}\lambda_{00}+\lambda_{12}\lambda_{02}
+\lambda_{12}\lambda_{10}+\lambda_{12}\lambda_{21}
+\lambda_{12}\lambda_{23}\nonumber\\
&&+\lambda_{12}\lambda_{31}
+\lambda_{12}\lambda_{33}-(\lambda_{11}\lambda_{00}
+\lambda_{10}\lambda_{01}
+\lambda_{32}\lambda_{23}+\lambda_{33}\lambda_{22}
+\lambda_{12}\lambda_{01}+\lambda_{12}\lambda_{03}
+\lambda_{12}\lambda_{11}+\lambda_{12}\lambda_{13}
+\lambda_{12}\lambda_{20}\nonumber\\
&&+\lambda_{12}\lambda_{22}
+\lambda_{12}\lambda_{30}+\lambda_{12}\lambda_{32}
+3\lambda_{12}\lambda_{12})]
+d_{1}^{2}[2(\lambda_{13}\lambda_{01}+\lambda_{11}\lambda_{03}
+\lambda_{32}\lambda_{20}+\lambda_{30}\lambda_{22})
+\lambda_{12}\lambda_{00}+\lambda_{12}\lambda_{02}\nonumber\\
&&+\lambda_{12}\lambda_{10}+\lambda_{12}\lambda_{21}
+\lambda_{12}\lambda_{23}+\lambda_{12}\lambda_{31}+\lambda_{12}\lambda_{33}
-(\lambda_{13}\lambda_{00}+\lambda_{10}\lambda_{03}
+\lambda_{31}\lambda_{22}+\lambda_{32}\lambda_{21}
+\lambda_{10}\lambda_{00}+\lambda_{11}\lambda_{01}\nonumber\\
&&
+\lambda_{12}\lambda_{02}+\lambda_{13}\lambda_{03}
+\lambda_{12}\lambda_{01}+\lambda_{12}\lambda_{03}
+\lambda_{12}\lambda_{11}+\lambda_{12}\lambda_{13}
+\lambda_{12}\lambda_{20}+\lambda_{12}\lambda_{22}
+\lambda_{12}\lambda_{30}+\lambda_{12}\lambda_{32}
+3\lambda_{12}\lambda_{12})]\},\label{Eq_RG_lambda12}\\
\frac{d\lambda_{22}}{dl}&=&\frac{1}{4\pi(d_{3}^{2}+d_{1}^{2})^{\frac{3}{2}}}
\{2d_{3}^{2}[\lambda_{23}\lambda_{01}+\lambda_{21}\lambda_{03}
+\lambda_{12}\lambda_{30}+\lambda_{10}\lambda_{32}
+\lambda_{22}\lambda_{00}+\lambda_{22}\lambda_{02}
+\lambda_{22}\lambda_{11}+\lambda_{22}\lambda_{13}
+\lambda_{22}\lambda_{20}\nonumber\\
&&+\lambda_{22}\lambda_{31}+\lambda_{22}\lambda_{33}
-(\lambda_{21}\lambda_{00}+\lambda_{20}\lambda_{01}
+\lambda_{12}\lambda_{33}+\lambda_{13}\lambda_{32}
+\lambda_{22}\lambda_{01}+\lambda_{22}\lambda_{03}
+\lambda_{22}\lambda_{10}+\lambda_{22}\lambda_{12}
+\lambda_{22}\lambda_{21}\nonumber\\
&&+\lambda_{22}\lambda_{23}+\lambda_{22}\lambda_{30}
+\lambda_{22}\lambda_{32}+3\lambda_{22}\lambda_{22})]
+d_{1}^{2}[2(\lambda_{23}\lambda_{01}+\lambda_{21}\lambda_{03}
+\lambda_{12}\lambda_{30}+\lambda_{10}\lambda_{32})
+\lambda_{22}\lambda_{00}+\lambda_{22}\lambda_{02}\nonumber\\
&&+\lambda_{22}\lambda_{11}+\lambda_{22}\lambda_{13}
+\lambda_{22}\lambda_{20}+\lambda_{22}\lambda_{31}
+\lambda_{22}\lambda_{33}-(\lambda_{23}\lambda_{00}
+\lambda_{20}\lambda_{03}
+\lambda_{11}\lambda_{32}+\lambda_{12}\lambda_{31}
+\lambda_{20}\lambda_{00}+\lambda_{21}\lambda_{01}\nonumber\\
&&+\lambda_{22}\lambda_{02}+\lambda_{23}\lambda_{03}
+\lambda_{22}\lambda_{01}+\lambda_{22}\lambda_{03}
+\lambda_{22}\lambda_{10}+\lambda_{22}\lambda_{12}
+\lambda_{22}\lambda_{21}+\lambda_{22}\lambda_{23}
+\lambda_{22}\lambda_{30}
+\lambda_{22}\lambda_{32}+3\lambda_{22}\lambda_{22})]\}
,\label{Eq_RG_lambda22}\\
\frac{d\lambda_{32}}{dl}&=&\frac{1}{4\pi(d_{3}^{2}+d_{1}^{2})^{\frac{3}{2}}}
\{2d_{3}^{2}[\lambda_{33}\lambda_{01}+\lambda_{31}\lambda_{03}
+\lambda_{22}\lambda_{10}+\lambda_{20}\lambda_{12}
+\lambda_{32}\lambda_{00}+\lambda_{32}\lambda_{02}
+\lambda_{32}\lambda_{11}+\lambda_{32}\lambda_{13}
+\lambda_{32}\lambda_{21}\nonumber\\
&&+\lambda_{32}\lambda_{23}+\lambda_{32}\lambda_{30}
-(\lambda_{31}\lambda_{00}+\lambda_{30}\lambda_{01}
+\lambda_{22}\lambda_{13}+\lambda_{23}\lambda_{12}
+\lambda_{32}\lambda_{01}+\lambda_{32}\lambda_{03}
+\lambda_{32}\lambda_{10}+\lambda_{32}\lambda_{12}
+\lambda_{32}\lambda_{20}\nonumber\\
&&+\lambda_{32}\lambda_{22}
+\lambda_{32}\lambda_{31}+\lambda_{32}\lambda_{33}
+3\lambda_{32}\lambda_{32})]
+d_{1}^{2}[2(\lambda_{33}\lambda_{01}+\lambda_{31}\lambda_{03}
+\lambda_{22}\lambda_{10}+\lambda_{20}\lambda_{12})
+\lambda_{32}\lambda_{00}+\lambda_{32}\lambda_{02}\nonumber\\
&&+\lambda_{32}\lambda_{11}
+\lambda_{32}\lambda_{13}+\lambda_{32}\lambda_{21}
+\lambda_{32}\lambda_{23}+\lambda_{32}\lambda_{30}
-(\lambda_{33}\lambda_{00}+\lambda_{30}\lambda_{03}
+\lambda_{21}\lambda_{12}+\lambda_{22}\lambda_{11}
+\lambda_{30}\lambda_{00}+\lambda_{31}\lambda_{01}\nonumber\\
&&+\lambda_{32}\lambda_{02}+\lambda_{33}\lambda_{03}
+\lambda_{32}\lambda_{01}+\lambda_{32}\lambda_{03}
+\lambda_{32}\lambda_{10}+\lambda_{32}\lambda_{12}
+\lambda_{32}\lambda_{20}+\lambda_{32}\lambda_{22}
+\lambda_{32}\lambda_{31}+\lambda_{32}\lambda_{33}
+3\lambda_{32}\lambda_{32})]\}
,\label{Eq_RG_lambda32}\\
\frac{d\lambda_{03}}{dl}&=&\frac{1}{4\pi(d_{3}^{2}+d_{1}^{2})^{\frac{3}{2}}}
\{d_{3}^{2}[2(\lambda_{02}\lambda_{01}+\lambda_{12}\lambda_{11}
+\lambda_{22}\lambda_{21}+\lambda_{32}\lambda_{31})
-(\lambda_{00}\lambda_{00}+\lambda_{10}\lambda_{10}
+\lambda_{20}\lambda_{20}+\lambda_{30}\lambda_{30}
+\lambda_{01}\lambda_{01}\nonumber\\
&&+\lambda_{11}\lambda_{11}
+\lambda_{21}\lambda_{21}+\lambda_{31}\lambda_{31}
+\lambda_{02}\lambda_{02}+\lambda_{12}\lambda_{12}
+\lambda_{22}\lambda_{22}+\lambda_{32}\lambda_{32}
+\lambda_{03}\lambda_{03}+\lambda_{13}\lambda_{13}
+\lambda_{23}\lambda_{23}+\lambda_{33}\lambda_{33})]\nonumber\\
&&+d_{1}^{2}[2(\lambda_{02}\lambda_{01}+\lambda_{12}\lambda_{11}
+\lambda_{22}\lambda_{21}+\lambda_{32}\lambda_{31}
+\lambda_{03}\lambda_{00}
+\lambda_{03}\lambda_{10}+\lambda_{03}\lambda_{13}
+\lambda_{03}\lambda_{20}+\lambda_{03}\lambda_{23}
+\lambda_{03}\lambda_{30}+\lambda_{03}\lambda_{33})\nonumber\\
&&-(\lambda_{01}\lambda_{00}+\lambda_{11}\lambda_{10}
+\lambda_{21}\lambda_{20}+\lambda_{31}\lambda_{30}
+\lambda_{02}\lambda_{00}+\lambda_{12}\lambda_{10}
+\lambda_{22}\lambda_{20}+\lambda_{32}\lambda_{30})
-2(\lambda_{03}\lambda_{01}+\lambda_{03}\lambda_{02}
+\lambda_{03}\lambda_{11}\nonumber\\
&&+\lambda_{03}\lambda_{12}
+\lambda_{03}\lambda_{21}+\lambda_{03}\lambda_{22}
+\lambda_{03}\lambda_{31}+\lambda_{03}\lambda_{32}
+3\lambda_{03}\lambda_{03})]\},\label{Eq_RG_lambda32}\\
\frac{d\lambda_{13}}{dl}&=&\frac{1}{4\pi(d_{3}^{2}+d_{1}^{2})^{\frac{3}{2}}}
\{2d_{3}^{2}[\lambda_{11}\lambda_{02}+\lambda_{12}\lambda_{01}
+\lambda_{33}\lambda_{20}+\lambda_{30}\lambda_{23}
-(\lambda_{10}\lambda_{00}+\lambda_{11}\lambda_{01}
+\lambda_{12}\lambda_{02}+\lambda_{13}\lambda_{03})]
+d_{1}^{2}[2(\lambda_{11}\lambda_{02}\nonumber\\
&&+\lambda_{12}\lambda_{01}
+\lambda_{33}\lambda_{20}+\lambda_{30}\lambda_{23}
+\lambda_{13}\lambda_{00}+\lambda_{13}\lambda_{03}
+\lambda_{13}\lambda_{10}+\lambda_{13}\lambda_{21}
+\lambda_{13}\lambda_{22}+\lambda_{13}\lambda_{31}
+\lambda_{13}\lambda_{32})-(\lambda_{12}\lambda_{00}\nonumber\\
&&
+\lambda_{10}\lambda_{02}+\lambda_{33}\lambda_{21}+\lambda_{31}\lambda_{23}
+\lambda_{11}\lambda_{00}+\lambda_{10}\lambda_{01}
+\lambda_{32}\lambda_{23}+\lambda_{33}\lambda_{22})
-2(\lambda_{13}\lambda_{01}+\lambda_{13}\lambda_{02}
+\lambda_{13}\lambda_{11}+\lambda_{13}\lambda_{12}\nonumber\\
&&+\lambda_{13}\lambda_{20}+\lambda_{13}\lambda_{23}
+\lambda_{13}\lambda_{30}+\lambda_{13}\lambda_{33}
+3\lambda_{13}\lambda_{13})]\},\label{Eq_RG_lambda13}\\
\frac{d\lambda_{23}}{dl}&=&\frac{1}{4\pi(d_{3}^{2}+d_{1}^{2})^{\frac{3}{2}}}
\{2d_{3}^{2}[\lambda_{21}\lambda_{02}+\lambda_{22}\lambda_{01}
+\lambda_{13}\lambda_{30}+\lambda_{10}\lambda_{33}
-(\lambda_{20}\lambda_{00}+\lambda_{21}\lambda_{01}
+\lambda_{22}\lambda_{02}+\lambda_{23}\lambda_{03})]
+d_{1}^{2}[2(\lambda_{21}\lambda_{02}\nonumber\\
&&+\lambda_{22}\lambda_{01}
+\lambda_{13}\lambda_{30}+\lambda_{10}\lambda_{33}
+\lambda_{23}\lambda_{00}+\lambda_{23}\lambda_{03}
+\lambda_{23}\lambda_{11}+\lambda_{23}\lambda_{12}+\lambda_{23}\lambda_{20}
+\lambda_{23}\lambda_{31}+\lambda_{23}\lambda_{32})
-(\lambda_{22}\lambda_{00}\nonumber\\
&&+\lambda_{20}\lambda_{02}+\lambda_{13}\lambda_{31}+\lambda_{11}\lambda_{33}
+\lambda_{21}\lambda_{00}+\lambda_{20}\lambda_{01}
+\lambda_{12}\lambda_{33}+\lambda_{13}\lambda_{32})
-2(\lambda_{23}\lambda_{01}+\lambda_{23}\lambda_{02}
+\lambda_{23}\lambda_{10}+\lambda_{23}\lambda_{13}\nonumber\\
&&
+\lambda_{23}\lambda_{21}+\lambda_{23}\lambda_{22}
+\lambda_{23}\lambda_{30}+\lambda_{23}\lambda_{33}
+3\lambda_{23}\lambda_{23})]\},\label{Eq_RG_lambda23}\\
\frac{d\lambda_{33}}{dl}&=&\frac{1}{4\pi(d_{3}^{2}+d_{1}^{2})^{\frac{3}{2}}}
\{2d_{3}^{2}[\lambda_{31}\lambda_{02}+\lambda_{32}\lambda_{01}
+\lambda_{23}\lambda_{10}+\lambda_{20}\lambda_{13}
-(\lambda_{30}\lambda_{00}+\lambda_{31}\lambda_{01}
+\lambda_{32}\lambda_{02}+\lambda_{33}\lambda_{03})]
+d_{1}^{2}[2(\lambda_{31}\lambda_{02}\nonumber\\
&&
+\lambda_{32}\lambda_{01}+\lambda_{23}\lambda_{10}
+\lambda_{20}\lambda_{13}+\lambda_{33}\lambda_{00}
+\lambda_{33}\lambda_{03}+\lambda_{33}\lambda_{11}
+\lambda_{33}\lambda_{12}+\lambda_{33}\lambda_{21}
+\lambda_{33}\lambda_{22}+\lambda_{33}\lambda_{30})
-(\lambda_{32}\lambda_{00}\nonumber\\
&&+\lambda_{30}\lambda_{02}
+\lambda_{23}\lambda_{11}+\lambda_{21}\lambda_{13}
+\lambda_{31}\lambda_{00}+\lambda_{30}\lambda_{01}
+\lambda_{22}\lambda_{13}+\lambda_{23}\lambda_{12})
-2(\lambda_{33}\lambda_{01}+\lambda_{33}\lambda_{02}
+\lambda_{33}\lambda_{10}+\lambda_{33}\lambda_{13}\nonumber\\
&&+\lambda_{33}\lambda_{20}+\lambda_{33}\lambda_{23}
+\lambda_{33}\lambda_{31}+\lambda_{33}\lambda_{32}
+3\lambda_{33}\lambda_{33})]\}.\label{Eq_RG_lambda33}
\end{eqnarray}
\end{small}

\section{RG equations of source terms}\label{Appendix_source}

After collecting all the one-loop corrections to source terms and performing
RG analysis~\cite{Murray2014PRB,Wang2017PRB,Roy2020-arxiv}, we can obtain RG flow
equations of the strengths $\Delta_{i}^{c/s}$ and $\Delta_{i}^{\mathrm{PP}}$
corresponding to fermion-source terms in particle-hole and particle-particle situations,
\begin{small}
\begin{eqnarray}
\frac{d\Delta^{c}_{1}}{dl}
&=&2\Delta^{c}_{1},\\
\frac{d\Delta^{c}_{2}}{dl}
&=&\Bigl[2+(\lambda_{00}-7\lambda_{01}-\lambda_{02}-\lambda_{03}
+\lambda_{10}+\lambda_{11}-\lambda_{12}-\lambda_{13}
+\lambda_{20}+\lambda_{21}-\lambda_{22}-\lambda_{23}
+\lambda_{30}+\lambda_{31}\nonumber\\
&&-\lambda_{32}-\lambda_{33})
\frac{(2d_{3}^{2}+d_{1}^{2})}{16\pi(d_{3}^{2}+d_{1}^{2})
^{\frac{3}{2}}}\Bigr]\Delta^{c}_{2},\\
\frac{d\Delta^{c}_{3}}{dl}
&=&\Bigl[2+(\lambda_{00}
-\lambda_{01}-7\lambda_{02}-\lambda_{03}
+\lambda_{10}-\lambda_{11}+\lambda_{12}-\lambda_{13}
+\lambda_{20}-\lambda_{21}+\lambda_{22}-\lambda_{23}
+\lambda_{30}-\lambda_{31}\nonumber\\
&&+\lambda_{32}-\lambda_{33})
\frac{(2d_{3}^{2}+d_{1}^{2})}{16\pi(d_{3}^{2}+d_{1}^{2})
^{\frac{3}{2}}}\Bigr]\Delta^{c}_{3},\\
\frac{d\Delta^{c}_{4}}{dl}
&=&\Bigl[2+(\lambda_{00}-\lambda_{01}-\lambda_{02}-7\lambda_{03}
+\lambda_{10}-\lambda_{11}-\lambda_{12}+\lambda_{13}
+\lambda_{20}-\lambda_{21}-\lambda_{22}+\lambda_{23}
+\lambda_{30}-\lambda_{31}\nonumber\\
&&-\lambda_{32}+\lambda_{33})
\frac{d_{1}^{2}}{8\pi(d_{3}^{2}+d_{1}^{2})^{\frac{3}{2}}}\Bigr]
\Delta^{c}_{4},\\
\frac{d\Delta^{s}_{1}}{dl}
&=&2\Delta^{s}_{1},\\
\frac{d\Delta^{s}_{2-x}}{dl}
&=&\Bigl[2+(\lambda_{00}+\lambda_{01}-\lambda_{02}-\lambda_{03}
+\lambda_{10}-7\lambda_{11}-\lambda_{12}-\lambda_{13}
-\lambda_{20}-\lambda_{21}+\lambda_{22}+\lambda_{23}
-\lambda_{30}-\lambda_{31}\nonumber\\
&&+\lambda_{32}+\lambda_{33})
\frac{(2d_{3}^{2}+d_{1}^{2})}{16\pi(d_{3}^{2}+d_{1}^{2})
^{\frac{3}{2}}}\Bigr]\Delta^{s}_{2-x},\\
\frac{d\Delta^{s}_{2-y}}{dl}
&=&\Bigl[2+(\lambda_{00}+\lambda_{01}-\lambda_{02}-\lambda_{03}
-\lambda_{10}-\lambda_{11}+\lambda_{12}+\lambda_{13}
+\lambda_{20}-7\lambda_{21}-\lambda_{22}-\lambda_{23}
-\lambda_{30}-\lambda_{31}\nonumber\\
&&+\lambda_{32}+\lambda_{33})
\frac{(2d_{3}^{2}+d_{1}^{2})}{16\pi(d_{3}^{2}+d_{1}^{2})
^{\frac{3}{2}}}\Bigr]\Delta^{s}_{2-y},\\
\frac{d\Delta^{s}_{2-z}}{dl}
&=&\Bigl[2+(\lambda_{00}+\lambda_{01}-\lambda_{02}-\lambda_{03}
-\lambda_{10}-\lambda_{11}+\lambda_{12}+\lambda_{13}
-\lambda_{20}-\lambda_{21}+\lambda_{22}+\lambda_{23}
+\lambda_{30}-7\lambda_{31}\nonumber\\
&&-\lambda_{32}-\lambda_{33})
\frac{(2d_{3}^{2}+d_{1}^{2})}{16\pi(d_{3}^{2}+d_{1}^{2})
^{\frac{3}{2}}}\Bigr]\Delta^{s}_{2-z},\\
\frac{d\Delta^{s}_{3-x}}{dl}
&=&\Bigl[2+(\lambda_{00}-\lambda_{01}+\lambda_{02}-\lambda_{03}
+\lambda_{10}-\lambda_{11}-7\lambda_{12}-\lambda_{13}
-\lambda_{20}+\lambda_{21}-\lambda_{22}+\lambda_{23}
-\lambda_{30}+\lambda_{31}\nonumber\\
&&-\lambda_{32}+\lambda_{33})
\frac{(2d_{3}^{2}+d_{1}^{2})}
{16\pi(d_{3}^{2}+d_{1}^{2})^{\frac{3}{2}}}\Bigr]
\Delta^{s}_{3-x},\\
\frac{d\Delta^{s}_{3-y}}{dl}
&=&\Bigl[2+(\lambda_{00}-\lambda_{01}+\lambda_{02}-\lambda_{03}
-\lambda_{10}+\lambda_{11}-\lambda_{12}+\lambda_{13}
+\lambda_{20}-\lambda_{21}-7\lambda_{22}-\lambda_{23}
-\lambda_{30}+\lambda_{31}\nonumber\\
&&-\lambda_{32}+\lambda_{33})
\frac{(2d_{3}^{2}+d_{1}^{2})}
{16\pi(d_{3}^{2}+d_{1}^{2})^{\frac{3}{2}}}\Bigr]
\Delta^{s}_{3-y},\\
\frac{d\Delta^{s}_{3-z}}{dl}
&=&\Bigl[2+(\lambda_{00}-\lambda_{01}+\lambda_{02}-\lambda_{03}
-\lambda_{10}+\lambda_{11}-\lambda_{12}+\lambda_{13}
-\lambda_{20}+\lambda_{21}-\lambda_{22}+\lambda_{23}
+\lambda_{30}-\lambda_{31}\nonumber\\
&&-7\lambda_{32}-\lambda_{33})
\frac{(2d_{3}^{2}+d_{1}^{2})
}{16\pi(d_{3}^{2}+d_{1}^{2})^{\frac{3}{2}}}\Bigr]
\Delta^{s}_{3-z},\\
\frac{d\Delta^{s}_{4-x}}{dl}
&=&\Bigl[2+(\lambda_{00}-\lambda_{01}-\lambda_{02}+\lambda_{03}
+\lambda_{10}-\lambda_{11}-\lambda_{12}-7\lambda_{13}
-\lambda_{20}+\lambda_{21}+\lambda_{22}-\lambda_{23}
-\lambda_{30}+\lambda_{31}\nonumber\\
&&+\lambda_{32}-\lambda_{33})
\frac{d_{1}^{2}}{8\pi(d_{3}^{2}+d_{1}^{2})^{\frac{3}{2}}}\Bigr]
\Delta^{s}_{4-x},\\
\frac{d\Delta^{s}_{4-y}}{dl}
&=&\Bigl[2+(\lambda_{00}-\lambda_{01}-\lambda_{02}+\lambda_{03}
-\lambda_{10}+\lambda_{11}+\lambda_{12}-\lambda_{13}
+\lambda_{20}-\lambda_{21}-\lambda_{22}-7\lambda_{23}
-\lambda_{30}+\lambda_{31}\nonumber\\
&&+\lambda_{32}-\lambda_{33})
\frac{d_{1}^{2}}{8\pi(d_{3}^{2}+d_{1}^{2})^{\frac{3}{2}}}\Bigr]
\Delta^{s}_{4-y},\\
\frac{d\Delta^{s}_{4-z}}{dl}
&=&\Bigl[2+(\lambda_{00}-\lambda_{01}-\lambda_{02}+\lambda_{03}
-\lambda_{10}+\lambda_{11}+\lambda_{12}-\lambda_{13}
-\lambda_{20}+\lambda_{21}+\lambda_{22}-\lambda_{23}
+\lambda_{30}-\lambda_{31}\nonumber\\
&&-\lambda_{32}-7\lambda_{33})
\frac{d_{1}^{2}}{8\pi(d_{3}^{2}+d_{1}^{2})^{\frac{3}{2}}}\Bigr]
\Delta^{s}_{4-z},
\end{eqnarray}
and
\begin{eqnarray}
\frac{d\Delta^{\mathrm{PP}}_{1}}{dl}
&=&\Bigl\{2+\Bigl[\lambda_{01}
+\lambda_{10}+\lambda_{12}+\lambda_{13}
+\lambda_{20}+\lambda_{22}+\lambda_{23}
+\lambda_{30}+\lambda_{32}+\lambda_{33}
-(\lambda_{00}+\lambda_{02}+\lambda_{03}+\lambda_{11}\nonumber\\
&&+\lambda_{21}+\lambda_{31})\Bigr]\frac{d_{3}^{2}}
{8\pi(d_{3}^{2}+d_{1}^{2})^{\frac{3}{2}}}\Bigr\}
\Delta^{\mathrm{PP}}_{1},\\
\frac{d\Delta^{\mathrm{PP}}_{2}}{dl}
&=&\Bigl\{2+\Bigl[\lambda_{03}+\lambda_{10}+\lambda_{11}+\lambda_{12}
+\lambda_{20}+\lambda_{21}+\lambda_{22}
+\lambda_{30}+\lambda_{31}+\lambda_{32}
-(\lambda_{00}+\lambda_{01}+\lambda_{02}
+\lambda_{13}\nonumber\\
&&+\lambda_{23}+\lambda_{33})\Bigr]
\frac{d_{1}^{2}}
{16\pi(d_{3}^{2}+d_{1}^{2})^{\frac{3}{2}}}\Bigr\}
\Delta^{\mathrm{PP}}_{2},\\
\frac{d\Delta^{\mathrm{PP}}_{3}}{dl}
&=&\Bigl\{2+\Bigl[\lambda_{02}+\lambda_{10}+\lambda_{11}+\lambda_{13}
+\lambda_{20}+\lambda_{21}+\lambda_{23}
+\lambda_{30}+\lambda_{31}+\lambda_{33}
-(\lambda_{00}+\lambda_{01}+\lambda_{03}+\lambda_{12}\nonumber\\
&&+\lambda_{22}+\lambda_{32})\Bigr]
\frac{1}
{8\pi(d_{3}^{2}+d_{1}^{2})^{\frac{1}{2}}}\Bigr\}
\Delta^{\mathrm{PP}}_{3},\\
\frac{d\Delta^{\mathrm{PP}}_{4-x}}{dl}
&=&\Bigl\{2+\Bigl[\lambda_{01}+\lambda_{02}+\lambda_{03}
+\lambda_{11}+\lambda_{12}+\lambda_{13}
+\lambda_{20}+\lambda_{31}+\lambda_{32}+\lambda_{33}
-(\lambda_{00}+\lambda_{10}+\lambda_{21}+\lambda_{22}\nonumber\\
&&+\lambda_{23}
+\lambda_{30})\Bigr]
\frac{d_{1}^{2}}
{16\pi(d_{3}^{2}+d_{1}^{2})^{\frac{3}{2}}}\Bigr\}
\Delta^{\mathrm{PP}}_{4-x},\\
\frac{d\Delta^{\mathrm{PP}}_{4-y}}{dl}
&=&\Bigl\{2+\Bigl[\lambda_{01}+\lambda_{02}+\lambda_{03}+\lambda_{11}
+\lambda_{12}+\lambda_{13}
+\lambda_{21}+\lambda_{22}+\lambda_{23}+\lambda_{30}
-(\lambda_{00}+\lambda_{10}+\lambda_{20}
+\lambda_{31}\nonumber\\
&&+\lambda_{32}+\lambda_{33})\Bigr]
\frac{d_{1}^{2}}
{16\pi(d_{3}^{2}+d_{1}^{2})^{\frac{3}{2}}}\Bigr\}
\Delta^{\mathrm{PP}}_{4-y},\\
\frac{d\Delta^{\mathrm{PP}}_{4-z}}{dl}
&=&\Bigl\{2+\Bigl[\lambda_{01}+\lambda_{02}+\lambda_{03}
+\lambda_{10}
+\lambda_{21}+\lambda_{22}+\lambda_{23}
+\lambda_{31}+\lambda_{32}+\lambda_{33}
-(\lambda_{00}+\lambda_{11}+\lambda_{12}+\lambda_{13}\nonumber\\
&&+\lambda_{20}+\lambda_{30})\Bigr]
\frac{d_{1}^{2}}
{16\pi(d_{3}^{2}+d_{1}^{2})^{\frac{3}{2}}}\Bigr\}
\Delta^{\mathrm{PP}}_{4-z},
\end{eqnarray}
\end{small}
where the right hand sides of these equations are designated as
$\mathcal{G}^{c/s,\mathrm{PP}}_i\Delta_{i}^{c/s,\mathrm{PP}}$ in Eq.~(\ref{RG_Eqs_source}).
\end{widetext}



\end{document}